\documentclass[doc]{apa7}

\usepackage[table]{xcolor}
\usepackage{amsmath,amsfonts,bm}
\usepackage{graphicx}
\usepackage{comment}
\usepackage{multicol,multirow,array,longtable, blkarray}
\usepackage{nicematrix}
\usepackage{enumitem}
\usepackage{float}
\usepackage{kbordermatrix}
\usepackage{booktabs}
\usepackage{placeins} 
\usepackage[style=apa]{biblatex}

\renewcommand{\arraystretch}{1.35}
\newcommand{\fMatrix}{%
  \kbordermatrix{
      & Y_{[p/m]1} & Y_{[p/m]2} \\
    F_{o1} & f_{11} & f_{12} \\
    F_{o2} & f_{21} & f_{22}
  }%
}

\newcommand{\thetaMatrix}{%
  \kbordermatrix{
      & [N]T_{[m]1} & [N]T_{[m]2} \\
    Y_{o1} & \theta_{11} & \theta_{12} \\
    Y_{o2} & \theta_{21} & \theta_{22}
  }%
}

\newcommand{\phiMatrix}{%
  \kbordermatrix{
      & [N]T_{[p/m]1} & [N]T_{[p/m]2} \\
    Y_{o1} & \frac{1}{2}\phi_{11} & \frac{1}{2}\phi_{12} \\
    Y_{o2} & \frac{1}{2}\phi_{21} & \frac{1}{2}\phi_{22}
  }%
}

\newcommand{\rhoMatrix}{%
  \kbordermatrix{
      & L[N]T_{[m]1} & L[N]T_{[m]2} \\
    Y_{o1} & \frac{1}{2}\rho_{11} & \frac{1}{2}\rho_{12} \\
    Y_{o2} & \frac{1}{2}\rho_{21} & \frac{1}{2}\rho_{22}
  }%
}

\newcommand{\muMatrix}{%
  \kbordermatrix{
      & Y_{m1} & Y_{m2} \\
    Y_{p1} & \mu_{11} & \mu_{12} \\
    Y_{p2} & \mu_{21} & \mu_{22}
  }%
}

\newcommand{\mateMatrix}{ \[ \begin{blockarray}{ccccc} & Y_{p1} & Y_{p2}& Y_{m1} & Y_{m2} \\ \begin{block}{c[cccc]} Y_{p1} & V_{Yp11} & V_{Yp12} & COV_{p1m1} & COV_{p1m2} \\ Y_{p2} & V_{Yp12} & V_{Yp22} & COV_{p2m1} & COV_{p2m2} \\ Y_{m1} & COV_{p1m1} & COV_{p2m1} & V_{Ym11} & V_{Ym12} \\ Y_{m2} & COV_{p1m2} & COV_{p2m2} & V_{Ym12} & V_{Ym22} \\ \end{block} \end{blockarray} \] }

\newcommand{\deltaMatrix}{%
  \kbordermatrix{
      & [N]T_{[m/p]1} & [N]T_{[m/p]2} \\
    Y_{m/p1} & \delta_{11} & 0 \\
    Y_{m/p2} & 0 & \delta_{22}
  }%
}

\newcommand{\aMatrix}{%
  \kbordermatrix{
      & L[N]T_{[m/p]1} & L[N]T_{[m/p]2} \\
    Y_{[m/p]1} & a_{11} & 0 \\
    Y_{[m/p]2} & 0 & a_{22}
  }%
}

\newcommand{\VYMatrix}{%
  \kbordermatrix{
     & Y_{1} & Y_{2} \\
   Y_{1} & V_{Y11} & V_{Y12} \\
   Y_{2} & V_{Y12} & V_{Y22}
  }%
}

\newcommand{\VFMatrix}{%
  \kbordermatrix{
     & F_{1} & F_{2} \\
   F_{1} & V_{F11} & V_{F12} \\
   F_{2} & V_{F12} & V_{F22}
  }%
}

\newcommand{\VeMatrix}{%
  \kbordermatrix{
     & \varepsilon_{1} & \varepsilon_{2} \\
   \varepsilon_{1} & V_{\epsilon11} & V_{\epsilon12} \\
   \varepsilon_{2} & V_{\epsilon12} & V_{\epsilon22}
  }%
}

\newcommand{\VAMatrix}{%
  \kbordermatrix{
     & Y_{1} & Y_{2} \\
   Y_{1} & V_{A11} & V_{A12} \\
   Y_{2} & V_{A12} & V_{A22}
  }%
}

\newcommand{\wMatrix}{%
  \kbordermatrix{
        & [N]T_{[p/m]1} & [N]T_{[p/m]2} \\
    F_{1} & \frac{1}{2}w_{11} & \frac{1}{2}w_{12} \\
    F_{2} & \frac{1}{2}w_{21} & \frac{1}{2}w_{22}
  }%
}

\newcommand{\vMatrix}{%
  \kbordermatrix{
        & L[N]T_{[p/m]1} & L[N]T_{[p/m]2} \\
    F_{1} & \frac{1}{2}v_{11} & \frac{1}{2}v_{12} \\
    F_{2} & \frac{1}{2}v_{21} & \frac{1}{2}v_{22}
  }%
}

\newcommand{\kMatrix}{%
  \kbordermatrix{
        & [N]T_{[p/m]1} & [N]T_{[p/m]2} \\
  {[N]T}_{[p/m]1} & k_{11} & k_{12} \\
  {[N]T}_{[p/m]2} & k_{12} & k_{22}
  }%
}

\newcommand{\jMatrix}{%
  \kbordermatrix{
            & L[N]T_{[p/m]1} & L[N]T_{[p/m]2} \\
  L[N]T_{[p/m]1} & j_{11} & j_{12} \\
  L[N]T_{[p/m]2} & j_{12} & j_{22}
  }%
}

\newcommand{\gtMatrix}{%
  \kbordermatrix{
        & [N]T_{[p/m]1} & [N]T_{[p/m]2} \\
  {[N]T}_{[p/m]1} & g_{\boldsymbol{t}11} & g_{\boldsymbol{t}12} \\
  {[N]T}_{[p/m]2} & g_{\boldsymbol{t}21} & g_{\boldsymbol{t}22}
  }%
}

\newcommand{\gcMatrix}{%
  \kbordermatrix{
        & [N]T_{[p/m]1} & [N]T_{[p/m]2} \\
  {[N]T}_{[p/m]1} & g_{\boldsymbol{c}11} & g_{\boldsymbol{c}12} \\
  {[N]T}_{[p/m]2} & g_{\boldsymbol{c}12} & g_{\boldsymbol{c}22}
  }%
}

\newcommand{\htMatrix}{%
  \kbordermatrix{
            & L[N]T_{[p/m]1} & L[N]T_{[p/m]2} \\
  L[N]T_{[p/m]1} & h_{\boldsymbol{t}11} & h_{\boldsymbol{t}12} \\
  L[N]T_{[p/m]2} & h_{\boldsymbol{t}21} & h_{\boldsymbol{t}22}
  }%
}

\newcommand{\hcMatrix}{%
  \kbordermatrix{
            & L[N]T_{[p/m]1} & L[N]T_{[p/m]2} \\
  L[N]T_{[p/m]1} & h_{\boldsymbol{c}11} & h_{\boldsymbol{c}12} \\
  L[N]T_{[p/m]2} & h_{\boldsymbol{c}12} & h_{\boldsymbol{c}22}
  }%
}

\newcommand{\icMatrix}{%
  \kbordermatrix{
        & [N]T_{[p/m]1} & [N]T_{[p/m]2} \\
  L[N]T_{[p/m]1} & i_{\boldsymbol{c}11} & i_{\boldsymbol{c}12} \\
  L[N]T_{[p/m]2} & i_{\boldsymbol{c}21} & i_{\boldsymbol{c}22}
  }%
}

\title{The Multivariate SEM-PGS Model: Using Polygenic Scores to Investigate Cross-Trait Genetic Nurture and Assortative Mating}
\shorttitle{Multivariate SEM-PGS Approach}
\authorsnames[{1,2},3,1,{1,2}]{Xuanyu Lyu, Jared Balbona, Tong Chen, Matthew C. Keller}
\authorsaffiliations{{Institute for Behavioural Genetics, University of Colorado Boulder},
{Department of Psychology and Neuroscience, University of Colorado Boulder},
{Virginia Institute for Psychiatric and Behavioral Genetics, Richmond, VA} }

\abstract{Genetic nurture effects and assortative mating (AM) occur across a range of human behaviors and can bias estimates from traditional genetic models. These influences are typically studied univariately, within the same trait. However, estimation of cross-trait genetic nurture effects and cross-trait AM remains underexplored due to the absence of suitable approaches. To address this, we developed a multivariate extension of the SEM–PGS model for datasets with genotyped and phenotyped parents and offspring, enabling joint estimation of within-trait and cross-trait genetic and environmental influences. By integrating haplotypic polygenic scores (PGSs) into a structural equation modeling framework, the model simultaneously estimates same-trait and cross-trait direct effects, genetic nurture, vertical transmission, and AM. We also provide the first formal description of how copaths can be used to model multivariate AM, and we derive the corresponding parameter expectations in matrix form. Forward-time Monte Carlo simulations under varying conditions of $r^2_{pgs}$ and $N_{trio}$ demonstrate that the model yields unbiased estimates of both within-trait and cross-trait effects when assumptions were met. The precision of estimates was adequate with large sample sizes ($N_{trio} > 16k$) and improved as PGS predictive power increased. In addition, our simulation results show that failing to model cross-trait effects biases within-trait estimates, underscoring the importance of incorporating cross-trait effects. The multivariate SEM-PGS model offers a powerful and flexible tool for disentangling gene-environment interplay and advancing the understanding of familial influences on human traits.}

\begin{document}

\maketitle

Individuals within a family are phenotypically more similar to each other than to a random group of people \autocite{kendler_family_2015,polderman_meta-analysis_2015}. This resemblance has long been recognized and arises from multiple sources, including shared rearing environment, parental influences, cultural values, and genetic inheritance \autocite{polderman_meta-analysis_2015, engzell_heritability_2019, abdellaoui_geneenvironment_2022}. Quantifying the reasons for similarity in specific traits remains challenging because genetic influences are deeply intertwined with environmental factors. For example, in nuclear families, parent-offspring covariance is caused by direct additive genetic effects of the genotype, half of which is shared between parent and offspring, by environmental influences shared between parent and offspring, and by vertical transmission (VT), in which parental phenotypes causally influence offspring through the rearing environment. \textit{Genetic nurture} refers to the influence of parental genetic effects—including alleles not transmitted to offspring—on parental traits that in turn affect offspring outcomes through VT \autocite{kong_nature_2018,balbona_estimation_2021}. Because parental phenotypes often reflect both genetic and environmental factors, VT also induces a covariance between direct genetic effects and the familial environment \autocite{cloninger_interpretation_1980}, further increasing parent-offspring covariance. Additional covariance can also arise from assortative mating (AM), which increases the similarity between parents’ and extended family members' genetic and environmental influences. Disentangling these processes is challenging because they often generate overlapping patterns of phenotypic covariance among family members.

Human genetic researchers have a long history of using structural equation modeling (SEM) to disentangle the sources of familial resemblance \autocite{wright_method_1934, cloninger_interpretation_1980, keller_modeling_2009, neale_methodology_2013, lyu_detecting_2025}. SEM is a statistical framework used to test a hypothesized causal model using observed data. SEM finds the variances and covariances among variables that are implied by a hypothesized data-generating mechanism and compares these to observed data, typically using maximum likelihood methods \autocite{kaplan_structural_2008}. By operationalizing constructs such as genetic and environmental factors as latent variables within an SEM, complex intrafamilial processes can be explicitly modeled. This approach makes it possible to partition variation in traits into genetic and environmental components, to model parental influences such as VT, and to test and account for processes like AM or gene-environment (G-E) covariance \autocite{keller_modeling_2009}. More broadly, SEM is valuable because it forces explicit specification of causal assumptions, directs attention to effect sizes rather than significance tests, and allows for complex extensions---such as the modeling of recursive relationships or incorporating multivariate data---that would otherwise be mathematically intractable \autocite{heath_resolution_1985, neale_methodology_2013, balbona_estimation_2021}.

Classical twin designs have traditionally been used to partition phenotypic variation into additive and dominant genetic effects, shared environmental influences, and unique environmental influences, but their estimates can be biased when both non-additive genetic and shared environmental influences simultaneously influence the trait \autocite{keller_quantifying_2005}, or when VT, genetic nurture, G-E covariance, or AM are present but not explicitly modeled. Family designs incorporating additional relatives, such as adoption and extended twin family designs, have improved estimation of these effects \autocite{lyu_detecting_2025, keller_modeling_2009}. Adoption studies can isolate direct genetic effects and VT, but additional biases can be introduced if prenatal effects, selective placement, and/or ongoing contact between biological parents and adoptees are not appropriately accounted for \autocite{horn_intellectual_1979,rutter_testing_2001,shih_review_2004}. Extended twin family designs allow estimation of genetic nurture and AM alongside direct genetic effects, using data from twins’ parents, spouses, and children to account for and mutually estimate these various influences \autocite{keller_modeling_2009, heath_resolution_1985,maes_genetic_1997}. A hallmark of extended twin family models is the use of non-linear constraints, which describe and constrain recursive relationships between parameters in a way that keeps the overall model internally consistent and identified (e.g., AM and G–E covariance; \nptextcite{keller_modeling_2009}). This feature illustrates one of the key advantages of using SEM for modeling familial effects, as it allows recursive processes—such as cases where A influences B, which in turn influences A—to be represented explicitly and estimated within a coherent framework. However, extended twin family designs require large samples to estimate latent effects with precision; obtaining such samples is difficult, and the models rely on strong assumptions about the sources of phenotypic covariance; violations of these assumptions can lead to biased estimates. The growing availability of genomic data offers a complementary and potentially less assumption-dependent alternative: by combining parent and offspring phenotypes with measured parental polygenic scores (PGSs)---the sum of trait-associated alleles weighted by their effect sizes from independent genome-wide association studies (GWASs)---it is possible to directly model the relative contributions of genetic sharing and VT \autocite{balbona_estimation_2021}. This general approach also facilitates genetically informed research in large, publicly available family genomic datasets, which are more widely available than extended twin family data.

A number of models incorporating PGSs into family models have now been developed and tested across a range of traits \autocite{mcadams_annual_2023}. For example, PGSs were incorporated in twin designs to investigate G-E covariance (\nptextcite{dolan_incorporating_2021}) and to test causal influences between traits \autocite{castro-de-araujo_mr-doc2_2023}. \textcite{demange_estimating_2022} used parental PGSs as measured genetic predictors to estimate the environmental, non-genetic effects of parents’ cognitive and non-cognitive skills on their offspring’s educational outcomes, thereby controlling for genetic transmission. \textcite{kong_nature_2018} provided the first empirical evidence for genetic nurture, showing that parental PGSs based on non-transmitted alleles, estimated using haplotype-based PGSs from both parents, predict offspring traits via the environment provided by the parent. Building on this approach, the SEM-PGS model of \textcite{balbona_estimation_2021} quantifies and accounts for confounding from AM and estimates the total effect of VT, defined as the environmental influence of a parental trait on an offspring trait. Thus, genomic data from large family samples, along with increasingly predictive PGSs, provide information that enables estimation of quantities previously inaccessible or allows them to be estimated in novel ways.

A key limitation of existing PGS–family models is that, with few exceptions—such as the MR-DoC2 model of \textcite{castro-de-araujo_mr-doc2_2023}, which estimates bidirectional causation—they are univariate, focusing on a single phenotype. The lack of multivariate models creates two challenges. First, it limits the ability to investigate cross-trait effects. In intergenerational transmission, for example, one might ask how parental mental health influences offspring educational attainment, accounting for direct genetic effects, genetic nurture, AM, and G-E covariance of each trait. Univariate approaches cannot address such questions adequately; modeling them as univariate problems, as has sometimes been done \autocite{kong_nature_2018}, ignores pleiotropy as well as within- and cross-trait AM and VT, and can therefore produce seriously biased cross-trait estimates, as we show below. Second, ignoring cross-trait effects can also bias within-trait estimates. For instance, when estimating the genetic nurture effect of parental BMI on offspring BMI, failing to account for parental educational attainment—which also affects offspring BMI—may upwardly bias the estimated genetic nurture pathway. Given the well-documented phenotypic, genetic, and environmental correlations between traits \autocite{bulik-sullivan_atlas_2015, cross-disorder_group_of_the_psychiatric_genomics_consortium_genetic_2013} and the long tradition of multivariate research in behavioral genetics \autocite{hart_nurture_2021, mcadams_accounting_2014}, it is essential to develop PGS–family models that can estimate and account for cross-trait influences, as univariate models are inherently vulnerable to bias from unmodeled pathways. To address this gap, we introduce the multivariate SEM-PGS model—an extension of our prior univariate SEM-PGS framework \autocite{balbona_estimation_2021, kim_bias_2021}—designed to estimate cross-trait familial influences both within and across generations. Below, we present the structure of the model and introduce novel multivariate path-tracing rules. We then use simulations to demonstrate the importance of accounting for multivariate influences and to evaluate the statistical properties of the model’s estimates under two different approaches to fitting the model.

\section{Multivariate SEM-PGS Model}

\begin{figure}
    \centering
    \includegraphics[width=\linewidth]{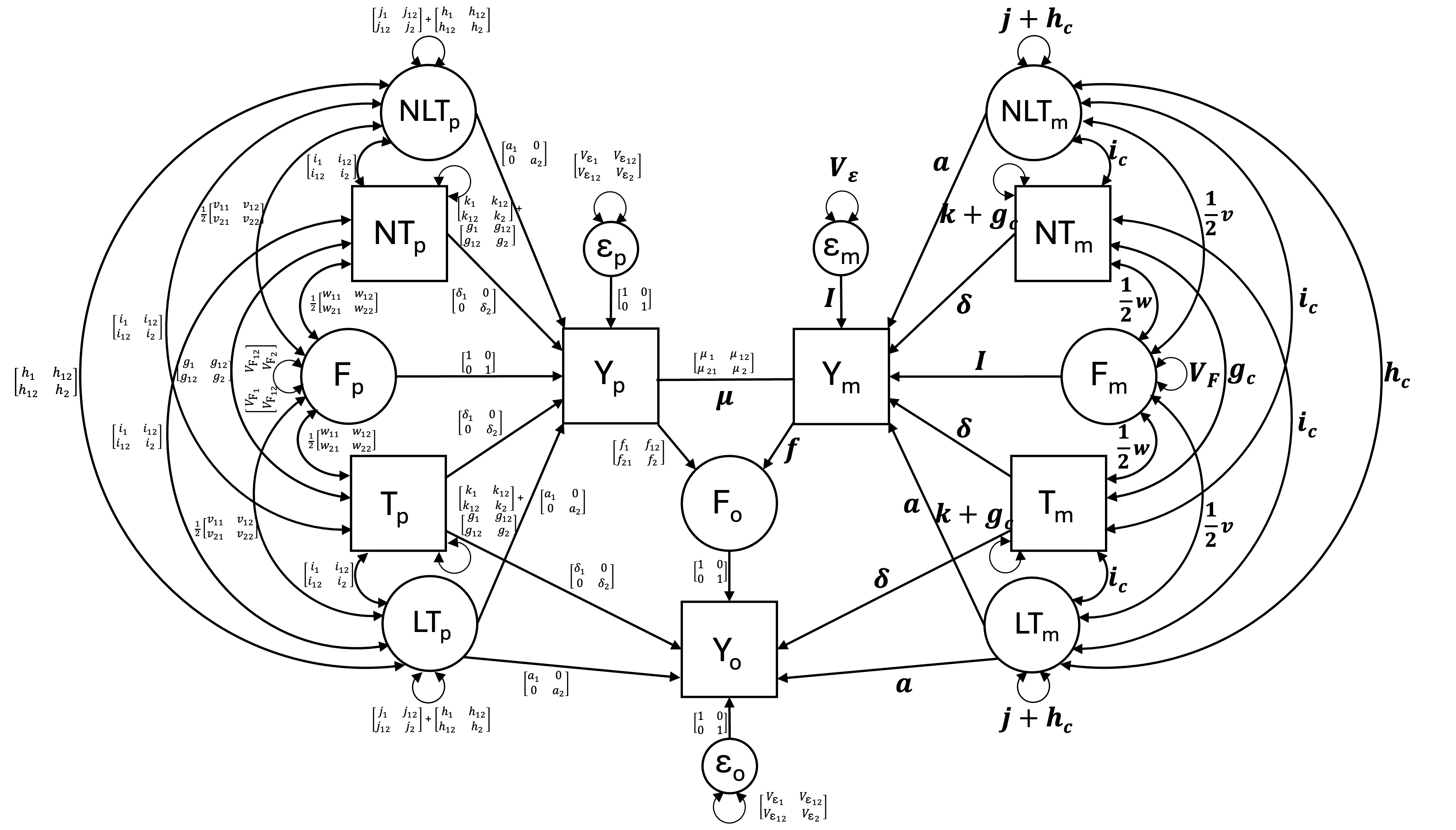}
    \caption{Path Diagram of Bivariate SEM-PGS Model. Parameters on the left side of the model are displayed in their matrix form.}
    \label{fig:BiSEMPGS}
\end{figure}

\begin{table}[!ht]
    %\centering
    \caption{Variable notations in bivariate SEM-PGS model}
    
    \begin{NiceTabular}{m{4.8cm} m{10cm}}
    \CodeBefore
    \rowcolors{2}{white}{gray!20}
    \Body
        \toprule
        \textbf{Parameter} & \textbf{Interpretation} \\
        \midrule
        $\overrightarrow{Y}_{p/m/o} = \begin{bmatrix}
            Y_{1} & Y_{2} 
        \end{bmatrix}$ & Observed: Phenotypic scores of two paternal, maternal, or offspring traits \\
        
        $\overrightarrow\varepsilon_{p/m/o} = \begin{bmatrix}
            \varepsilon_{1} & \varepsilon_{2} 
        \end{bmatrix}$ & Latent Factor: Residual scores of two paternal, maternal, or offspring traits \\  

        $\overrightarrow{F}_{p/m/o} = \begin{bmatrix}
            F_{1} & F_{2} 
        \end{bmatrix}$ & Latent Factor: Family environmental scores of the two paternal, maternal, or offspring traits, arising from $Y_{p/m} \rightarrow F_o$, vertical transmission (VT) \\

        $\overrightarrow{T}_{p/m} = \begin{bmatrix}
            T_{1} & T_{2} 
        \end{bmatrix}$ & Observed: Haplotypic PGSs of the two traits, constructed from paternal or maternal transmitted alleles \\  
        
        $\overrightarrow{NT}_{p/m} = \begin{bmatrix}
            NT_{1} & NT_{2} 
        \end{bmatrix}$ & Observed: Haplotypic PGSs of the two traits, constructed from paternal or maternal non-transmitted alleles \\ 

        $\overrightarrow{LT}_{p/m} = \begin{bmatrix}
            LT_{1} & LT_{2} 
        \end{bmatrix}$ & Latent Factor: Haplotypic latent genetic scores (LGS) of the two traits from paternal or maternal transmitted alleles \\ 
        
        $\overrightarrow{LNT}_{p/m} = \begin{bmatrix}
            LNT_{1} & LNT_{2} 
        \end{bmatrix}$ & Latent Factor: Haplotypic LGS of the two traits from paternal or maternal transmitted alleles \\

    \bottomrule
    \end{NiceTabular}
    \label{tab:notations}
\end{table}

\begin{table}[ht]
    %\centering
    \caption{Path coefficient matrices in bivariate SEM-PGS model}
    \label{tab:path}
    \begin{NiceTabular}{m{7cm} m{9cm}}
    \CodeBefore
    \rowcolors{2}{white}{gray!20}
    \Body
        \toprule
        \textbf{Parameter} & \textbf{Interpretation} \\ 
        \midrule        
      $\boldsymbol{\mu} = \muMatrix$ &
      Primary phenotypic assortative mating (AM) copath coefficients. 
      $\mu_{11}$ and $\mu_{22}$ indicate assortment on trait 1 and trait 2, respectively.
      $\mu_{21}$ indicates the assortment between maternal trait 1 and paternal trait 2.
      $\mu_{12}$ indicates the assortment between paternal trait 1 and maternal trait 2. \\
        
        $\mathbf{f} = \fMatrix$ & Vertical transmission (VT) path coefficients; the causal effect of $Y_{p/m} \rightarrow F_o$. $f_{11}$ and $f_{22}$ indicate VT within trait 1 and trait 2, respectively.  $f_{21}$ indicates VT from parental trait 1 to offspring trait 2. $f_{12}$ indicates VT from parental trait 2 to offspring trait 1.\\

        $\boldsymbol{\delta} = \deltaMatrix$ & $\delta_{1}$ and $\delta_{2}$ indicate the direct additive genetic effect of the haplotypic PGS on $Y_1$ and $Y_2$  respectively. Note: $[N]T$ denotes either $NT$ or $T$, as both share the same matrix content.\\
        
        $\mathbf{a} = \aMatrix$ & $a_{1}$ and $a_{2}$ indicate the direct additive genetic effect of haplotypic LGS on $Y_1$ and $Y_2$ respectively. \\

        \bottomrule
    \end{NiceTabular}

\end{table}

\newpage
\FloatBarrier
% zebra stripes for the body
\rowcolors{2}{white}{gray!20}
\begin{longtable}{>{\centering\arraybackslash}m{7cm} m{10cm}}
\caption{Variance and covariance matrices in bivariate SEM-PGS Model}\label{tab:var}\\
\toprule
\textbf{Parameter} & \textbf{Interpretation} \\
\midrule
\endfirsthead

\multicolumn{2}{c}{\tablename~\thetable{} (continued)}\\
\toprule
\textbf{Parameter} & \textbf{Interpretation} \\
\midrule
\endhead

\midrule
\multicolumn{2}{r}{\small Continued on next page}\\
\endfoot

\bottomrule
\endlastfoot

\showrowcolors
$\boldsymbol{V}_{\boldsymbol{Y}} = \VYMatrix$ &
  Phenotypic variance-covariance matrix for the two traits. \\

$\boldsymbol{V}_{\boldsymbol{F}} = \VFMatrix$ &
  Family environmental variance-covariance matrix due to VT. \\

 $\boldsymbol{V}_{\boldsymbol{A}} = \VAMatrix$ &
  Genetic variance-covariance matrix due to direct genetic effects. \\

$\boldsymbol{V}_{\boldsymbol{\epsilon}} = \VeMatrix$ &
  Residual variance-covariance matrix of residual factors not explained by other parameters. \\

$\frac{1}{2}\boldsymbol{w} = \wMatrix$ &
  Gene-environment (G-E) covariance between the haplotypic PGS and $F$. $\frac{1}{2}w_{11}$ and $\frac{1}{2}w_{22}$: within-trait G-E covariance.
  $\frac{1}{2}w_{21}$: Covariance between the haplotypic PGS of trait 1 and the family environment of trait 2.
  $\frac{1}{2}w_{12}$: Covariance between the haplotypic PGS of trait 2 and the family environment of trait 1. \\

$\frac{1}{2}\boldsymbol{v} = \vMatrix$ &
  G-E covariance between the haplotypic LGS and $F$. $\frac{1}{2}v_{11}$ and $\frac{1}{2}v_{22}$: within-trait G-E covariance effects.
  $\frac{1}{2}v_{21}$: Covariance between the haplotypic LGS of trait 1 and the familial environment of trait 2.
  $\frac{1}{2}v_{12}$: Covariance between the haplotypic LGS of trait 2 and the familial environment of trait 1. \\

$\frac{1}{2}\boldsymbol{\phi} = \phiMatrix$ &
  Genetic nurture pathways captured by haplotypic PGSs. The pathway represents the effects of parental PGSs on offspring phenotypes mediated by parental phenotypic effects on the family environments. $\frac{1}{2}\phi_{11}$ and $\frac{1}{2}\phi_{22}$: within-trait genetic nurture effects.
  $\frac{1}{2}\phi_{21}$: Genetic nurture effect of the haplotypic PGS of trait 1 on the offspring trait 2.
  $\frac{1}{2}\phi_{12}$: Genetic nurture effect of the haplotypic PGS of trait 2 on the offspring trait 1. \\

$\frac{1}{2}\boldsymbol{\rho} = \rhoMatrix$ &
  Genetic nurture pathways captured with haplotypic LGSs. The pathway represents the effects of parental LGSs on offspring phenotypes mediated by parental phenotypes and family environments. $\frac{1}{2}\rho_{11}$ and $\frac{1}{2}\rho_{22}$: within-trait genetic nurture effects.
  $\frac{1}{2}\rho_{21}$: Genetic nurture effect of the haplotypic LGS of trait 1 on the offspring trait 2.
  $\frac{1}{2}\rho_{12}$: Genetic nurture effect of the haplotypic LGS of trait 2 on the offspring trait 1. \\

$\boldsymbol{k} = \kMatrix$ &
  Covariance matrix of the haplotypic PGSs for two traits in the base population. $k_{11}$ and $k_{22}$ are the variance of
  haplotypic PGSs in the base population for trait 1 and trait 2, respectively. $k_{12}$ covariance (pleiotropy) between two traits in the base population captured by the haplotypic PGS. \\

$\boldsymbol{j} = \jMatrix$ &
  Covariance matrix of the latent haplotypic genetic scores for two traits in the base population. $j_{11}$ and $j_{22}$ are the
  variance of latent genetic scores in the base population for trait 1 and trait 2, respectively. $j_{12}$ is the covariance (pleiotropy) between two traits in the base population captured by the LGS. \\

$\boldsymbol{g}_{\boldsymbol{t}} = \gtMatrix$ &
  The AM-induced increase in the cross-mate "trans" covariance of the haplotypic PGSs of two traits.\\

$\boldsymbol{g}_{\boldsymbol{c}} = \gcMatrix$ &
    The AM-induced increase in the within-person "cis" (co)variance of the haplotypic PGSs of two traits. \\

$\boldsymbol{h}_{\boldsymbol{t}} = \htMatrix$ &
  The AM-induced increase in the cross-mate "trans" covariance of the haplotypic LGSs of two traits. \\

$\boldsymbol{h}_{\boldsymbol{c}} = \hcMatrix$ &
  The AM-induced increase in the within-person "cis" (co)variance of the haplotypic LGSs of two traits. \\

$\boldsymbol{i}_{\boldsymbol{c}} = \icMatrix$ &
  The AM-induced increase in the within-person covariance between the haplotypic PGS and the haplotypic LGSs of two traits.  \\
\end{longtable}

The multivariate SEM-PGS model jointly estimates and accounts for within-trait and cross-trait effects—including direct genetic effects, VT, genetic nurture, and AM—within a unified structural equation modeling (SEM) framework. For clarity of presentation, we focus on a model involving two traits throughout this paper. While extending the model to include three or more traits is mathematically straightforward, given that expectations are derived using matrices rather than scalar equations, such extensions substantially increase model complexity and the number of parameters to be estimated. This added complexity makes interpretation and communication of results much more challenging. Nevertheless, the qualitative conclusions drawn from the bivariate simulations generalize to higher-dimensional models. 

The path diagram of the bivariate model is displayed in Figure \ref{fig:BiSEMPGS}. The identification of model parameters requires seven vectors of observed components (Table \ref{tab:notations}): three vectors of phenotypic scores ($\overrightarrow{Y}_p, \overrightarrow{Y}_m, \overrightarrow{Y}_o$) and four vectors of haplotypic PGSs ($\overrightarrow{T}_{p}, \overrightarrow{T}_{m}, \overrightarrow{NT}_{p}, \overrightarrow{NT}_{m}$), all of length two. To simplify notation, we write $[N]T_{p/m}$ to indicate any of $NT_p$, $T_p$, $NT_m$, or $T_m$, where subscripts $p$ and $m$ refer to paternal and maternal haplotypes, respectively, and where $NT$ stands for the haplotypic PGS not transmitted and $T$ for the haplotypic PGS transmitted to offspring. For a bivariate model, there are therefore fourteen observed statistics, forming a 14x14 observed variance-covariance matrix. The definitions of all estimated variance and covariance parameters involved in the multivariate SEM are listed in Table \ref{tab:var}, and the definitions of path coefficients are listed in Table \ref{tab:path}. 

\subsubsection{Multivariate Path-Tracing Rules}

We derived the expected variance-covariance equations using multivariate path-tracing rules \autocite{vogler_multivariate_1985}. Multivariate path-tracing in SEM is based on the well-known univariate rules \autocite{cloninger_interpretation_1980,wright_method_1934}, with some additional rules regarding matrix transposition that are necessary to obtain the correct expectations. The 'copath' concept was introduced in univariate SEM to model selection processes such as AM, in which two variables covary without sharing a common antecedent cause \autocite{cloninger_interpretation_1980}. Formal path-tracing rules for copaths in multivariate SEM have not yet been described and require special considerations. Here, we review the multivariate path-tracing rules and extend them to incorporate copaths. Our focus is on the multivariate case, which builds directly on the univariate rules. Readers unfamiliar with the univariate path-tracing rules are encouraged to review them first, as this background will clarify the logic and interpretation of the multivariate extensions \autocite{balbona_estimation_2021}.

\noindent \textbf{Multivariate path tracing rules:}
To aid comprehension of the rules, we first introduce two heuristic definitions: \textit{upstream variable} and \textit{downstream variable}. For variables connected by a single-headed arrow, the variable the arrow points to is defined as downstream, and the variable from which the arrow originates is defined as upstream, as this convention is intuitively consistent with how causation flows from one variable to the next. Unlike "exogenous" and "endogenous" terms in SEM, which are qualities of variables themselves, "upstream" and "downstream" are relative terms that refer to how variables relate to one another: a variable may be upstream of one variable and downstream of another. With these definitions in mind, we list below the multivariate path tracing rules for all components in an SEM model:
\begin{itemize}
    \item \underline{Rules for path coefficient matrices}: A path coefficient matrix is transposed when traversing from the upstream variable to the downstream variable (with the direction of the arrow) \autocite{wright_method_1934}. It is untransposed when traversing from the downstream variable to the upstream variable (against the direction of the arrow). If a path coefficient matrix is a full matrix, the upstream variable should be placed along the columns of the matrix, and the downstream variable should be placed along the row of the matrix. For example, $\mathbf{f}$ is a full path coefficient matrix that connects the upstream $Y_{p/m}$ variable to the downstream $F_o$ variable ($Y_{p/m} \xrightarrow{\mathbf{f}} F_o$), and thus
    \[
    \mathbf{f}  = \fMatrix
    \]
    \item \underline{Rules for covariance matrices}: A covariance matrix is represented by a double-headed arrow in path diagrams.Variables connected by double-headed arrows are not truly upstream or downstream of one another—they occupy the same level in a causal pathway. However, because covariance matrices can sometimes be full rather than symmetric (due to them being a subsection of a larger symmetric variance-covariance matrix), a transposition rule is needed. Thus, as with other matrices, a full covariance matrix is transposed when traversing from the upstream variable to the downstream variable. As long as users apply a consistent convention when arbitrarily designating such variables as upstream or downstream, the path-tracing rules described here will yield the correct variance–covariance expectations. For example, one possible approach, and the one we recommend, is to consistently assign the parental variables as upstream of offspring ones and female spouses upstream of male spouses. As with other path coefficient matrices, upstream variables should be placed along the columns and downstream variables along the rows. For example, if we define the covariance between $\overrightarrow{T}_{m}$ (upstream) and $\overrightarrow{Y}_o$ (downstream) as $\mathbf{\boldsymbol{\theta}_{Tm}}$, then  
    \[
    \\\boldsymbol{\theta}_{Tm}  = \thetaMatrix
    \]
    so when doing the path tracing, we have a direct genetic effect path $\boldsymbol{\delta}\mathbf{k}$, four paths through the increased genetic covariance from AM $2\boldsymbol{\delta}\mathbf{g_c} + 2\mathbf{a}\mathbf{i_c}$ and a path through G-E covariance  $\frac{1}{2}\mathbf{w}$. Adding up all paths lead us to the correct covariance $\boldsymbol{\theta}_{Tm} = \boldsymbol{\delta}\mathbf{k} + 2\boldsymbol{\delta}\mathbf{g_c} + 2\mathbf{a}\mathbf{i_c} + \frac{1}{2}\mathbf{w}$. 
    \item \underline{Rules for copath matrices}: The copath models multivariate AM and is represented by a straight line, without arrows, connecting two variables (here, $Y_p$ and $Y_m$). It is a full matrix, and is transposed when traversing from the upstream variable to the downstream variable. As with a covariance matrix, one variable connected by the copath should be designated as the upstream variable and the other as the downstream variable, and this convention should be applied consistently throughout the derivation of all parameter expectations. Here, we adopt the convention that the maternal spouse is upstream of the paternal spouse. As with covariance matrices, the upstream variable should be placed along the columns and the downstream variable along the rows. For example, we define $\mathbf{Y}_m$ as the upstream variable and $\mathbf{Y}_p$ as the downstream variable, and thus the $\boldsymbol{\mu}$ matrix will be
    \[
    \boldsymbol{\mu} = \muMatrix
    \]
    A copath can only be traversed once within a given chain, and a chain must be legitimate prior to traversing the copath. In essence, a copath serves to connect two legitimate chains, thereby creating one longer chain. For example, $COV(\mathbf{Y}_p, \mathbf{Y}_m) = \mathbf{V}_{Yp} \boldsymbol{\mu} \mathbf{V}_{Ym}$
    \item \underline{Rules for path-tracing multiplication}: Matrices are multiplied in the sequence in which they are encountered while tracing a path. When doing the path tracing, it is crucial to make sure the multiplied matrices are conformant to the matrix multiplication rules. Specifically, the upstream variable of the premultiplied matrix should be the same variable as the downstream variable of the post-multiplied matrix.
\end{itemize}

Although all matrices involved in the path tracing are of the same $n \times n$ dimensions (for example, they are all $2 \times 2$ matrices in a bivariate model), it is helpful to determine whether each matrix is diagonal, full, or symmetric. Distinguishing among these forms is important for verifying the correctness of mathematical expectations and interpreting their conceptual meaning. A $n \times n$ matrix will be full if the designated upstream and downstream variables are not the same variables in the model—for example, when the $2 \times 2$ covariance matrix is an off-diagonal submatrix of a larger symmetric $4 \times 4$ covariance matrix that includes all variables. Consider the following $4 \times 4$ symmetric phenotypic variance-covariance matrix between two paternal and two maternal phenotypic scores:
\mateMatrix
This matrix can be partitioned into four $2 \times 2$ covariance submatrices. The two $\mathbf{V}_Y$ matrices along the diagonal represent the within-person phenotypic variance–covariance matrices between the paternal and two maternal phenotypes, respectively, and are symmetric. In contrast, the two off-diagonal $\text{mate}$ submatrices are transpositions of one another and are full matrices, because their own off-diagonal elements correspond to conceptually different constructs. Moreover, given the equation for mate covariance, $COV(\mathbf{Y}_p, \mathbf{Y}_m) = \mathbf{V}_{Yp} \boldsymbol{\mu} \mathbf{V}_{Ym}$, and the fact that both $\mathbf{V}_Y$ matrices are symmetric, it follows that $\boldsymbol{\mu}$ must also be a full matrix. This is intuitive: the assortment between males on trait 1 and females on trait 2 is not necessarily the same as the assortment between males on trait 2 and females on trait 1.

\subsubsection{Parameters and Their Interpretations}

Using multivariate path-tracing rules, we derived the expectations of all the variance and covariance listed in Table \ref{tab:var} in the Supplement. In this section, we outline the conceptual meaning of key estimated parameters. 

Direct additive genetic effects are modeled through the path coefficients $\boldsymbol{\delta}$ and $\mathbf{a}$. Both are diagonal matrices, as pleiotropy between the two traits in the base population is captured in the off-diagonal elements of the $\mathbf{k}$ and $\mathbf{j}$ matrices, while the increase in additive genetic (co)variance induced by AM is quantified in the $\mathbf{g}$, $\mathbf{h}$, and $\mathbf{i}$ matrices. The coefficient $\boldsymbol{\delta}$ indexes the additive genetic effects of haplotypic PGSs on $Y$, whereas $\mathbf{a}$ indexes residual additive genetic effects not captured by haplotypic PGSs. For almost all traits currently, $\mathbf{a} \gg \boldsymbol{\delta}$. Information for estimating $\boldsymbol{\delta}$ is easily estimable from the covariance between the haplotypic PGSs and $\mathbf{Y}$, after correcting for the effects of AM and genetic nurture. Information for estimating $\mathbf{a}$ can come from the residual covariance between $\mathbf{Y}_o$ and $\mathbf{Y}_{p/m}$ after accounting for all other parameters estimated in the model, though because other factors besides additive direct and indirect genetic effects, VT, and AM can increase parent–offspring covariance, it may be preferable to obtain this parameter from external sources such as RDR regression \autocite{young_relatedness_2018}. The total additive genetic variance of a trait is then derived by tracing all paths that begin with $\boldsymbol{\delta}$ and $\mathbf{a}$ and return to the phenotype ($\mathbf{Y}_{p/m}$).

The $\mathbf{k}$ matrix represents the variance–covariance of the haplotypic PGSs in the base population (before AM). Similarly, $\mathbf{j}$ is defined as the genetic variance of the haplotypic latent genetic score (LGS) in the base population. The values of $\mathbf{k}$ depend on the scaling of the haplotypic PGSs. Below are several ways to fix the value of ${k}_{ii}$ - the diagonal entries of trait $i$ in the $\mathbf{k}$ matrix:
\begin{itemize}
    \item ${k}_{ii} = \frac{1}{2}$, when the full PGS is standardized in the base population (which is only possible with simulated data).
    \item ${k}_{ii} = \frac{1}{2} - 2{g}_{cii}$, when the full PGS is standardized in the current generation, which is the approach we recommend in real data.
    \item ${k}_{ii} = \frac{1}{2} - {g}_{cii}$, when the haplotypic PGS is standardized in the current generation.
\end{itemize}
However, unlike $\mathbf{k}$, $\mathbf{j}$ is the variance of a latent construct and can take any arbitrary value. The simplest choice is to set ${j}_{ii} = {k}_{ii}$ and we recommend doing so because this allows the estimates of $\mathbf{a}$ and $\boldsymbol{\delta}$ to be comparable.  ${k}_{12}$ and ${j}_{12}$ are freely estimated parameters that represent the scaled genetic covariance between the two haplotypic PGSs (${k}_{12}$) or the two haplotypic LGSs (${j}_{12}$) , respectively, after accounting for AM and G-E covariance. In the version of the model described here, the coefficients $\boldsymbol{\delta}$ and $\mathbf{a}$ for parents are constrained to equal those for offspring. This assumption may be violated when the measured phenotype reflects a different construct, or when its genetic architecture differs across age or cohort (e.g., substance use)---a point we return to in the Discussion. We also assume ${k}_{12} = {j}_{12}$, implying that the correlation of allelic effects between traits is the same for variants captured and not captured by the PGS. This assumption is probably typically met, but could fail, for instance, if the genetic correlation between traits differs for rare versus common variants, given that common variant effects are currently better captured by PGSs.

The VT effect is modeled by the path coefficient matrix $\mathbf{f}$. VT represents the direct influence of parental phenotypes on offspring phenotypes. Most often this reflects the impact of parental traits on offspring traits through parental behavior, though other pathways (e.g., in utero effects) are also possible. $\mathbf{f}$ is a full matrix: the effects of parental trait 1 on offspring trait 2 can be different than the effects of parental trait 2 on offspring trait 1. The estimation of $\mathbf{f}$ depends most strongly on the ratio $COV(\overrightarrow{NT}_{p/m}, \overrightarrow{Y_o}) / COV(\overrightarrow{T}_{p/m}, \overrightarrow{Y_o})$ \autocite{balbona_estimation_2021}. Put simply, when the association of non-transmitted alleles with offspring outcomes grows relative to that of transmitted alleles, this pattern indicates VT, because the non-transmitted alleles can affect offspring only indirectly by shaping parental behavior and environments. By leveraging parental PGSs based on transmitted and non-transmitted alleles in this way, $\mathbf{f}$ can be estimated without confounding from shared parent–offspring genes. Embedding this approach within the SEM framework further reduces bias from AM and from genetic variance not captured by the PGS. 

\textit{Genetic nurture} refers to parental genetic effects—including non-transmitted alleles—that influence offspring outcomes indirectly via the parental environment, a process arising through VT\autocite{kong_nature_2018,balbona_estimation_2021,balbona_estimation_2022}. Our model quantifies genetic nurture through two mediated pathways, $\boldsymbol{\phi}$ and $\boldsymbol{\rho}$. Specifically, $\boldsymbol{\phi}$ represents the effect of full PGSs (the sum of two haplotypic PGSs) on offspring outcomes as mediated through parental traits, while $\boldsymbol{\rho}$ represents the corresponding effect of full LGSs. These pathways are defined as:
\[
\boldsymbol{\phi} = \boldsymbol{2f \delta k} \quad \text{and} \quad \boldsymbol{\rho}=\boldsymbol{2f a j}
\]
Here, $\boldsymbol{\phi}$ and $\boldsymbol{\rho}$ represent genetic nurture that stems from founder-generation additive genetic effects—the "pure" genetic nurture effect not distorted by AM. Because $\mathbf{f}$ is a full matrix, both $\boldsymbol{\phi}$ and $\boldsymbol{\rho}$ are also full, allowing asymmetric cross-trait genetic nurture in which the genetic nurture effect of trait 1 on trait 2 can differ from that of trait 2 on trait 1. The $\boldsymbol{\phi}$ and $\boldsymbol{\rho}$ are defined in a way that is independent of the influence of AM, so they target different estimands compared to other available methods, such as $\eta$ and regression coefficients in regression-based methods \autocite{kong_nature_2018, young_mendelian_2022, mcadams_annual_2023}.

% We estimate the genetic nurture effects with the covariance between $\overrightarrow{Y}_o$ and $\overrightarrow{NTm} + \overrightarrow{NTp}$ by deriving $\boldsymbol{\theta}_{NT} = 4\boldsymbol{\delta}\mathbf{g_c} + 4\mathbf{a}\mathbf{i_c} + \mathbf{w}$. Genetic nurture is a type of passive G-E correlation as a consequence of VT \autocite{kong_nature_2018, balbona_estimation_2022}. Both transmitted and non-transmitted alleles directly affect parental phenotypes, which in turn influence offspring phenotypes through VT effects. $\mathbf{w}$, which is the covariance between parental full PGS ($\overrightarrow{T} + \overrightarrow{NT}$) and $\overrightarrow{F}_o$, is the key component to obtaining genetic nurture effects. Since $\mathbf{f}$ is a full matrix, $\mathbf{w}$ is also a full matrix, and in turn  which implies the genetic nurture effects of trait 1 on trait 2 are different from the effects of trait 2 on trait 1. By constructing genetic nurture in this way, the current model manages to elucidate cross-trait genetic nurture effects with genomic data while controlling for within-trait genetic nurture and AM. 

As noted above, we model AM using a copath $\boldsymbol{\mu}$. AM refers to nonrandom mating, in which individuals preferentially pair with others who share similar or dissimilar traits at rates exceeding chance \autocite{vandenberg_assortative_1972}. The copath was introduced as an elegant way to model AM in path diagrams \autocite{cloninger_interpretation_1980}, but has rarely been applied in multivariate family models—with only one exception we are aware of \autocite{maes_genetic_1997}. In this paper, we formally lay out the multivariate copath rules for SEM and use them to model multivariate AM. Copaths provide a path tracing framework that links the expected covariance between parental phenotypes and their haplotypic PGSs to the observed data. In the present model, mate covariance is expressed as $COV(\mathbf{Y}_p, \mathbf{Y}_m) = \mathbf{V}_{Yp}\boldsymbol{\mu}\mathbf{V}_{Ym}$, allowing estimation of $\boldsymbol{\mu}$. Empirically, copath estimates correspond to mate phenotypic covariance scaled by phenotypic variance, given by $\boldsymbol{\mu} = \mathbf{V}_{Yp}^{-1}COV(\mathbf{Y}_p, \mathbf{Y}_m)\mathbf{V}_{Ym}^{-1}$.

AM increases the genetic variance in the population and the genetic covariance between relatives and mates. In our model, this is captured with the $\mathbf{g}$, $\mathbf{h}$, and $\mathbf{i}$ parameters. Specifically, $\mathbf{g}$ generally represents the increase in the genetic (co)variance of haplotypic PGSs between two traits under AM. We used different subscripts "t" and "c" to denote trans-individual (i.e., between mate) and cis-individual (within-person) haplotypic PGS covariances, respectively. For example, $\mathbf{g}_t$ is a full matrix representing the AM-induced covariance between maternal and paternal haplotypic PGSs. $\mathbf{g}_t$ is a full matrix for the same reason that $\boldsymbol{\mu}$ is full. In contrast, $\mathbf{g}_c$ is a symmetric matrix representing (a) the AM-induced increase in the covariance between the two haplotypic PGSs within-person or, equivalently, (b) the AM-induced increase in their variance. Although $\mathbf{g}_c$ is the covariance matrix between two distinct variables, $\mathbf{T}_{p/m}$ and $\mathbf{NT}_{p/m}$, and would be expected to be full according to our previously described rules, it must be symmetric for biological reasons. This symmetry arises from the randomness of haplotype formation during meiosis, as haplotypes are defined relative to the offspring genotype. Analogous to $\mathbf{g}$, $\mathbf{h}_t$ and $\mathbf{h}_c$ represent increases in latent genetic covariance for trans- and cis- covariances, respectively. Finally, unlike the matrices $\mathbf{g}_c$ and $\mathbf{h}_c$, $\mathbf{i}_c$ is the full within-individual covariance matrix between $\overrightarrow{L[N]T}_{p/m}$ and $\overrightarrow{[N]T}_{p/m}$. Because $\mathbf{i}_c$ models the covariance between two distinct sets of variables, the matrix is full. To illustrate this, consider an extreme scenario where trait 1 is determined entirely by its PGS and trait 2 entirely by its LGS; here, the term $i_{c21}$ will be non-zero while $i_{c12}$ would be zero.

\textit{G-E covariance} rises as a function of VT and AM.  We quantify this total G-E covariance using the matrices $\mathbf{w}$ and $\mathbf{v}$, which are defined as the covariance between the familial environment ($\overrightarrow{F}_o$) and the combined non-transmitted parental alleles ($\overrightarrow{NTm} + \overrightarrow{NTp}$). Unlike $\boldsymbol{\phi}$ and $\boldsymbol{\rho}$, which represent pure genetic nurture pathways, $\mathbf{w}$ and $\mathbf{v}$ capture the full G-E covariance—both within and across traits—that results from both AM and VT. Because they are a function of $\mathbf{f}$, $\mathbf{w}$ and $\mathbf{v}$ are inherently full matrices. The choice between $\boldsymbol{\phi}$ and $\mathbf{w}$ as the focal parameter depends on whether the research question pertains to a specific genetic nurture pathway or the total G-E covariance.

%  For the first time, this framework integrates within-trait and between-trait genetic influences, G-E covariance, VT, and AM into a unified analytical approach. In contrast, previous family-based models have typically focused on one or two of these factors in isolation, risking biased estimates due to unaccounted correlations between these processes. By jointly modeling these interdependencies, the Multivariate SEM-PGS model enhances the accuracy and robustness of parameter estimates. 

In the following sections, we present simulations evaluating the bias and precision of the parameter estimates, and then discuss the simulation results, the model’s assumptions, and recommendations for its application.

\section{Methods}

We conducted simulations to evaluate the performance of the Multivariate SEM-PGS model, focusing on two questions. First, we assessed bias by testing whether the model's median estimates were systematically different from the true simulated parameters. Second, we evaluated the precision of these estimates as a function of sample size and PGS effect size. To address the first question, we simulated evolutionary processes across generations using an R implementation of the GeneEvolve software \autocite{tahmasbi_geneevolve_2017}. Unlike methods that draw phenotypes directly from a multivariate normal distribution, GeneEvolve uses a forward-time simulation approach by generating a baseline population that evolves according to predefined mating and reproductive rules. Thus, GeneEvolve simulates the underlying causal processes instead of relying on model-implied covariance matrices, thereby providing an independent way to validate our derived equations and to examine how parameter estimates behave when model assumptions deviate from the true causal processes. To address the second question regarding estimate precision, we simulated two traits and their haplotypic PGSs directly from the model-implied multivariate normal distribution. The covariance matrix, $\Sigma$, for this distribution was derived from multivariate path-tracing from our model diagram. Because the forward-time simulations already established that the model produced unbiased estimates, we did not need to revisit bias here. Instead, this approach isolates estimation precision by avoiding the additional stochasticity inherent to forward-time simulators, which can inflate the variance of parameter estimates \autocite{chen_comparing_2025}. All R scripts for this article are available on our GitHub repository (https://github.com/Xuanyu-Lyu/BiSEMPGS).

\subsection{Forward-Time Simulation Design}
We first explain the process for simulating genetic, environmental, and phenotypic scores in the baseline population. Each individual had two phenotypes, which were both standardized to have a total variance of 1. Heritability of each trait (the proportion of phenotypic variance ($\mathbf{V_Y}$) attributable to genetic factors ($\mathbf{V_G}$) and genetic effects in the base population were simulated by drawing $m = 50$ causal variants (CVs) from a binomial distribution with minor allele frequencies uniformly distributed as $p \sim U(0.1, 0.5)$. The effect sizes of the CVs were drawn from a multivariate normal distribution $\sim N([0,0], \begin{pmatrix}
 \frac{1}{2mp(1-p)} & \frac{r_g}{2mp(1-p)} \\ \frac{r_g}{2mp(1-p)}  & \frac{1}{2mp(1-p)}
\end{pmatrix})$, where $r_g$ represents pleiotropy (the genetic correlation) in the base population. We then computed the haplotypic PGS and haplotypic LGS from these variants, ensuring that each score had a variance of 1 (thereby using the first method of standardized PGSs where ${k}_{ii} = \frac{1}{2}$) and that their genetic correlation aligned with the specified $r_g$ in first generation. These scores were subsequently scaled using the $\boldsymbol{\delta}$ and $\mathbf{a}$ matrices to achieve the desired proportions $\mathbf{r}^2_{pgs}$ and $\mathbf{r}^2_{lgs}$ ($\mathbf{V_G} = \mathbf{r^2_{pgs}} + \mathbf{r^2_{lgs}}$), thereby generating the PGSs and LGSs for the base population (note that these were by definition uncorrelated in the base population). We then simulated residual environmental scores ($\boldsymbol{\epsilon}$) to account for residuals in the phenotypic variance unexplained by $\mathbf{V_G}$. Phenotypic variance was scaled to be 1 in the based population and therefore the distribution of $\boldsymbol{\epsilon}$ was $\sim N([0,0], \mathbf{I} - diag(\mathbf{V}_G))$. Phenotypic scores for each individual were obtained by summing their genetic and environmental scores.

After generating the base population, the GeneEvolve algorithm paired males and females such that their phenotypic correlation matched the specified mate correlation matrix ($\mathbf{r}_{mate}$). Each pair of mates produced a number of offspring that followed a Poisson distribution with a mean of 2. We simulated genetic scores of offspring (including both PGS and LGS) using randomly sampled CVs from the mother and separately from the father. Family environmental scores were created for offspring by $\overrightarrow{F}_{o,gen(n+1)} = \mathbf{f}\overrightarrow{F}_{p,gen(n)} + \mathbf{f}\overrightarrow{F}_{m,gen(n)}$, reflecting the VT process. We created phenotypic scores of offspring by summing genetic scores, family environmental scores, and residual environmental scores ($\boldsymbol{\epsilon}$). The algorithms for mating and reproduction iterated through several generations. We used PGSs and phenotypic scores from the final generation as inputs for model fitting.

A key challenge in forward-time simulations is modeling cross-trait AM and its genetic consequences \autocite{border_rbahadur_2023}. To impose the target mate correlation matrix ($\mathbf{r}_{mate}$) across two traits, we first split the generation's phenotypic data into males and females and constructed a four-dimensional template of correlated scores, $X_{\text{sim}}=(m_1,m_2,f_1,f_2)$, by drawing from $\mathcal{N}(\mathbf{0}, \mathbf{\Sigma})$. The off-diagonal symmetric entries of $\mathbf{\Sigma}$ were set to the empirical correlations between the two traits in that generation, while the 2-by-2 block linking males and females was defined by the user-specified $\mathbf{r}_{mate}$. We then computed Euclidean distance matrices between observed, standardized phenotypes $\{Y_1,Y_2\}$ and their corresponding template vectors for males and females, and greedily paired individuals using a duplicate-removal/matching routine that minimized total distance, thereby aligning the realized cross-partner correlation structure with $\mathbf{\Sigma}$. This template-guided matching roughly approach achieved both within-trait and cross-trait AM patterns specified by $\mathbf{r}_{mate}$ while preserving the observed phenotype distributions within each sex.Nevertheless, this approach consistently produced $\mathbf{r}_{mate}$ values slightly (e.g., < .01) below the target correlation in each generation. This discrepancy resulted in small deviations in the estimated $\boldsymbol{\mu}$ and $\mathbf{g}_c$ values from their true population values (see details in Results). To address this limitation, we also fit the model using data generated directly from a covariance matrix derived from mathematical expectations. 

\begin{table}[ht]
    %\centering
    \caption{Parameter Setup of Simulations}
    \label{tab:simSetup}
    \begin{NiceTabular}{cw{c}{12cm}}
    \CodeBefore
    \rowcolors{2}{white}{gray!20}
    \Body
        \toprule
        \textbf{Parameter} & \textbf{Setup} \\
        \midrule
        $N_{trio}$ & 4k, 8k, 16k, 32k, 48k, 64k \\
        $r^2_{pgs1}$ & 1\%, 2\%, 4\%, 8\%, 16\% \\
        $r^2_{pgs2}$ & 7.2\% \\
        $\mathbf{f}$ & $
            \mathbf{f} = \begin{bmatrix}
                f_{11} = 0.15 & f_{12} = 0.10 \\
                f_{21} = 0.05 & f_{22} = 0.10
            \end{bmatrix}
        $ \\
        $\mathbf{r}_{mate}$ & $
            \mathbf{r}_{mate} = \begin{bmatrix}
                r_{11} = 0.4 & r_{12} = 0.2 \\
                r_{21} = 0.1 & r_{22} = 0.3
            \end{bmatrix}
        $ \\
        $\mathbf{V_G}$ & $
            \mathbf{V_G} = \begin{bmatrix}
                V_{G11} = 0.64 & V_{G12} = 0.10 \\
                V_{G21} = 0.10 & V_{G22} = 0.36
            \end{bmatrix}
        $ \\     
        \bottomrule
    \end{NiceTabular}
    \\
    \vspace{0.5em} % Adds a small vertical space between the table and the note
    \footnotesize
    \textbf{Note:} $N_{trio}$ represents the number of trios (father, mother, offspring) in the simulation. $\mathbf{V_G}$ is the genetic covariance matrix of the base generation. The parameters $\mathbf{f}$, $\mathbf{r}_{mate}$, and $\mathbf{V_G}$ were held constant across all conditions. 
\end{table}

\subsection{Multivariate Normal Simulations}

We used multivariate normal simulations to assess the precision of parameter estimates. These simulations were performed by sampling PGSs and phenotypic scores from the model-implied covariance matrix derived from Figure \ref{fig:BiSEMPGS}. This task was not straightforward because, while some parameters can be pre-defined, others are nonlinear functions of multiple parameters. Such dependencies arise from the recursive interrelationships among variables: for example, AM and VT alter G-E covariance across generations, which in turn modifies genetic variances, further feeding back into G-E covariance. Therefore, to find the model-implied covariance, we implemented an iterative algorithm that specified baseline values and then simulated how parameters evolved across generations under the evolutionary processes represented in our model. Our models assumed equilibrium—that is, a population state in which phenotypic and genetic variances remain constant across generations due to a balance between variance-increasing forces (AM and VT) and variance-reducing forces such as recombination \autocite{bulmer_effect_1971}. Thus, parameter values were retained once they stabilized across generations, indicating equilibrium. Specifically, in the multivariate SEM–PGS model (Table \ref{tab:simSetup}), $\mathbf{f}$, $\mathbf{r_{mate}}$, $\mathbf{V_G}$, $\mathbf{V_E}$, $r^2_{pgs1}$, and $r^2_{pgs2}$ were fixed across generations, while all other parameters were algebraic functions of these pre-defined values and obtained from the iterative procedure. Using these equilibrium values, we then computed the full model-implied covariance matrix.

We conducted both the bias analysis using forward-time simulation and the precision analysis using multivariate normal simulations under a range of conditions detailed in Table \ref{tab:simSetup}. To assess model precision, we systematically varied the sample size of trios ($N \in \{4\text{k}, 8\text{k}, 16\text{k}, 32\text{k}, 48\text{k}, 64\text{k}\}$) and the variance explained by the PGS for the first trait ($\mathbf{r}^2_{pgs1}$) across five levels. The variance explained by the second trait's PGS ($\mathbf{r}^2_{pgs2}$), the VT parameter ($\mathbf{f}$), and the additive genetic variance in the base generation ($\mathbf{V}_G$) were held constant across all conditions. Both the forward-time simulations and the analytical derivation of the expected equilibrium covariance were iterated for 20 generations to ensure that parameters influenced by AM and VT reached equilibrium. For each parameter combination, we generated 500 replicate datasets. The simulation code is available at https://github.com/Xuanyu-Lyu/BiSEMPGS.

\begin{table}[ht]
    %\centering
    \caption{Parameter Setup of Simulations for Bias Analysis of Univariate Model}
    \label{tab:uniBiasSetup}
    \begin{NiceTabular}{cw{c}{7cm}c}
    \CodeBefore
    \rowcolors{2}{white}{gray!20}
    \Body
        \toprule
        \textbf{Parameter} & \textbf{Condition 1} & \textbf{Condition 2} \\
        \midrule
        $N_{trio}$ & 64k & 64k \\
        $r^2_{pgs1}$ & 4\% & 4\% \\
        $r^2_{pgs2}$ & 7.2\% & 7.2\% \\
        $\mathbf{f}$ & $
            \mathbf{f} = \begin{bmatrix}
                f_{11} = 0.15 & f_{12} = 0.10 \\
                f_{21} = 0.05 & f_{22} = 0.10
            \end{bmatrix}
        $ & $
            \mathbf{f} = \begin{bmatrix}
                f_{11} = 0 & f_{12} = 0.30 \\
                f_{21} = 0.25 & f_{22} = 0.10
            \end{bmatrix}
        $ \\
        $\mathbf{r_{mate}}$ & $
            \mathbf{r_{mate}} = \begin{bmatrix}
                r_{11} = 0.4 & r_{12} = 0.2 \\
                r_{21} = 0.2 & r_{22} = 0.3
            \end{bmatrix}
        $ &  $
            \mathbf{r_{mate}} = \begin{bmatrix}
                r_{11} = 0.05 & r_{12} = 0.4 \\
                r_{21} = 0.4 & r_{22} = 0.1
            \end{bmatrix}
        $ \\
        $\mathbf{V_G}$ & $
            \mathbf{V_G} = \begin{bmatrix}
                V_{G11} = 0.64 & V_{G12} = 0 \\
                V_{G21} = 0 & V_{G22} = 0.36
            \end{bmatrix}
        $ & $
            \mathbf{V_G} = \begin{bmatrix}
                V_{G11} = 0.64 & V_{G12} = 0 \\
                V_{G21} = 0 & V_{G22} = 0.36
            \end{bmatrix}
        $ \\     
        \bottomrule
    \end{NiceTabular}
    \\
    \vspace{0.5em} % Adds a small vertical space between the table and the note
    \footnotesize
    \textbf{Note:} See Table \ref{tab:simSetup} for description of terms.
\end{table}

\subsection{Model Fitting and Estimation Analyses}
We fit models using OpenMx with the NPSOL optimizer to address the non-linear constraints in the model \autocite{neale_openmx_2016}. We used the median estimate across 500 models run on forward-time simulated data to evaluate bias, and the median absolute deviation (MAD) on 500 models run on multivariate normally distributed data to evaluate precision.  We report median and MAD as primary statistics because they are robust to outliers, which complex SEM models with non-linear constraints can occasionally produce, even with large sample sizes. Based on our experience, approximately 3–4\% of models yielded outlier estimates when $N_{trio} = 8k$, decreasing to around 1\% when $N_{trio} = 64k$. Because the number of estimates in the bivariate SEM-PGS model is very large, we limited those presented to keep the paper manageable and accessible. Specifically, we report two sets of results: eight within-trait estimates for trait 1 and eight cross-trait estimates between traits 1 and 2 (Figure \ref{fig:within_estimates}). We show the bias of parameter estimates as a function of $N_{trio}$ and their precision as a function of both $N_{trio}$ and $\mathbf{r}^2_{pgs}$.

The estimation of the latent genetic path coefficient $\mathbf{a}$ is one of the most challenging aspects of the SEM-PGS model, as the only information to estimate it comes from the covariance between $ \overrightarrow{Y}_{p/m} $ and $ \overrightarrow{Y}_o $ after accounting for the other estimated factors in the model. In empirical studies, researchers can choose to fix $\mathbf{a}$ by calculating genetic variance of the base population ($V_{G-Base}$) using the RDR or sibling regression approach \autocite{young_relatedness_2018, visscher_assumption-free_2006} in their own dataset. To evaluate the effect of fixing $\mathbf{a}$ on the precision of other parameter estimates, we fitted each simulated MVN dataset in two ways: (1) with the latent genetic path coefficient $\mathbf{a}$ fixed to its true value, and (2) with $\mathbf{a}$ freely estimated. As a follow-up, we performed a sensitivity analysis investigating the bias introduced by fixing $\mathbf{a}$ to incorrect values. To do this, we fit the bivariate SEM-PGS model where $a_{11}$ was fixed across a range of $\pm 0.125$ from its true value ($\approx 0.7746$) in increments of 0.025, with $N_{trio} = 32,000$ and $r^2_{pgs1} = 4\%$.

Finally, we investigated the effect on parameter estimates when a univariate model is applied to a truly bivariate data-generating phenotypic architecture. We simulated two scenarios (Table \ref{tab:uniBiasSetup}): (1) a phenotypic architecture characterized by strong within-trait and weak cross-trait effects; and (2) a phenotypic architecture dominated by weak within-trait and strong cross-trait effects. In both scenarios, the genetic correlation in the founder generation was set to zero to simplify interpretation. To quantify the bias, we fit a univariate SEM-PGS model on trait 1. For each condition, we generated 500 replicate datasets with $N_{trio} = 64,000$. 

\section{Results}

\subsection{Unbiasedness of the Parameters}

\begin{figure}[!htp]
    \centering
    \includegraphics[width=1\linewidth]{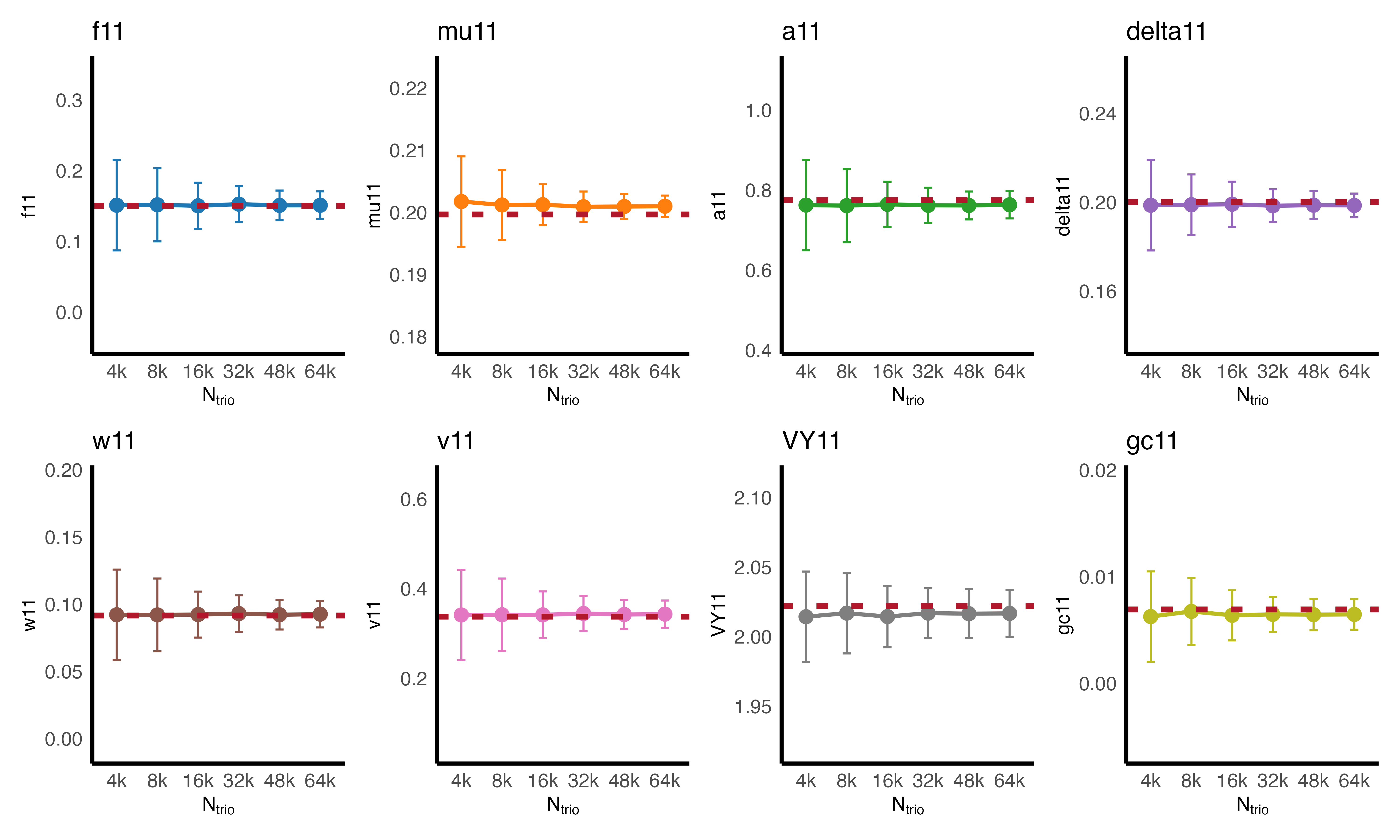}
    \caption{Median within-trait parameter estimates (+/- median absolute deviation, or MAD) as a function of $N_{trios}$ when $r^2_{pgs1} = .04$. The red dashed line represents the true parameter values. Minor deviations between the median estimates and the true values were likely due to small deviations in $\mathbf{r}_{mate}$ introduced during the bivariate foward-time simulation.}
    \label{fig:within_estimates}
\end{figure}

\begin{figure}[!htp]
    \centering
    \includegraphics[width=1\linewidth]{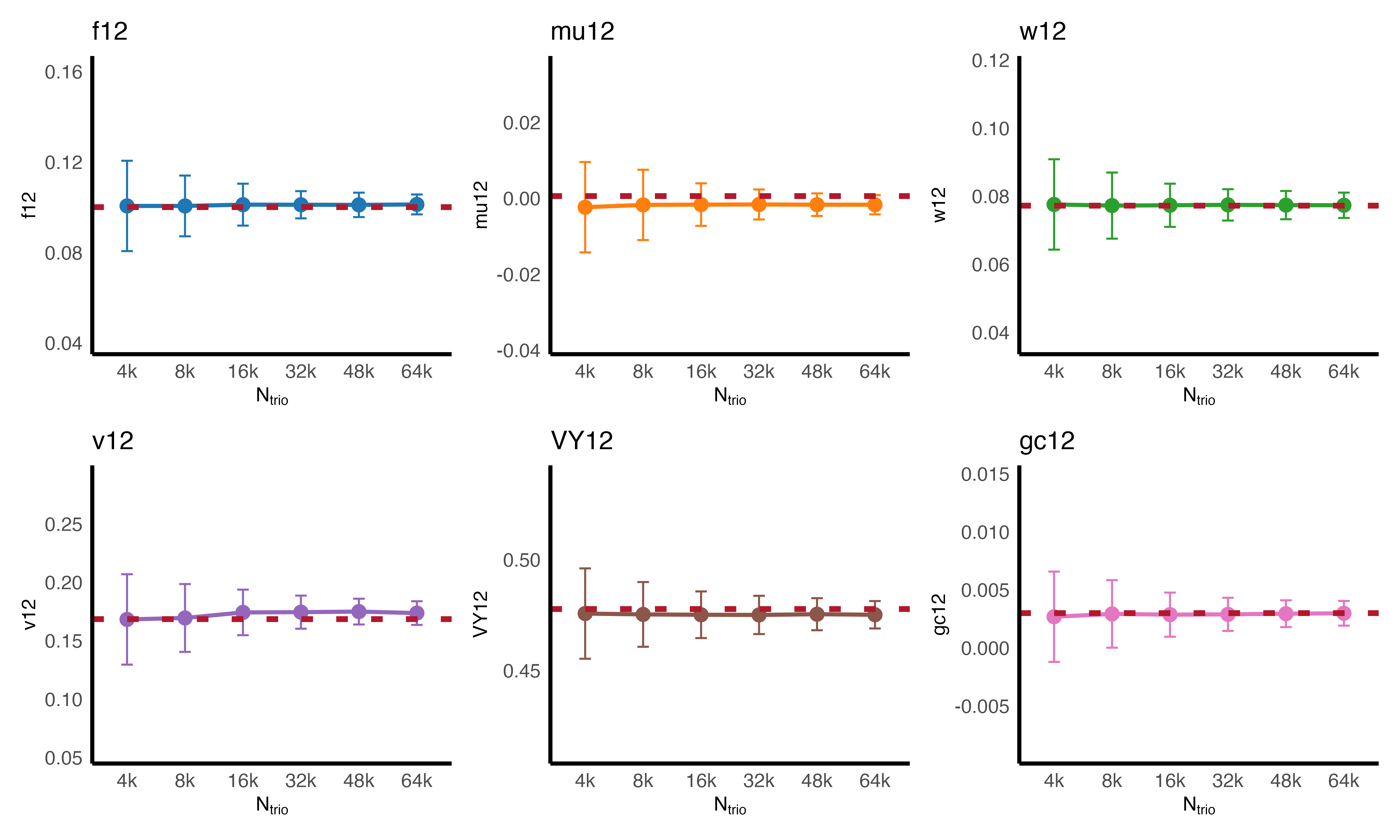}
    \caption{Median cross-trait parameter estimates and their MAD as a function of the $N_{trios}$, when $r^2_{pgs1} = .04$. See Figure 2 caption for additional details.}
    \label{fig:cross_estimates}
\end{figure}

Figures \ref{fig:within_estimates} and \ref{fig:cross_estimates} show the median and MAD of Bivariate SEM-PGS model estimates from forward-time simulation as a function of $N_{trios}$. The results indicate that parameters were unbiased or nearly unbiased across all sample sizes, with the bootstrap tests suggesting some of them significantly deviated from their true value (see Supplement for p-values). For the within trait estimates (Figure \ref{fig:within_estimates}), very small deviations in the median values of $\mathbf{V}_Y$, $\mathbf{a}$, $\boldsymbol{\mu}$, and $\mathbf{g}_c$ from their true values were likely due to the imperfect simulation of bivariate AM and genetic effects in the forward-time simulation algorithms. Similarly, the cross-trait estimates (Figure \ref{fig:cross_estimates}) were unbiased or nearly so, with slight deviations observed in $\boldsymbol{\mu}$ and $\mathbf{V}_Y$. All significant biases were small relative to the sampling error of the estimates, accounting for $<10\%$ of the total variability in estimated parameters, regardless of sample size (see Appendix I). Similar patterns were observed across all other conditions of $N_{trios}$ and $\mathbf{r}^2_{pgs}$. Using the alternative multivariate normal simulation approach with covariances based on the model math, no estimates were biased (no estimate median was statistically different from the true parameter values; Figures S1 and S2 in the Supplement).

\subsection{Estimation Precision}
\begin{figure}[!htp]
    \centering
    \includegraphics[width=1\linewidth]{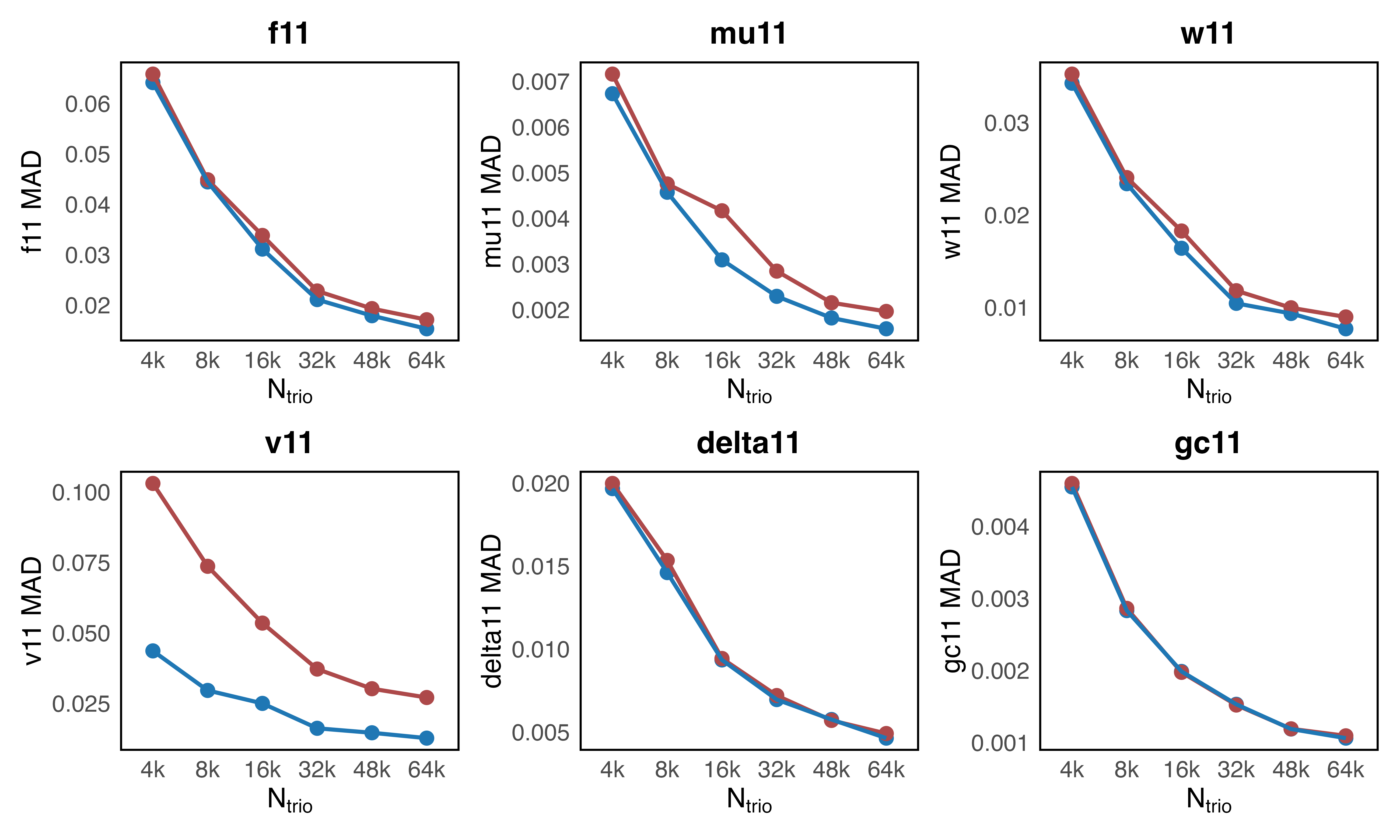}
    \caption{MAD of within-trait parameter estimates as a function of $N_{trio}$ when $r^2_{pgs1} = .04$. The red line represents the MAD when the latent genetic path coefficient $\mathbf{a}$ was freely estimated, while the blue line represents the MAD when $\mathbf{a}$ was fixed at its true population level.}
    \label{fig:within_samplesize}
\end{figure}

In Figure \ref{fig:within_samplesize}, we show the change in MAD of within-trait effect estimates of trait 1 (the [1,1] entry of each parameter estimate matrix) as a function of $N_{trio}$ when fitting the model with fixed $\mathbf{a}$ compared to estimating $\mathbf{a}$, given $r^2_{pgs1} = .04$. As expected, larger sample sizes led to better precision for all the within-trait estimates. For most within-trait estimates, the MAD showed a sharp decline as $N_{trio}$ increased from 4k to 32k and decreased more gradually thereafter. Fixing $\mathbf{a}$ elements to their true values resulted in more precise latent parameter estimates across the board except for $\mathbf{g_c}$. Among the within-trait parameters, the latent G-E covariance $v_{11}$ benefited most from fixing $\mathbf{a}$, because it is a direct function of $a_{11}$ and $f_{11}$. For other parameters, the precision gains from fixing $\mathbf{a}$ were modest compared to those achieved by simply increasing $N_{trio}$. The pattern of changes for trait 1 within-trait estimates ([1,1] entry) was qualitatively similar to that for trait 2 ([2,2] entry).

\begin{figure}[!htp]
    \centering
    \includegraphics[width=1\linewidth]{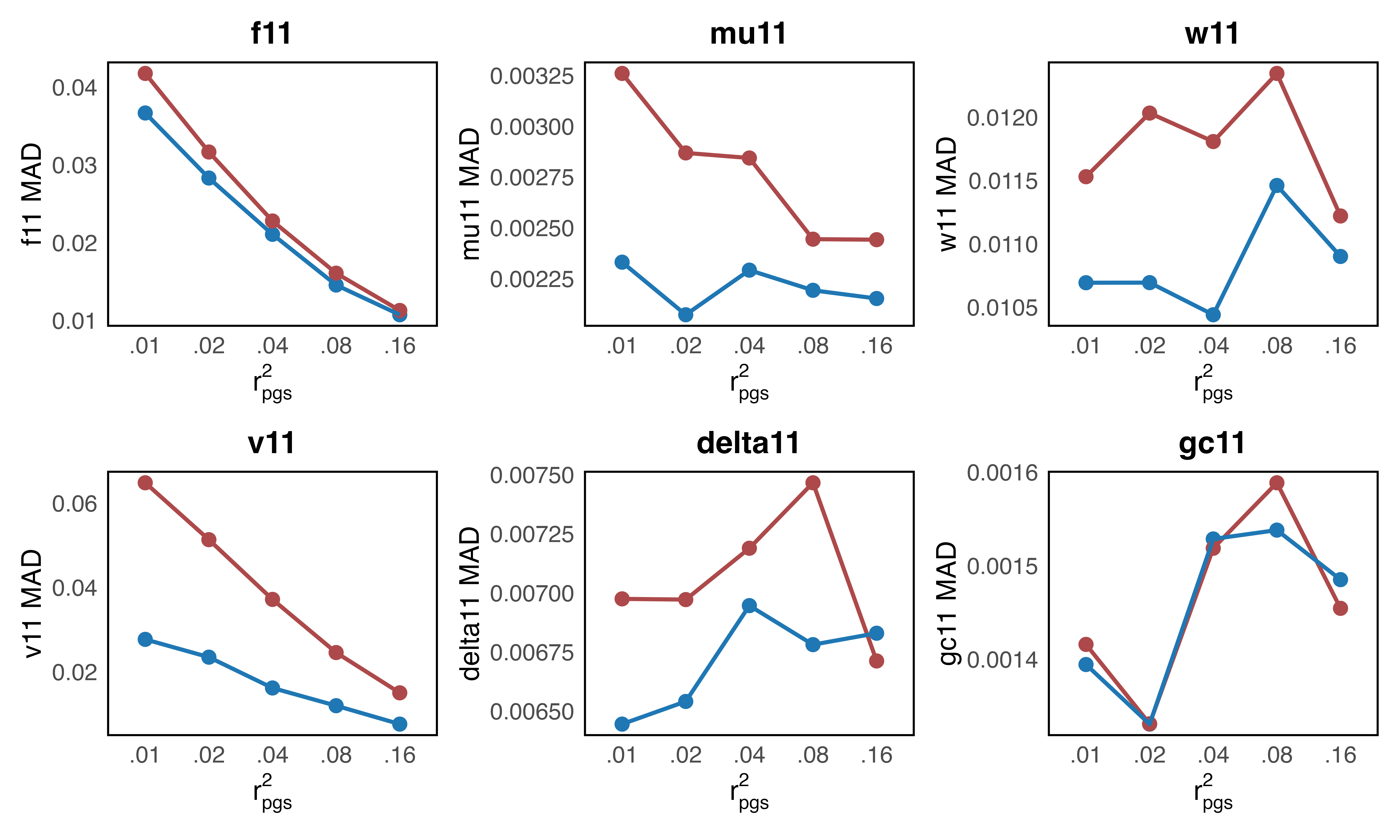}
    \caption{MAD of within-trait parameter estimates are shown as a function of $r^2_{pgs}$ for trait 1, when $N = 32k$. See the caption of Figure 4 for additional details.}
    \label{fig:within_SE_r2pgs}
\end{figure}

Figure \ref{fig:within_SE_r2pgs} shows how within-trait parameter precision changes with $r^2_{pgs1}$ at fixed $N=32k$. Larger $\mathbf{r}^2_{pgs}$ values yielded much more precise estimates of $\mathbf{f}$ and $\mathbf{v}$, because more predictive PGSs provide more reliable information about both direct and indirect genetic effects, thereby reducing sampling error of estimates that depend on these effects. In contrast, estimates of $\boldsymbol{\mu}$, $\boldsymbol{\delta}$, $\mathbf{w}$, and $\mathbf{g}_c$ showed little dependence on $r^2_{pgs1}$. The small fluctuations observed were minor relative to the scale of the y-axes and likely reflected sampling error. Consistent with Figure \ref{fig:within_samplesize}, fixing $\mathbf{a}$ improved the precision of all parameters except $\mathbf{g}_c$, with the largest benefit for $\mathbf{v}$ when PGSs are weak. For example, when $r^2_{pgs1}=.16$, the gap in SEs for $v_{11}$ between fixed and estimated $\mathbf{a}$ was smaller than when $r^2_{pgs1}=.01$, illustrating that fixing $\mathbf{a}$ is most valuable when PGSs explain little variance. Patterns for trait 2 mirrored those for trait 1.

\begin{figure}[!htp]
    \centering
    \includegraphics[width=1\linewidth]{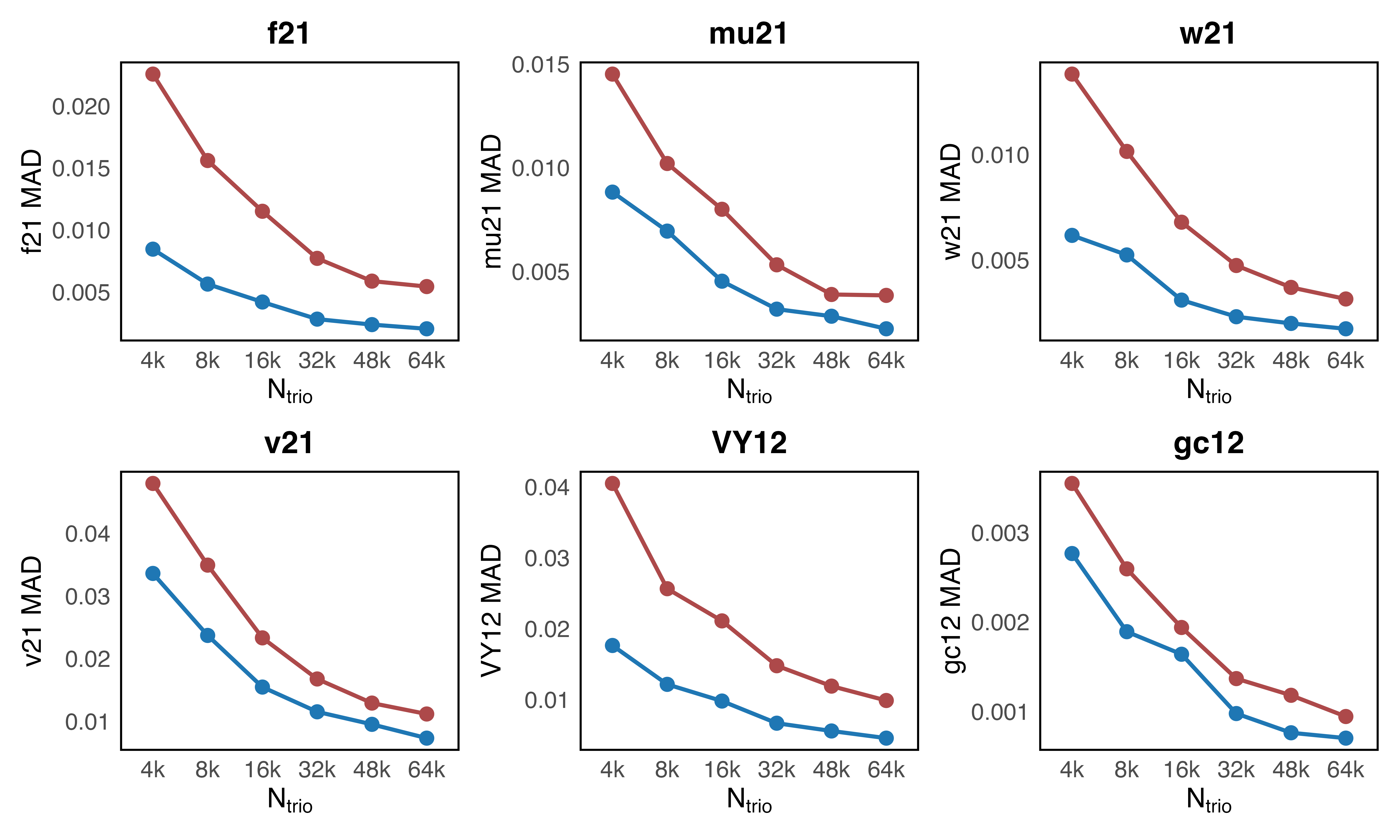}
    \caption{MAD of cross-trait parameter estimates are shown as a function of number of trios ($N$), when the $r^2_{pgs1} = .04$. See Figure 4 caption for additional details.}
    \label{fig:cross_samplesize}
\end{figure}

Figure \ref{fig:cross_samplesize} illustrates the MAD of cross-trait parameter estimates as a function of $N_{trio}$. Consistent with the within-trait findings, larger sample sizes led to improved estimation precision for all parameters, with the most pronounced gains occurring for sample sizes up to 32K trios. Unlike in the within-trait analysis, fixing the direct genetic effect ($\mathbf{a}$) benefited all cross-trait estimates, though the magnitude of the improvement varied by parameter. This benefit was particularly substantial for the VT ($\mathbf{f}$) and G-E covariance ($\mathbf{w}$) parameters at smaller sample sizes. For these same parameters ($f_{21}$ and $w_{21}$), precision increased more rapidly with sample size when $\mathbf{a}$ was freely estimated, but the fixed-$\mathbf{a}$ model maintained lower MAD across all sample sizes. For the remaining parameters, increasing sample size had a comparable effect on precision regardless of whether $\mathbf{a}$ was fixed or estimated.

\begin{figure}[!htp]
    \centering
    \includegraphics[width=1\linewidth]{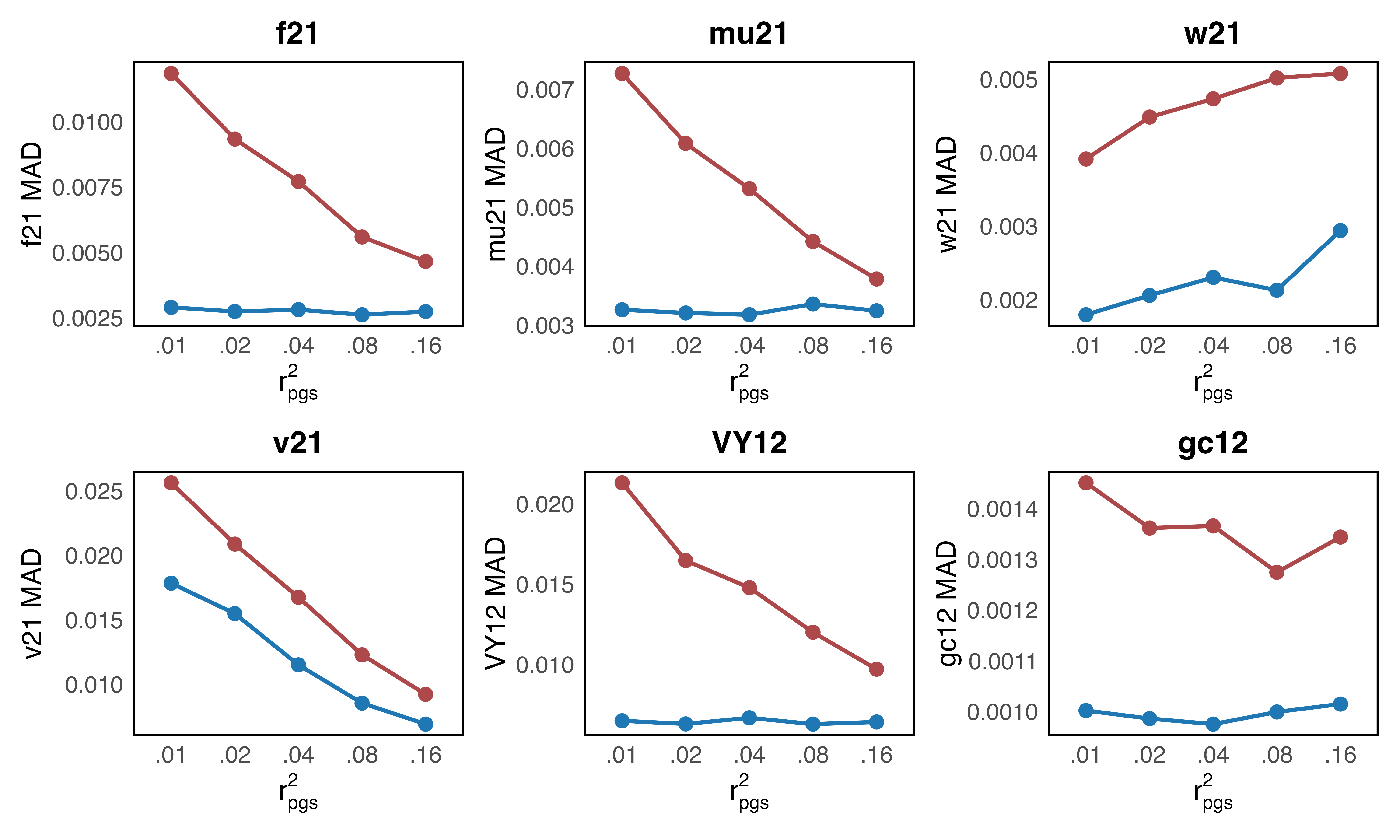}
    \caption{MAD of cross-trait parameter estimates are shown as a function of $r^2_{pgs}$ for trait 1, when $N = 32k$. See the Figure 4 caption for additional details.}
    \label{fig:cross_SE_r2pgs}
\end{figure}

Figure \ref{fig:cross_SE_r2pgs} also reveals that fixing $\mathbf{a}$ consistently improved the precision of cross-trait estimates, but the effect of $r^2_{pgs1}$ depended on whether $\mathbf{a}$ was fixed or freely estimated. Higher $r^2_{pgs1}$ led to more precise estimates only when $\mathbf{a}$ was freely estimated; when $\mathbf{a}$ was fixed, the precision of most cross-trait estimates showed little dependence on the PGS’s predictive power. The only exception was the cross-trait latent G-E covariance parameter, $v_{21}$, which still benefited from higher $r^2_{pgs1}$. Overall, these results suggest that fixing $\mathbf{a}$ provides more precise estimates of cross-trait effects, particularly when PGSs are weak—a dynamic not observed for within-trait estimates.

\subsection{Parameter Bias When Fixing $\mathbf{a}$ at Incorrect Values}
\begin{figure}[!htp]
    \centering
    \includegraphics[width=1\linewidth]{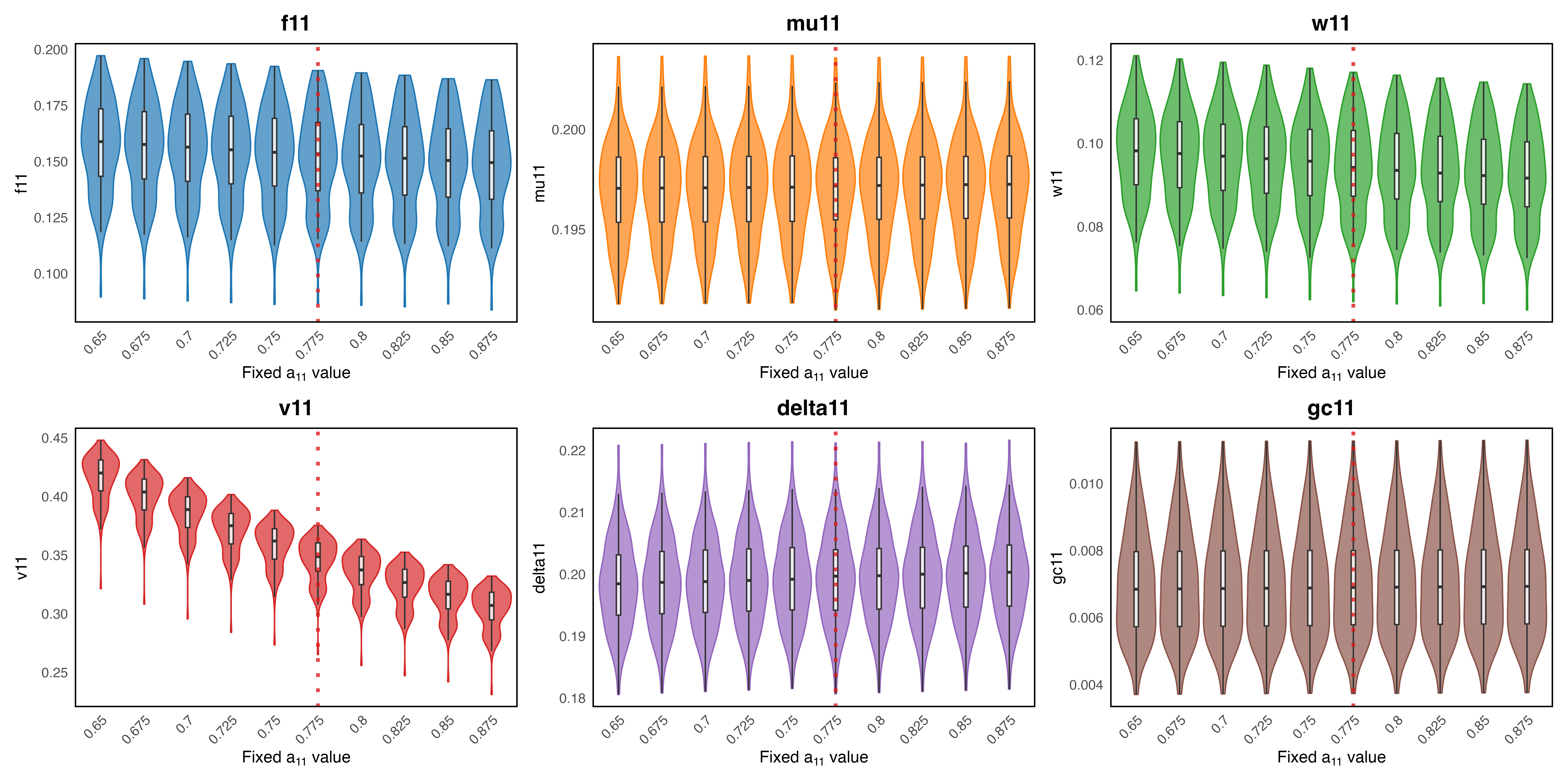}
    \caption{Distribution of six key within-trait parameter estimates from a sensitivity analysis where the direct genetic effect, $a_{11}$, was fixed to various values ($N_{trio} = 32k$ and $r^2_{pgs1} = 4\%$). Each violin plot shows the distribution of estimates across 100 replications fitted in OpenMx. The red dashed line highlights the results obtained when $a_{11}$ was fixed to its true simulated value.}
    \label{fig:sensi11}
\end{figure}

\begin{figure}[!htp]
    \centering
    \includegraphics[width=1\linewidth]{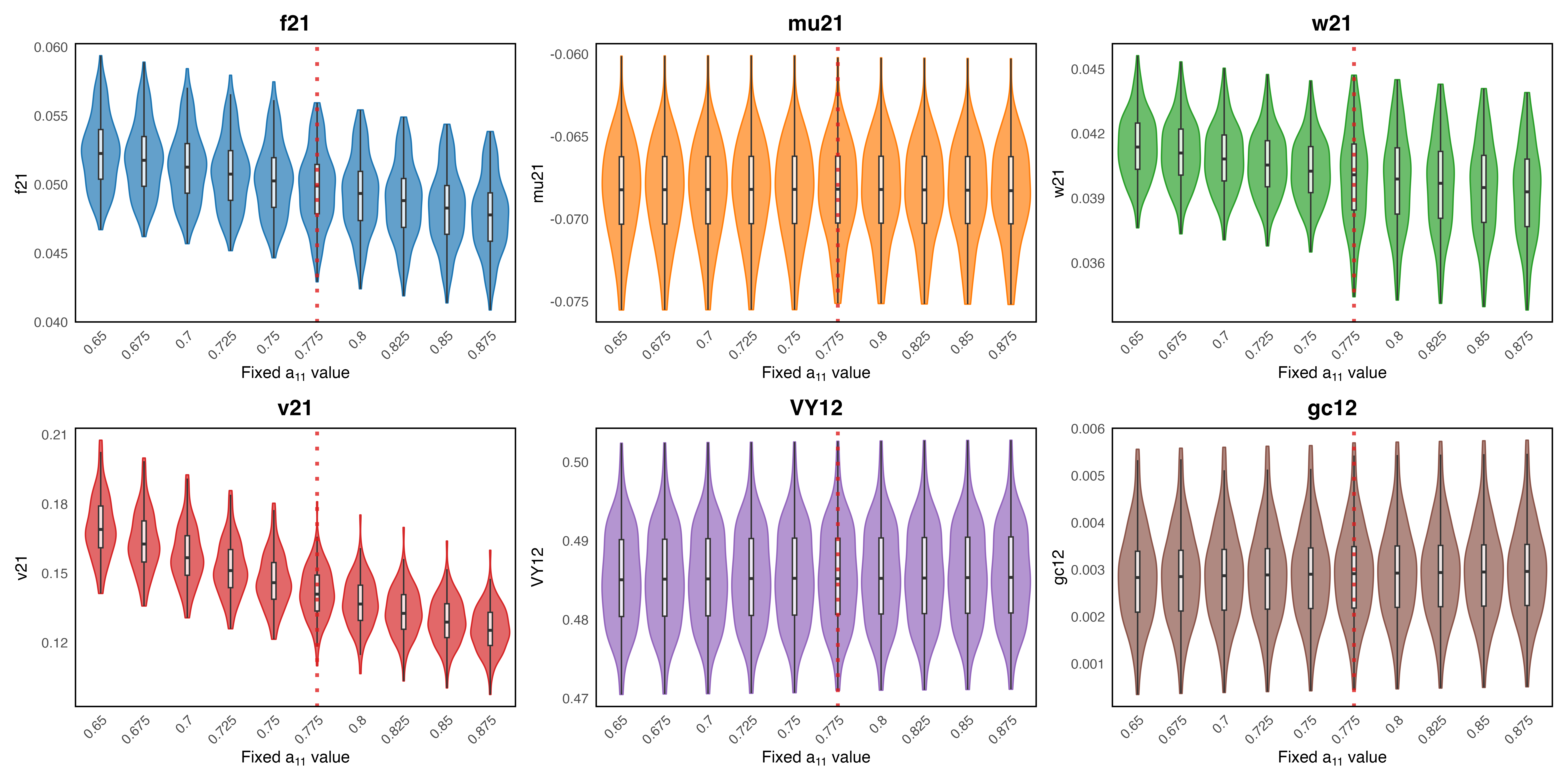}
    \caption{Distribution of six key cross-trait parameter estimates from a sensitivity analysis where the direct genetic effect, $a_{11}$, was fixed to various values. See the caption of Figure \ref{fig:sensi12} for additional details.}
    \label{fig:sensi12}
\end{figure}

Figures \ref{fig:sensi11} and \ref{fig:sensi12} show the results of a sensitivity analysis on within-trait and cross-trait parameter estimates, respectively, when $a_{11}$ was fixed to incorrect values. Overall, the VT ($\mathbf{f}$) and G-E covariance ($\mathbf{v}$) estimates were affected to varying degrees, whereas the estimates for $\boldsymbol{\mu}$, $\boldsymbol{\delta}$, $\mathbf{V_Y}$, and $\mathbf{g_c}$ were minimally impacted, with their variation dominated by stochastic noise. Among the affected parameters, $f_{11}$, $w_{11}$, and $w_{21}$ were influenced to a smaller extent than the cross-trait parameters $f_{21}$, $v_{11}$, and $v_{21}$, with the latent G-E covariance parameters ($v_{11}$ and $v_{21}$) most affected. Additionally, both the within- and between-trait familial variances (${V_{F11}}$ and ${V_{F12}}$) exhibited negative relationships with the fixed value of $a_{11}$, as both components are functions of $f_{11}$ and $f_{21}$ (Figure \ref{fig:sensiVF}).

\begin{figure}[!htp]
    \centering
    \includegraphics[width=1\linewidth]{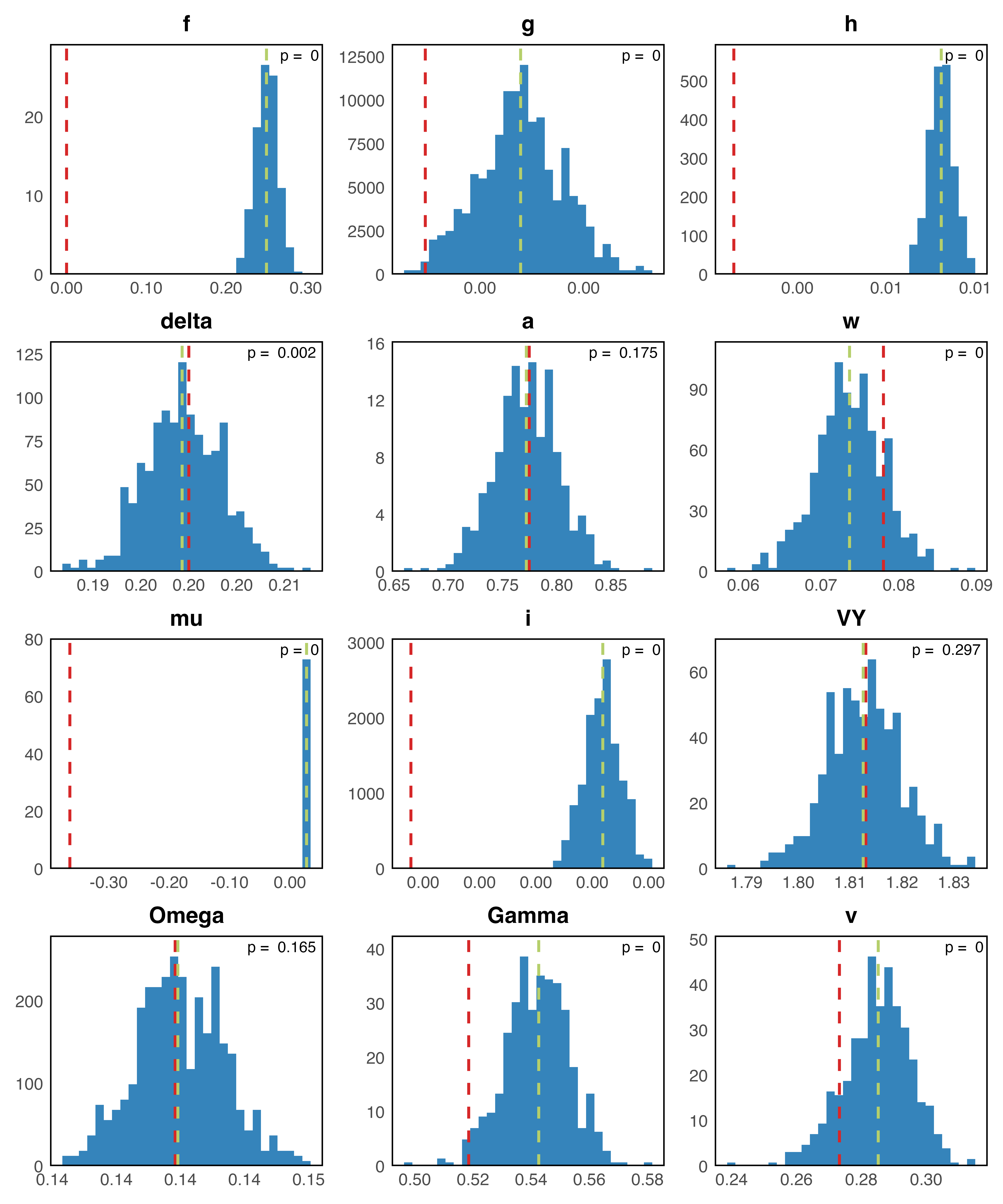}
    \caption{Histograms of 500 parameter estimates from a univariate SEM-PGS model when the data was simulated with strong cross-trait effects. The red dashed line indicates the true parameter value. The green dashed line indicates the median of the 500 estimates. P-values of the bootstrap test for each parameter are shown for all parameters. Omega is the covariance between parental phenotype and one haplotypic PGS. Gamma is the covariance between parental phenotype and one haplotypic LGS.}
    \label{fig:uni_bias_1}
\end{figure}

\subsection{Estimate Bias Using a Univariate Model When the Phenotypic Architecture Is Truly Bivariate}

Figure \ref{fig:uni_bias_1} illustrates the bias that arises when a univariate model for trait 1 is fit when the true data-generating process has weak within-trait but strong cross-trait VT and AM effects (Table \ref{tab:uniBiasSetup}, Condition 2). This model misspecification resulted in significant bias for nearly all parameters, with the exceptions of $\boldsymbol{\delta}$, $\mathbf{a}$, and $\mathbf{V}_Y$. Most notably, the within-trait VT parameter, $f_{11}$, was severely overestimated: despite a true value of zero, its estimates were centered around 0.25. At the same time, the familial variance estimate for trait 1, $V_{F11}$, was substantially underestimated (median = 0.2396 vs. the true value of 0.3665). These two results—an overestimate of $f_{11}$ alongside an underestimate of $V_{F11}$—are not contradictory. The true $V_{F11}$ arises entirely from cross-trait VT originating from trait 2. When both traits are modeled jointly, the bivariate model correctly identifies no VT from trait 1 to trait 1 but substantial VT from trait 2 to trait 1. In contrast, the univariate model misattributes the cross-trait influence as a spurious within-trait VT effect ($f_{11}$) while simultaneously underestimating the total familial variance for trait 1.  

A related practice is to study cross-trait VT or genetic nurture by fitting a univariate model that pairs different traits across generations—for example, parental education and offspring health \autocite{kong_nature_2018}. However, this approach also fails to account for pleiotropy or for both within- and cross-trait AM and VT, and therefore yields biased estimates. Just as unmodeled AM can masquerade as VT in a univariate setting, strong cross-trait AM can masquerade as cross-trait VT when only a single trait is modeled in each generation—for example, parental substance use and offspring educational attainment.

\section{Discussion}
In this study, we introduced the multivariate SEM-PGS model. This represents the first multivariate framework to simultaneously estimate the effects of both within- and cross-trait AM, genetic nurture, VT, G-E covariance, and direct genetic effects. To support the development of this model, we described a multivariate path-tracing rule for copaths, complementing the multivariate path-tracing rules introduced by \textcite{vogler_multivariate_1985}. Through Monte-Carlo simulations and model fitting with OpenMx, we validated the accuracy of these path-tracing rules, demonstrating that both within- and cross-trait estimates were unbiased. Larger sample sizes and more powerful PGSs ($\mathbf{r}^2_{pgs}$) improved estimation precision, with the steepest gains between 4k and 32k trios. Fixing $\mathbf{a}$ to its population value enhanced estimate precision for most estimates, with cross-trait ones benefiting more than within-trait ones and latent factors such as VT ($\mathbf{f}$) and G-E covariance ($\mathbf{w}$ and $\mathbf{v}$) benefiting more than estimates based on observed variables such as ($\mathbf{\delta}$ and $\mathbf{g}_c$). However, fixing $\mathbf{a}$ to incorrect values introduced bias for these latent parameters. Furthermore, we showed that analyzing data using a univariate model when the data-generating process was actually multivariate can introduce substantial bias to parameter estimates, especially when cross-trait effects are strong. In the following sections, we discuss the strengths and assumptions of the multivariate SEM–PGS model, interpret the results in greater detail, and offer best-practice guidance for its application.

\subsection{A Novel Tool for Gene-Environment Interplay}
By integrating parental and offspring phenotypes with transmitted and non-transmitted PGSs for multiple traits in a SEM framework, the multivariate SEM–PGS model introduces a novel approach to studying genetic and environmental processes within families. Unlike traditional family designs that rely on assumptions about the sources of phenotypic resemblance, this framework uses directly observable relationships between phenotypes and transmitted and non-transmitted PGSs to provide empirical evidence of VT and AM. Moreover, recent family-based models that utilize PGSs have estimated direct and indirect genetic effects for single traits \autocite{balbona_estimation_2021, kong_nature_2018}, but have not properly accounted for cross-trait influences. For instance, parental traits may often influence different offspring traits (i.e., cross-trait VT, such as parental education influencing offspring BMI) and parents may assort across different traits (e.g., across different psychiatric disorders). In this manuscript, we have demonstrated that within-trait estimates may also reflect unmodeled cross-trait effects, and not modeling these cross-trait influences, or modeling them in a univariate framework, can lead to incorrect inference.

Through forward-time simulations, we confirmed that the multivariate SEM-PGS model provides unbiased estimates of direct genetic effects ($\mathbf{a}$ and $\boldsymbol{\delta}$), VT effects ($\mathbf{f}$), genetic nurture effects ($\boldsymbol{\phi}$ and $\boldsymbol{\rho}$), G-E covariance $\boldsymbol{w}$ and $\boldsymbol{v}$, AM effects ($\boldsymbol{\mu}$), and other parameters when its assumptions are met. The forward-time simulation approach also provided an independent check on our mathematical derivations and path-tracing rules for copaths. Furthermore, we found that the precision of latent parameter estimates improved with both $N_{\text{trio}}$ and $r^2_{\text{PGS}}$; we recommend $N_{\text{trio}} > 32K$ and $r^2_{\text{PGS}} > 0.01$ as rough lower bounds for reliable estimation, although these trade off, so smaller sample sizes can be used as $r^2_{\text{PGS}}$ grows and vice-versa. Nevertheless, this highlights two limitations of the current model. First, it requires large family-based datasets with genomic information; at present, only a few resources, such as the Norwegian Mother, Father, and Child Cohort Study (MoBa), meet these criteria. However, the number of such datasets is expected to grow in the coming decade, given the distinct advantages of family-based genomic data over individual-level data \autocite{davies_importance_2024}. Second, the model is currently limited to traits with sufficiently high ${r}^2_{pgs}$. While this threshold is already met for many traits, the expansion of large-scale GWASs will likely increase the range of traits for which the model can be applied.

Furthermore, our analysis demonstrates that fitting a univariate SEM-PGS model to data with an underlying bivariate causal structure introduces bias in multiple estimates, particularly when there are strong cross-trait effects. For instance, a significant indirect genetic effect estimate for $\mathbf{f}$ in a univariate analysis of depressive symptoms could be an artifact from an unmodeled trait that is genetically correlated with depression. While these indirect effects are themselves real, the parameters must be interpreted with the caveat that they may represent integrated effects from multiple pathways. This same caution applies to the interpretation of multivariate SEM-PGS results, as it is practically infeasible to account for all potential inter-generational transmission pathways with a limited number of measured phenotypes. By the same token, our results suggest that attempting to estimate cross-trait effects using a univariate model can also lead to biased estimates, and such models should be interpreted with caution.

Building on our previous univariate model \autocite{balbona_estimation_2021}, the present framework provides a more refined parameterization of genetic nurture. In the univariate setting, the G-E covariance ($\boldsymbol{w}$) and genetic nurture (the effect of parental genes on offspring mediated through parental phenotype) are conceptually distinct but numerically identical. This equivalence arises because the same set of genes that influence the parental phenotype, and thereby affect the offspring phenotype through VT, also influence the offspring phenotype directly. This equivalence does not hold once cross-trait effects are introduced. For example, when parental trait 1 influences offspring trait 2 through VT, the genes that shape parental trait 1 are not identical to the genes directly influencing offspring trait 2. In such cases, genetic nurture still contributes to cross-trait G-E covariance, but it is no longer reducible to either of the within-trait G–E covariance terms or even to the cross-trait covariance term itself. In a multivariate context, therefore, G–E covariance and genetic nurture are related but not synonymous.  

To capture this distinction, we introduce two parameters in the multivariate SEM–PGS model, $\boldsymbol{\phi}$ and $\boldsymbol{\rho}$, that explicitly represent genetic nurture pathways. These parameters separate genetic nurture effects arising from both within-trait and cross-trait influences, and they are defined so as to be independent of AM. This framework clarifies the interpretation of genetic nurture in multivariate models and differentiates our approach from earlier measures such as $\eta$ in \textcite{kong_nature_2018} and the regression-based $\alpha$ estimates summarized in \textcite{young_mendelian_2022}, both of which can be conflated with AM.

\subsection{Model Assumptions}
Like all statistical models, the current model relies on several assumptions regarding genetic effects, AM, and environmental influences to provide accurate estimates. First, as currently implemented when estimating $\mathbf{a}$ within the model itself rather than using external estimates to fix it, the model assumes that the residual parent–offspring covariance after accounting for VT, AM, and direct PGS effects is purely due to additive genetic factors, and uses this residual covariance to estimate $\mathbf{a}$. However, in truth, this residual covariance is unlikely to be purely genetic in origin; for example, shared environmental influences can also contribute. As a result, estimating $\mathbf{a}$ directly within the model can lead to biased estimates of other quanities. Unbiased estimates of $\mathbf{a}$ can instead be obtained from external methods such as Relatedness Disequilibrium Regression (RDR) \autocite{young_relatedness_2018} or sibling-based regression \autocite{visscher_assumption-free_2006}, including multivariate extensions of these approaches. For example, in trio datasets with parental genotypes and offspring phenotypes, RDR can be used to estimate founder-generation genetic variance, from which $\mathbf{a}$ can be derived by imposing an additional constraint on the total additive genetic variance in the base population: $\mathbf{V}\mathbf{A}_{Base} = (\mathbf{a}^2 + \boldsymbol{\delta}^2)(\mathbf{a}^2 + \boldsymbol{\delta}^2 + \mathbf{V}_\epsilon)$. Fixing $\mathbf{a}$ in this way reduces the standard errors of other parameters, enabling analyses in smaller samples and with weaker PGSs. However, for this approach to provide valid standard errors of estimates, the uncertainty in the external estimate of $\mathbf{a}$ should be incorporated into the SEM–PGS model. One strategy (a parametric bootstrap approach)is to draw $\mathbf{a}$ repeatedly from a normal distribution with mean equal to its point estimate and standard deviation equal to its standard error, re-fitting the model at each draw. This propagates uncertainty from the independent method into the SEM–PGS framework. Alternatively, this same principle could be implemented in a Bayesian framework by placing a prior on $\mathbf{a}$ informed by its external estimate. Our sensitivity analyses suggest that small misspecifications of $\mathbf{a}$ have limited impact on most estimates, but cross-trait VT and G-E covariance parameters are more vulnerable. It is important to note that, if parental phenotypes are included in the model, an additional latent factor (e.g., a shared environmental factor across both generations) should also be specified to absorb any residual parent–offspring covariance, ensuring that the model fits well. Alternatively, it is possible to simply omit parental phenotypes altogether when fixing $\mathbf{a}$ and still achieve unbiased estimates of all other parameters. Overall, future SEM–PGS models should aim to better estimate $\mathbf{a}$ internally (e.g., by incorporating sibling or extended relative information), but in the meantime, we believe that drawing on external approaches such as RDR will generally yield nearly unbiased estimates.

Second, the model assumes that genetic effects are consistent between parents and offspring. Specifically, it posits that $\boldsymbol{\delta}_{p/m} = \boldsymbol{\delta}_o$ and $\mathbf{a}_{p/m} = \mathbf{a}_o$. The model also implicitly assumes a genetic correlation of unity between parental and offspring phenotypes. This assumption may be reasonable for traits such as height or weight, but may be violated in practice when social traits are measured at different ages or when the factors influencing the phenotype differ across cohorts. To address this limitation, we are working on extending the model by using informatio on additional relatives to allow parental genetic effects to differ from offspring genetic effects. 

Third, the model assumes equivalence in genetic correlation between latent and observed genetic variables ($k_{12} = j_{12}$). This assumption is generally plausible, since measured causal alleles are likely to resemble unmeasured (latent) alleles in their cross-trait genetic correlations. However, this assumption might be violated if common variants (more likely captured by PGSs) and rare variants differ in their contributions to cross-trait genetic correlations.

Fourth, the model assumes the absence of dominance or epistasis effects. This assumption is common in most statistical genetic models, as estimating dominance or epistasis at the level of PGSs is challenging and rarely attempted. However, since dominance and epistasis genetic effects are statistically orthogonal components to additive genetic effects, ignoring them is only likely to inflate $\mathbf{V_\epsilon}$, which alters the interpretation of $\mathbf{V_\epsilon}$ but should not bias or otherwise alter interpretations of other parameter estimates.

Fifth, in its present form, it assumes that VT effects are identical for both father and mother ($\mathbf{f}_{p} = \mathbf{f}_{m}$). This assumption may not hold true for certain traits where VT from mothers differs from fathers. It would be simple to reparameterize the model to allow for sex-different VT paths, although doing so would reduce estimates' precision. 

Finally, the SEM–PGS model makes two assumptions about AM. One is that assortment is primarily phenotypic, even though in practice it may also arise through social homogamy or genetic homogamy \autocite{robinson_genetic_2017}. Because these mechanisms are not mutually exclusive, misclassifying the source of AM can bias parameter estimates by leaving residual spousal similarity unaccounted for. For example, treating social homogamy as phenotypic assortment could lead to underestimation of environmental influences such as $\mathbf{V}_F$ and $\mathbf{w}$. In our simulations, however, misspecifying social or genetic homogamy as phenotypic assortment produced only limited bias in the final parameter estimates. Furthermore, a major advantage of incorporating genomic data is that the type of AM can be tested empirically before fitting the full bivariate SEM–PGS model. This is done by comparing the observed spousal phenotypic correlation ($\mathbf{g}$) to the correlation implied by haplotypic PGSs: if the latter exceeds the observed correlation, it suggests genetic homogamy, whereas if the observed correlation is larger, it implies social homogamy. The SEM–PGS framework has sufficient data to estimate these correlations and can be extended to model each contribution directly by adding latent AM factors. We have previously implemented such modifications in the Cascade model, which is based on extended twin family data, and adapting them here would be straightforward \autocite{keller_modeling_2009}. Alternatively, a shared environmental factor between parents could be added to the model to soak up residual spousal covariance not captured by that implied from the spousal PGS covariance. 

Another assumption is that AM has reached equilibrium, meaning that the level of assortment in the parental generation matches that in the offspring generation. Again, an advantage of using genomic data is that this assumption can be evaluated by comparing the within-trait increase in the covariance of within-individual haplotypic PGSs ($\mathbf{g}_{c}$), which reflects AM in the grandparental generation and before, to the covariance of cross-individual haplotypic PGSs ($.5(\mathbf{g}_{t} + \mathbf{g}_{t}^T)$), which reflects AM in the parental generation. When disequilibrium is present, the mathematical expectations of the parameters can be modified accordingly, and we provide a derivation in the supplement for when AM is only present in the parental generation.

\section{Conclusion}
In summary, we introduced the multivariate SEM–PGS model, a flexible statistical framework that integrates within- and cross-trait direct genetic effects, genetic nurture, G-E covariance, VT, and AM into a single structural equation modeling approach. A key innovation is the development of multivariate path-tracing rules, which enable rigorous modeling of cross-trait AM. Simulations demonstrated that the model yields unbiased estimates and increasing precision with larger trio samples and more predictive polygenic scores, while a complementary strategy of borrowing information from external methods highlights its adaptability. We provide practical suggestions for fitting the model in the Supplement, but emphasize that the exemplar specification presented here should be adapted to the unique properties of different datasets. Although the framework rests on assumptions and requires relatively large family-genomic samples, it offers a general foundation that can be tailored for different data and questions. By providing tools to disentangle complex intergenerational pathways, the multivariate SEM–PGS model advances the study of how genetic and environmental influences combine to shape human traits and lays the groundwork for future extensions to more nuanced and biologically informed models.

\newpage
\printbibliography

@article{keller_quantifying_2005,
	title = {Quantifying and Addressing Parameter Indeterminacy in the Classical Twin Design},
	volume = {8},
	issn = {1839-2628, 1832-4274},
	url = {https://www.cambridge.org/core/journals/twin-research-and-human-genetics/article/quantifying-and-addressing-parameter-indeterminacy-in-the-classical-twin-design/43A336C6B40FEFC3063488C509BA58ED},
	doi = {10.1375/twin.8.3.201},
	pages = {201--213},
	number = {3},
	journaltitle = {Twin Research and Human Genetics},
	author = {Keller, Matthew C. and Coventry, William L.},
	urldate = {2025-09-15},
	date = {2005-06},
	langid = {english},
}

@article{chen_comparing_2025,
	title = {Comparing Forward-Time and Model-Implied Covariance-Based Simulations for Evaluating Family-Based Genetic Models},
	journaltitle = {Manuscript in Preparation},
	author = {Chen, Tong and Lyu, Xuanyu and Keller, Matthew C.},
	date = {2025},
}

@article{maes_genetic_1997,
	title = {Genetic and Environmental Factors in Relative Body Weight and Human Adiposity},
	volume = {27},
	issn = {1573-3297},
	url = {https://doi.org/10.1023/A:1025635913927},
	doi = {10.1023/A:1025635913927},
	pages = {325--351},
	number = {4},
	journaltitle = {Behavior Genetics},
	shortjournal = {Behav Genet},
	author = {Maes, Hermine H. M. and Neale, Michael C. and Eaves, Lindon J.},
	urldate = {2025-09-10},
	date = {1997-07-01},
	langid = {english},
}

@article{heath_resolution_1985,
	title = {The resolution of cultural and biological inheritance: Informativeness of different relationships},
	volume = {15},
	issn = {1573-3297},
	url = {https://doi.org/10.1007/BF01066238},
	doi = {10.1007/BF01066238},
	shorttitle = {The resolution of cultural and biological inheritance},
	pages = {439--465},
	number = {5},
	journaltitle = {Behavior Genetics},
	shortjournal = {Behav Genet},
	author = {Heath, A. C. and Kendler, K. S. and Eaves, L. J. and Markell, D.},
	urldate = {2025-09-10},
	date = {1985-09-01},
	langid = {english},
}

@book{kaplan_structural_2008,
	title = {Structural Equation Modeling: Foundations and Extensions},
	isbn = {978-1-4522-4512-6},
	shorttitle = {Structural Equation Modeling},
	pagetotal = {273},
	publisher = {{SAGE} Publications},
	author = {Kaplan, David},
	date = {2008-07-23},
	langid = {english},
	note = {Google-Books-{ID}: {LwZ}1AwAAQBAJ},
}

@book{neale_methodology_2013,
	title = {Methodology for Genetic Studies of Twins and Families},
	isbn = {978-94-015-8018-2},
	pagetotal = {503},
	publisher = {Springer Science \& Business Media},
	author = {Neale, M. and Cardon, L. R.},
	date = {2013-03-09},
	langid = {english},
	note = {Google-Books-{ID}: {EKYyBwAAQBAJ}},
}

@article{young_mendelian_2022,
	title = {Mendelian imputation of parental genotypes improves estimates of direct genetic effects},
	volume = {54},
	rights = {2022 The Author(s)},
	issn = {1546-1718},
	url = {https://www.nature.com/articles/s41588-022-01085-0},
	doi = {10.1038/s41588-022-01085-0},
	pages = {897--905},
	number = {6},
	journaltitle = {Nature Genetics},
	shortjournal = {Nat Genet},
	author = {Young, Alexander I. and Nehzati, Seyed Moeen and Benonisdottir, Stefania and Okbay, Aysu and Jayashankar, Hariharan and Lee, Chanwook and Cesarini, David and Benjamin, Daniel J. and Turley, Patrick and Kong, Augustine},
	urldate = {2025-09-03},
	date = {2022-06},
	langid = {english},
	note = {Publisher: Nature Publishing Group},
}

@article{lyu_detecting_2025,
	title = {Detecting {mtDNA} Effects with an Extended Pedigree Model: An Analysis of Statistical Power and Estimation Bias},
	volume = {55},
	issn = {1573-3297},
	url = {https://doi.org/10.1007/s10519-025-10225-1},
	doi = {10.1007/s10519-025-10225-1},
	shorttitle = {Detecting {mtDNA} Effects with an Extended Pedigree Model},
	pages = {320--337},
	number = {4},
	journaltitle = {Behavior Genetics},
	shortjournal = {Behav Genet},
	author = {Lyu, Xuanyu and Hunter, Michael D. and Burt, S. Alexandra and Good, Rachel and Carroll, Sarah L. and Mason Garrison, S.},
	urldate = {2025-08-26},
	date = {2025-07-01},
	langid = {english},
}

@article{bulmer_effect_1971,
	title = {The Effect of Selection on Genetic Variability},
	volume = {105},
	issn = {0003-0147},
	url = {https://www.journals.uchicago.edu/doi/10.1086/282718},
	doi = {10.1086/282718},
	pages = {201--211},
	number = {943},
	journaltitle = {The American Naturalist},
	author = {Bulmer, M. G.},
	urldate = {2025-07-28},
	date = {1971-05},
	note = {Publisher: The University of Chicago Press},
}

@article{wright_method_1934,
	title = {The Method of Path Coefficients},
	volume = {5},
	issn = {0003-4851},
	url = {https://www.jstor.org/stable/2957502},
	pages = {161--215},
	number = {3},
	journaltitle = {The Annals of Mathematical Statistics},
	author = {Wright, Sewall},
	urldate = {2025-07-21},
	date = {1934},
	note = {Publisher: Institute of Mathematical Statistics},
}

@article{young_relatedness_2018,
	title = {Relatedness disequilibrium regression estimates heritability without environmental bias},
	volume = {50},
	rights = {2018 The Author(s)},
	issn = {1546-1718},
	url = {https://www.nature.com/articles/s41588-018-0178-9},
	doi = {10.1038/s41588-018-0178-9},
	pages = {1304--1310},
	number = {9},
	journaltitle = {Nature Genetics},
	shortjournal = {Nat Genet},
	author = {Young, Alexander I. and Frigge, Michael L. and Gudbjartsson, Daniel F. and Thorleifsson, Gudmar and Bjornsdottir, Gyda and Sulem, Patrick and Masson, Gisli and Thorsteinsdottir, Unnur and Stefansson, Kari and Kong, Augustine},
	urldate = {2023-02-21},
	date = {2018-09},
	langid = {english},
	note = {Number: 9
Publisher: Nature Publishing Group},
}

@article{shih_review_2004,
	title = {A review of the evidence from family, twin and adoption studies for a genetic contribution to adult psychiatric disorders},
	volume = {16},
	issn = {0954-0261},
	url = {https://doi.org/10.1080/09540260400014401},
	doi = {10.1080/09540260400014401},
	pages = {260--283},
	number = {4},
	journaltitle = {International Review of Psychiatry},
	author = {Shih, Regina A. and Belmonte, Pamela L. and Zandi, Peter P.},
	urldate = {2025-03-12},
	date = {2004-11-01},
	note = {Publisher: Taylor \& Francis
\_eprint: https://doi.org/10.1080/09540260400014401},
}

@article{rutter_testing_2001,
	title = {Testing hypotheses on specific environmental causal effects on behavior},
	volume = {127},
	issn = {1939-1455},
	doi = {10.1037/0033-2909.127.3.291},
	pages = {291--324},
	number = {3},
	journaltitle = {Psychological Bulletin},
	author = {Rutter, Michael and Pickles, Andrew and Murray, Robin and Eaves, Lindon},
	date = {2001},
	note = {Place: {US}
Publisher: American Psychological Association},
}

@article{horn_intellectual_1979,
	title = {Intellectual resemblance among adoptive and biological relatives: The texas Adoption Project},
	volume = {9},
	issn = {1573-3297},
	url = {https://doi.org/10.1007/BF01071300},
	doi = {10.1007/BF01071300},
	shorttitle = {Intellectual resemblance among adoptive and biological relatives},
	pages = {177--207},
	number = {3},
	journaltitle = {Behavior Genetics},
	shortjournal = {Behav Genet},
	author = {Horn, Joseph M. and Loehlin, John C. and Willerman, Lee},
	urldate = {2025-03-12},
	date = {1979-05-01},
	langid = {english},
}

@article{abdellaoui_geneenvironment_2022,
	title = {Gene–environment correlations across geographic regions affect genome-wide association studies},
	volume = {54},
	rights = {2022 The Author(s)},
	issn = {1546-1718},
	url = {https://www.nature.com/articles/s41588-022-01158-0},
	doi = {10.1038/s41588-022-01158-0},
	pages = {1345--1354},
	number = {9},
	journaltitle = {Nature Genetics},
	shortjournal = {Nat Genet},
	author = {Abdellaoui, Abdel and Dolan, Conor V. and Verweij, Karin J. H. and Nivard, Michel G.},
	urldate = {2025-03-11},
	date = {2022-09},
	langid = {english},
	note = {Publisher: Nature Publishing Group},
}

@article{engzell_heritability_2019,
	title = {Heritability of education rises with intergenerational mobility},
	volume = {116},
	issn = {0027-8424},
	url = {https://www.ncbi.nlm.nih.gov/pmc/articles/PMC6926022/},
	doi = {10.1073/pnas.1912998116},
	pages = {25386--25388},
	number = {51},
	journaltitle = {Proceedings of the National Academy of Sciences of the United States of America},
	shortjournal = {Proc Natl Acad Sci U S A},
	author = {Engzell, Per and Tropf, Felix C.},
	urldate = {2025-03-11},
	date = {2019-12-17},
	pmid = {31792187},
	pmcid = {PMC6926022},
}

@article{kendler_family_2015,
	title = {Family environment and the malleability of cognitive ability: A Swedish national home-reared and adopted-away cosibling control study},
	volume = {112},
	issn = {0027-8424},
	url = {https://www.ncbi.nlm.nih.gov/pmc/articles/PMC4403216/},
	doi = {10.1073/pnas.1417106112},
	shorttitle = {Family environment and the malleability of cognitive ability},
	pages = {4612--4617},
	number = {15},
	journaltitle = {Proceedings of the National Academy of Sciences of the United States of America},
	shortjournal = {Proc Natl Acad Sci U S A},
	author = {Kendler, Kenneth S. and Turkheimer, Eric and Ohlsson, Henrik and Sundquist, Jan and Sundquist, Kristina},
	urldate = {2025-03-11},
	date = {2015-04-14},
	pmid = {25831538},
	pmcid = {PMC4403216},
}

@article{polderman_meta-analysis_2015,
	title = {Meta-analysis of the heritability of human traits based on fifty years of twin studies},
	volume = {47},
	issn = {1546-1718},
	doi = {10.1038/ng.3285},
	pages = {702--709},
	number = {7},
	journaltitle = {Nature Genetics},
	shortjournal = {Nat Genet},
	author = {Polderman, Tinca J. C. and Benyamin, Beben and de Leeuw, Christiaan A. and Sullivan, Patrick F. and van Bochoven, Arjen and Visscher, Peter M. and Posthuma, Danielle},
	date = {2015-07},
	pmid = {25985137},
}

@article{kong_nature_2018,
	title = {The nature of nurture: Effects of parental genotypes},
	volume = {359},
	url = {https://www.science.org/doi/10.1126/science.aan6877},
	doi = {10.1126/science.aan6877},
	shorttitle = {The nature of nurture},
	pages = {424--428},
	number = {6374},
	journaltitle = {Science},
	author = {Kong, Augustine and Thorleifsson, Gudmar and Frigge, Michael L. and Vilhjalmsson, Bjarni J. and Young, Alexander I. and Thorgeirsson, Thorgeir E. and Benonisdottir, Stefania and Oddsson, Asmundur and Halldorsson, Bjarni V. and Masson, Gisli and Gudbjartsson, Daniel F. and Helgason, Agnar and Bjornsdottir, Gyda and Thorsteinsdottir, Unnur and Stefansson, Kari},
	urldate = {2022-10-23},
	date = {2018-01-26},
	langid = {american},
	note = {Publisher: American Association for the Advancement of Science},
}

@article{border_rbahadur_2023,
	title = {{rBahadur}: efficient simulation of structured high-dimensional genotype data with applications to assortative mating},
	volume = {24},
	issn = {1471-2105},
	url = {https://doi.org/10.1186/s12859-023-05442-6},
	doi = {10.1186/s12859-023-05442-6},
	shorttitle = {{rBahadur}},
	pages = {314},
	number = {1},
	journaltitle = {{BMC} Bioinformatics},
	shortjournal = {{BMC} Bioinformatics},
	author = {Border, Richard and Malik, Osman Asif},
	urldate = {2025-01-13},
	date = {2023-08-18},
	langid = {american},
}

@article{neale_openmx_2016,
	title = {{OpenMx} 2.0: Extended Structural Equation and Statistical Modeling},
	volume = {81},
	issn = {1860-0980},
	url = {https://doi.org/10.1007/s11336-014-9435-8},
	doi = {10.1007/s11336-014-9435-8},
	shorttitle = {{OpenMx} 2.0},
	pages = {535--549},
	number = {2},
	journaltitle = {Psychometrika},
	shortjournal = {Psychometrika},
	author = {Neale, Michael C. and Hunter, Michael D. and Pritikin, Joshua N. and Zahery, Mahsa and Brick, Timothy R. and Kirkpatrick, Robert M. and Estabrook, Ryne and Bates, Timothy C. and Maes, Hermine H. and Boker, Steven M.},
	urldate = {2024-12-18},
	date = {2016-06-01},
	langid = {english},
}

@article{robinson_genetic_2017,
	title = {Genetic evidence of assortative mating in humans},
	volume = {1},
	rights = {2017 Springer Nature Limited},
	issn = {2397-3374},
	url = {https://www.nature.com/articles/s41562-016-0016},
	doi = {10.1038/s41562-016-0016},
	pages = {1--13},
	number = {1},
	journaltitle = {Nature Human Behaviour},
	shortjournal = {Nat Hum Behav},
	author = {Robinson, Matthew R. and Kleinman, Aaron and Graff, Mariaelisa and Vinkhuyzen, Anna A. E. and Couper, David and Miller, Michael B. and Peyrot, Wouter J. and Abdellaoui, Abdel and Zietsch, Brendan P. and Nolte, Ilja M. and van Vliet-Ostaptchouk, Jana V. and Snieder, Harold and Medland, Sarah E. and Martin, Nicholas G. and Magnusson, Patrik K. E. and Iacono, William G. and {McGue}, Matt and North, Kari E. and Yang, Jian and Visscher, Peter M.},
	urldate = {2024-12-09},
	date = {2017-01-09},
	langid = {english},
	note = {Publisher: Nature Publishing Group},
}

@article{vandenberg_assortative_1972,
	title = {Assortative mating, or who marries whom?},
	volume = {2},
	issn = {1573-3297},
	url = {https://doi.org/10.1007/BF01065686},
	doi = {10.1007/BF01065686},
	pages = {127--157},
	number = {2},
	journaltitle = {Behavior Genetics},
	shortjournal = {Behav Genet},
	author = {Vandenberg, Steven G.},
	urldate = {2024-12-06},
	date = {1972-06-01},
	langid = {english},
}

@article{keller_modeling_2009,
	title = {Modeling Extended Twin Family Data I: Description of the Cascade Model},
	volume = {12},
	issn = {1832-4274, 1839-2628},
	url = {https://www.cambridge.org/core/product/identifier/S1832427400009543/type/journal_article},
	doi = {10.1375/twin.12.1.8},
	shorttitle = {Modeling Extended Twin Family Data I},
	pages = {8--18},
	number = {1},
	journaltitle = {Twin Research and Human Genetics},
	shortjournal = {Twin Res Hum Genet},
	author = {Keller, Matthew C. and Medland, Sarah E. and Duncan, Laramie E. and Hatemi, Peter K. and Neale, Michael C. and Maes, Hermine H. M. and Eaves, Lindon J.},
	urldate = {2023-08-31},
	date = {2009-02-01},
	langid = {english},
}

@article{balbona_estimation_2021,
	title = {Estimation of Parental Effects Using Polygenic Scores},
	volume = {51},
	issn = {1573-3297},
	url = {https://doi.org/10.1007/s10519-020-10032-w},
	doi = {10.1007/s10519-020-10032-w},
	pages = {264--278},
	number = {3},
	journaltitle = {Behavior Genetics},
	shortjournal = {Behav Genet},
	author = {Balbona, Jared V. and Kim, Yongkang and Keller, Matthew C.},
	urldate = {2023-08-31},
	date = {2021-05-01},
	langid = {english},
}

@article{balbona_estimation_2022,
	title = {The estimation of environmental and genetic parental influences},
	volume = {34},
	issn = {0954-5794, 1469-2198},
	url = {https://www.cambridge.org/core/product/identifier/S0954579422000761/type/journal_article},
	doi = {10.1017/S0954579422000761},
	pages = {1876--1886},
	number = {5},
	journaltitle = {Development and Psychopathology},
	shortjournal = {Dev Psychopathol},
	author = {Balbona, Jared V. and Kim, Yongkang and Keller, Matthew C.},
	urldate = {2024-09-18},
	date = {2022-12},
	langid = {english},
}

@article{kim_bias_2021,
	title = {Bias and Precision of Parameter Estimates from Models Using Polygenic Scores to Estimate Environmental and Genetic Parental Influences},
	volume = {51},
	issn = {1573-3297},
	url = {https://doi.org/10.1007/s10519-020-10033-9},
	doi = {10.1007/s10519-020-10033-9},
	pages = {279--288},
	number = {3},
	journaltitle = {Behavior Genetics},
	shortjournal = {Behav Genet},
	author = {Kim, Yongkang and Balbona, Jared V. and Keller, Matthew C.},
	urldate = {2023-08-31},
	date = {2021-05-01},
	langid = {english},
}

@article{hart_nurture_2021,
	title = {Nurture might be nature: cautionary tales and proposed solutions},
	volume = {6},
	rights = {2021 The Author(s)},
	issn = {2056-7936},
	url = {https://www.nature.com/articles/s41539-020-00079-z},
	doi = {10.1038/s41539-020-00079-z},
	shorttitle = {Nurture might be nature},
	pages = {1--12},
	number = {1},
	journaltitle = {npj Science of Learning},
	shortjournal = {npj Sci. Learn.},
	author = {Hart, Sara A. and Little, Callie and van Bergen, Elsje},
	urldate = {2024-02-21},
	date = {2021-01-08},
	langid = {english},
	note = {Number: 1
Publisher: Nature Publishing Group},
}

@article{visscher_assumption-free_2006,
	title = {Assumption-Free Estimation of Heritability from Genome-Wide Identity-by-Descent Sharing between Full Siblings},
	volume = {2},
	issn = {1553-7404},
	url = {https://journals.plos.org/plosgenetics/article?id=10.1371/journal.pgen.0020041},
	doi = {10.1371/journal.pgen.0020041},
	pages = {e41},
	number = {3},
	journaltitle = {{PLOS} Genetics},
	shortjournal = {{PLOS} Genetics},
	author = {Visscher, Peter M. and Medland, Sarah E. and Ferreira, Manuel A. R. and Morley, Katherine I. and Zhu, Gu and Cornes, Belinda K. and Montgomery, Grant W. and Martin, Nicholas G.},
	urldate = {2024-07-23},
	date = {2006-03-24},
	langid = {english},
	note = {Publisher: Public Library of Science},
}

@article{mcadams_annual_2023,
	title = {Annual Research Review: Towards a deeper understanding of nature and nurture: combining family-based quasi-experimental methods with genomic data},
	volume = {64},
	rights = {© 2022 The Authors. Journal of Child Psychology and Psychiatry published by John Wiley \& Sons Ltd on behalf of Association for Child and Adolescent Mental Health.},
	issn = {1469-7610},
	url = {https://onlinelibrary.wiley.com/doi/abs/10.1111/jcpp.13720},
	doi = {10.1111/jcpp.13720},
	shorttitle = {Annual Research Review},
	pages = {693--707},
	number = {4},
	journaltitle = {Journal of Child Psychology and Psychiatry},
	author = {{McAdams}, Tom A. and Cheesman, Rosa and Ahmadzadeh, Yasmin I.},
	urldate = {2024-05-22},
	date = {2023},
	langid = {english},
	note = {\_eprint: https://onlinelibrary.wiley.com/doi/pdf/10.1111/jcpp.13720},
}

@article{bulik-sullivan_atlas_2015,
	title = {An atlas of genetic correlations across human diseases and traits},
	volume = {47},
	rights = {2015 Springer Nature America, Inc.},
	issn = {1546-1718},
	url = {https://www.nature.com/articles/ng.3406},
	doi = {10.1038/ng.3406},
	pages = {1236--1241},
	number = {11},
	journaltitle = {Nature Genetics},
	shortjournal = {Nat Genet},
	author = {Bulik-Sullivan, Brendan and Finucane, Hilary K. and Anttila, Verneri and Gusev, Alexander and Day, Felix R. and Loh, Po-Ru and Duncan, Laramie and Perry, John R. B. and Patterson, Nick and Robinson, Elise B. and Daly, Mark J. and Price, Alkes L. and Neale, Benjamin M.},
	urldate = {2024-11-11},
	date = {2015-11},
	langid = {english},
	note = {Publisher: Nature Publishing Group},
}

@article{mcadams_accounting_2014,
	title = {Accounting for genetic and environmental confounds in associations between parent and child characteristics: a systematic review of children-of-twins studies},
	volume = {140},
	issn = {1939-1455},
	doi = {10.1037/a0036416},
	shorttitle = {Accounting for genetic and environmental confounds in associations between parent and child characteristics},
	pages = {1138--1173},
	number = {4},
	journaltitle = {Psychological Bulletin},
	shortjournal = {Psychol Bull},
	author = {{McAdams}, Tom A. and Neiderhiser, Jenae M. and Rijsdijk, Fruhling V. and Narusyte, Jurgita and Lichtenstein, Paul and Eley, Thalia C.},
	date = {2014-07},
	pmid = {24749497},
}

@article{demange_estimating_2022,
	title = {Estimating effects of parents’ cognitive and non-cognitive skills on offspring education using polygenic scores},
	volume = {13},
	rights = {2022 The Author(s)},
	issn = {2041-1723},
	url = {https://www.nature.com/articles/s41467-022-32003-x},
	doi = {10.1038/s41467-022-32003-x},
	pages = {4801},
	number = {1},
	journaltitle = {Nature Communications},
	shortjournal = {Nat Commun},
	author = {Demange, Perline A. and Hottenga, Jouke Jan and Abdellaoui, Abdel and Eilertsen, Espen Moen and Malanchini, Margherita and Domingue, Benjamin W. and Armstrong-Carter, Emma and de Zeeuw, Eveline L. and Rimfeld, Kaili and Boomsma, Dorret I. and van Bergen, Elsje and Breen, Gerome and Nivard, Michel G. and Cheesman, Rosa},
	urldate = {2024-10-01},
	date = {2022-08-23},
	langid = {english},
	note = {Publisher: Nature Publishing Group},
}

@article{dolan_incorporating_2021,
	title = {Incorporating Polygenic Risk Scores in the {ACE} Twin Model to Estimate A–C Covariance},
	volume = {51},
	issn = {1573-3297},
	url = {https://doi.org/10.1007/s10519-020-10035-7},
	doi = {10.1007/s10519-020-10035-7},
	pages = {237--249},
	number = {3},
	journaltitle = {Behavior Genetics},
	shortjournal = {Behav Genet},
	author = {Dolan, Conor V. and Huijskens, Roel C. A. and Minică, Camelia C. and Neale, Michael C. and Boomsma, Dorret I.},
	urldate = {2024-10-01},
	date = {2021-05-01},
	langid = {english},
}

@article{cross-disorder_group_of_the_psychiatric_genomics_consortium_genetic_2013,
	title = {Genetic relationship between five psychiatric disorders estimated from genome-wide {SNPs}},
	volume = {45},
	issn = {1546-1718},
	doi = {10.1038/ng.2711},
	pages = {984--994},
	number = {9},
	journaltitle = {Nature Genetics},
	shortjournal = {Nat Genet},
	author = {{Cross-Disorder Group of the Psychiatric Genomics Consortium} and Lee, S. Hong and Ripke, Stephan and Neale, Benjamin M. and Faraone, Stephen V. and Purcell, Shaun M. and Perlis, Roy H. and Mowry, Bryan J. and Thapar, Anita and Goddard, Michael E. and Witte, John S. and Absher, Devin and Agartz, Ingrid and Akil, Huda and Amin, Farooq and Andreassen, Ole A. and Anjorin, Adebayo and Anney, Richard and Anttila, Verneri and Arking, Dan E. and Asherson, Philip and Azevedo, Maria H. and Backlund, Lena and Badner, Judith A. and Bailey, Anthony J. and Banaschewski, Tobias and Barchas, Jack D. and Barnes, Michael R. and Barrett, Thomas B. and Bass, Nicholas and Battaglia, Agatino and Bauer, Michael and Bayés, Mònica and Bellivier, Frank and Bergen, Sarah E. and Berrettini, Wade and Betancur, Catalina and Bettecken, Thomas and Biederman, Joseph and Binder, Elisabeth B. and Black, Donald W. and Blackwood, Douglas H. R. and Bloss, Cinnamon S. and Boehnke, Michael and Boomsma, Dorret I. and Breen, Gerome and Breuer, René and Bruggeman, Richard and Cormican, Paul and Buccola, Nancy G. and Buitelaar, Jan K. and Bunney, William E. and Buxbaum, Joseph D. and Byerley, William F. and Byrne, Enda M. and Caesar, Sian and Cahn, Wiepke and Cantor, Rita M. and Casas, Miguel and Chakravarti, Aravinda and Chambert, Kimberly and Choudhury, Khalid and Cichon, Sven and Cloninger, C. Robert and Collier, David A. and Cook, Edwin H. and Coon, Hilary and Cormand, Bru and Corvin, Aiden and Coryell, William H. and Craig, David W. and Craig, Ian W. and Crosbie, Jennifer and Cuccaro, Michael L. and Curtis, David and Czamara, Darina and Datta, Susmita and Dawson, Geraldine and Day, Richard and De Geus, Eco J. and Degenhardt, Franziska and Djurovic, Srdjan and Donohoe, Gary J. and Doyle, Alysa E. and Duan, Jubao and Dudbridge, Frank and Duketis, Eftichia and Ebstein, Richard P. and Edenberg, Howard J. and Elia, Josephine and Ennis, Sean and Etain, Bruno and Fanous, Ayman and Farmer, Anne E. and Ferrier, I. Nicol and Flickinger, Matthew and Fombonne, Eric and Foroud, Tatiana and Frank, Josef and Franke, Barbara and Fraser, Christine and Freedman, Robert and Freimer, Nelson B. and Freitag, Christine M. and Friedl, Marion and Frisén, Louise and Gallagher, Louise and Gejman, Pablo V. and Georgieva, Lyudmila and Gershon, Elliot S. and Geschwind, Daniel H. and Giegling, Ina and Gill, Michael and Gordon, Scott D. and Gordon-Smith, Katherine and Green, Elaine K. and Greenwood, Tiffany A. and Grice, Dorothy E. and Gross, Magdalena and Grozeva, Detelina and Guan, Weihua and Gurling, Hugh and De Haan, Lieuwe and Haines, Jonathan L. and Hakonarson, Hakon and Hallmayer, Joachim and Hamilton, Steven P. and Hamshere, Marian L. and Hansen, Thomas F. and Hartmann, Annette M. and Hautzinger, Martin and Heath, Andrew C. and Henders, Anjali K. and Herms, Stefan and Hickie, Ian B. and Hipolito, Maria and Hoefels, Susanne and Holmans, Peter A. and Holsboer, Florian and Hoogendijk, Witte J. and Hottenga, Jouke-Jan and Hultman, Christina M. and Hus, Vanessa and Ingason, Andrés and Ising, Marcus and Jamain, Stéphane and Jones, Edward G. and Jones, Ian and Jones, Lisa and Tzeng, Jung-Ying and Kähler, Anna K. and Kahn, René S. and Kandaswamy, Radhika and Keller, Matthew C. and Kennedy, James L. and Kenny, Elaine and Kent, Lindsey and Kim, Yunjung and Kirov, George K. and Klauck, Sabine M. and Klei, Lambertus and Knowles, James A. and Kohli, Martin A. and Koller, Daniel L. and Konte, Bettina and Korszun, Ania and Krabbendam, Lydia and Krasucki, Robert and Kuntsi, Jonna and Kwan, Phoenix and Landén, Mikael and Långström, Niklas and Lathrop, Mark and Lawrence, Jacob and Lawson, William B. and Leboyer, Marion and Ledbetter, David H. and Lee, Phil H. and Lencz, Todd and Lesch, Klaus-Peter and Levinson, Douglas F. and Lewis, Cathryn M. and Li, Jun and Lichtenstein, Paul and Lieberman, Jeffrey A. and Lin, Dan-Yu and Linszen, Don H. and Liu, Chunyu and Lohoff, Falk W. and Loo, Sandra K. and Lord, Catherine and Lowe, Jennifer K. and Lucae, Susanne and MacIntyre, Donald J. and Madden, Pamela A. F. and Maestrini, Elena and Magnusson, Patrik K. E. and Mahon, Pamela B. and Maier, Wolfgang and Malhotra, Anil K. and Mane, Shrikant M. and Martin, Christa L. and Martin, Nicholas G. and Mattheisen, Manuel and Matthews, Keith and Mattingsdal, Morten and McCarroll, Steven A. and McGhee, Kevin A. and McGough, James J. and McGrath, Patrick J. and McGuffin, Peter and McInnis, Melvin G. and McIntosh, Andrew and McKinney, Rebecca and McLean, Alan W. and McMahon, Francis J. and McMahon, William M. and McQuillin, Andrew and Medeiros, Helena and Medland, Sarah E. and Meier, Sandra and Melle, Ingrid and Meng, Fan and Meyer, Jobst and Middeldorp, Christel M. and Middleton, Lefkos and Milanova, Vihra and Miranda, Ana and Monaco, Anthony P. and Montgomery, Grant W. and Moran, Jennifer L. and Moreno-De-Luca, Daniel and Morken, Gunnar and Morris, Derek W. and Morrow, Eric M. and Moskvina, Valentina and Muglia, Pierandrea and Mühleisen, Thomas W. and Muir, Walter J. and Müller-Myhsok, Bertram and Murtha, Michael and Myers, Richard M. and Myin-Germeys, Inez and Neale, Michael C. and Nelson, Stan F. and Nievergelt, Caroline M. and Nikolov, Ivan and Nimgaonkar, Vishwajit and Nolen, Willem A. and Nöthen, Markus M. and Nurnberger, John I. and Nwulia, Evaristus A. and Nyholt, Dale R. and O'Dushlaine, Colm and Oades, Robert D. and Olincy, Ann and Oliveira, Guiomar and Olsen, Line and Ophoff, Roel A. and Osby, Urban and Owen, Michael J. and Palotie, Aarno and Parr, Jeremy R. and Paterson, Andrew D. and Pato, Carlos N. and Pato, Michele T. and Penninx, Brenda W. and Pergadia, Michele L. and Pericak-Vance, Margaret A. and Pickard, Benjamin S. and Pimm, Jonathan and Piven, Joseph and Posthuma, Danielle and Potash, James B. and Poustka, Fritz and Propping, Peter and Puri, Vinay and Quested, Digby J. and Quinn, Emma M. and Ramos-Quiroga, Josep Antoni and Rasmussen, Henrik B. and Raychaudhuri, Soumya and Rehnström, Karola and Reif, Andreas and Ribasés, Marta and Rice, John P. and Rietschel, Marcella and Roeder, Kathryn and Roeyers, Herbert and Rossin, Lizzy and Rothenberger, Aribert and Rouleau, Guy and Ruderfer, Douglas and Rujescu, Dan and Sanders, Alan R. and Sanders, Stephan J. and Santangelo, Susan L. and Sergeant, Joseph A. and Schachar, Russell and Schalling, Martin and Schatzberg, Alan F. and Scheftner, William A. and Schellenberg, Gerard D. and Scherer, Stephen W. and Schork, Nicholas J. and Schulze, Thomas G. and Schumacher, Johannes and Schwarz, Markus and Scolnick, Edward and Scott, Laura J. and Shi, Jianxin and Shilling, Paul D. and Shyn, Stanley I. and Silverman, Jeremy M. and Slager, Susan L. and Smalley, Susan L. and Smit, Johannes H. and Smith, Erin N. and Sonuga-Barke, Edmund J. S. and St Clair, David and State, Matthew and Steffens, Michael and Steinhausen, Hans-Christoph and Strauss, John S. and Strohmaier, Jana and Stroup, T. Scott and Sutcliffe, James S. and Szatmari, Peter and Szelinger, Szabocls and Thirumalai, Srinivasa and Thompson, Robert C. and Todorov, Alexandre A. and Tozzi, Federica and Treutlein, Jens and Uhr, Manfred and van den Oord, Edwin J. C. G. and Van Grootheest, Gerard and Van Os, Jim and Vicente, Astrid M. and Vieland, Veronica J. and Vincent, John B. and Visscher, Peter M. and Walsh, Christopher A. and Wassink, Thomas H. and Watson, Stanley J. and Weissman, Myrna M. and Werge, Thomas and Wienker, Thomas F. and Wijsman, Ellen M. and Willemsen, Gonneke and Williams, Nigel and Willsey, A. Jeremy and Witt, Stephanie H. and Xu, Wei and Young, Allan H. and Yu, Timothy W. and Zammit, Stanley and Zandi, Peter P. and Zhang, Peng and Zitman, Frans G. and Zöllner, Sebastian and Devlin, Bernie and Kelsoe, John R. and Sklar, Pamela and Daly, Mark J. and O'Donovan, Michael C. and Craddock, Nicholas and Sullivan, Patrick F. and Smoller, Jordan W. and Kendler, Kenneth S. and Wray, Naomi R. and {International Inflammatory Bowel Disease Genetics Consortium (IIBDGC)}},
	date = {2013-09},
	pmid = {23933821},
	pmcid = {PMC3800159},
}

@article{castro-de-araujo_mr-doc2_2023,
	title = {{MR}-{DoC}2: Bidirectional Causal Modeling with Instrumental Variables and Data from Relatives},
	volume = {53},
	issn = {1573-3297},
	url = {https://doi.org/10.1007/s10519-022-10122-x},
	doi = {10.1007/s10519-022-10122-x},
	shorttitle = {{MR}-{DoC}2},
	pages = {63--73},
	number = {1},
	journaltitle = {Behavior Genetics},
	shortjournal = {Behav Genet},
	author = {Castro-de-Araujo, Luis F. S. and Singh, Madhurbain and Zhou, Yi and Vinh, Philip and Verhulst, Brad and Dolan, Conor V. and Neale, Michael C.},
	urldate = {2024-10-04},
	date = {2023-02-01},
	langid = {english},
}

@article{tahmasbi_geneevolve_2017,
	title = {{GeneEvolve}: a fast and memory efficient forward-time simulator of realistic whole-genome sequence and {SNP} data},
	volume = {33},
	issn = {1367-4811},
	doi = {10.1093/bioinformatics/btw606},
	shorttitle = {{GeneEvolve}},
	pages = {294--296},
	number = {2},
	journaltitle = {Bioinformatics (Oxford, England)},
	shortjournal = {Bioinformatics},
	author = {Tahmasbi, Rasool and Keller, Matthew C.},
	date = {2017-01-15},
	pmid = {27659450},
	pmcid = {PMC6074839},
}

@article{cloninger_interpretation_1980,
	title = {Interpretation of intrinsic and extrinsic structural relations by path analysis: theory and applications to assortative mating},
	volume = {36},
	issn = {0016-6723, 1469-5073},
	url = {https://www.cambridge.org/core/product/identifier/S0016672300019765/type/journal_article},
	doi = {10.1017/S0016672300019765},
	shorttitle = {Interpretation of intrinsic and extrinsic structural relations by path analysis},
	pages = {133--145},
	number = {2},
	journaltitle = {Genetical Research},
	shortjournal = {Genet. Res.},
	author = {Cloninger, C. Robert},
	urldate = {2024-02-25},
	date = {1980-10},
	langid = {english},
}

@article{davies_importance_2024,
	title = {The importance of family-based sampling for biobanks},
	issn = {0028-0836},
	url = {http://www.nature.com/nature/},
	journaltitle = {Nature},
	author = {Davies, Neil and Hemani, Gibran and Neiderhiser, Jenae M. and Martin, Hilary C. and Mills, Melinda C. and Visscher, Peter M. and Yengo, Loïc and Young, Alexander S. and Keller, Matthew C.},
	urldate = {2024-08-01},
	date = {2024-10-31},
	note = {Publisher: Nature Publishing Group},
}

@article{vogler_multivariate_1985,
	title = {Multivariate path analysis of familial resemblance: Multivariate Path Analysis},
	volume = {2},
	issn = {07410395},
	url = {https://onlinelibrary.wiley.com/doi/10.1002/gepi.1370020105},
	doi = {10.1002/gepi.1370020105},
	shorttitle = {Multivariate path analysis of familial resemblance},
	pages = {35--53},
	number = {1},
	journaltitle = {Genetic Epidemiology},
	shortjournal = {Genet. Epidemiol.},
	author = {Vogler, George P.},
	urldate = {2023-09-08},
	date = {1985},
	langid = {english},
}

\newpage
{
\setlength{\parindent}{0pt}
\setlength{\parskip}{10pt}
\section{Supplement}
\subsection{Parameter Expectations}
This section provides a step-by-step guide to deriving the expected values of all estimated parameters. Throughout, symbols in the equations correspond to the matrices defined in Tables \ref{tab:path} and \ref{tab:var}. To simplify notation, we write $[N]T_{p/m}$ to indicate any of $NT_p$, $T_p$, $NT_m$, or $T_m$, where subscripts $p$ and $m$ refer to paternal and maternal haplotypes, respectively, and where $NT$ stands for the haplotypic PGS not transmitted and $T$ for the haplotypic PGS transmitted to offspring. The "$L$" in $LNT$ and $LT$ stands for "latent" (as opposed to the observed haplotypic PGSs). Finally, subscripts "$LO$" in, e.g., $i_{tLO}$ stand for "latent to observed". For each parameter, we first show the relevant path-tracing chain (given in parentheses for convenience). For example, $Y_{p/m}  \rightarrow  [N]T_{p/m}$ denotes the path tracing starts with  $Y_{p/m}$ and finishes at $[N]T_{p/m}$, which means $Y_{p/m}$ is the downstream variable and $[N]T_{p/m}$ is the upstream variable. We then present the corresponding mathematical derivation. This approach is intended to make the logic of each step transparent, so that readers can see how expectations arise from simple path-tracing rules and so that the model can be altered depending on the data and questions at hand.

\subsubsection{Assuming Equilibrium AM}

\vspace{1em}
We define the $\Omega$ parameters as "shortcuts" or shorthand covariances to facilitate the derivation of other parameter expectations. Specifically, they represent the covariance between haplotypic PGSs and parental phenotypes. We show both non-transposed and transposed forms, as each is useful for deriving subsequent expectations.

$ \Omega_{p/m} (Y_{p/m}  \rightarrow  [N]T_{p/m}) $ 
$ = cov(Y_{p/m}, [N]T_{p/m}) = 2a_{p/m} i_c + 2\delta_{p/m} g_c + \delta_{p/m} k + {1\over2}w_{p/m} $

$ \Omega_{p/m}^T ([N]T_{p/m} \rightarrow Y_{p/m}) $
$ = \text{cov}([N]T_{p/m}, Y_{p/m}) = 2i_c^T a_{p/m}^T  + 2g_c \delta_{p/m}^T + k \delta_{p/m}^T  + {1\over2}w_{p/m}^T $

\vspace{1em}
Similar to $\Omega$, $\Gamma$s are shortcut covariance between LGS and parental phenotypes. 

$ \Gamma_{p/m} (Y_{p/m} \rightarrow L[N]T_{p/m}) $
$ = \text{cov}(Y_{p/m}, L[N]T_{p/m}) = 2\delta_{p/m} i_c^T + 2a_{p/m} h_c + a_{p/m} j + {1\over2}v_{p/m} $

$ \Gamma_{p/m}^T (L[N]T_{p/m} \rightarrow Y_{p/m}) $
$ = \text{cov}(L[N]T_{p/m}, Y_{p/m}) = 2i_c \delta_{p/m}^T  + 2h_c a_{p/m}^T + j a_{p/m}^T + {1\over2}v_{p/m}^T $

\vspace{1em}
With $\Omega$ and $\Gamma$ defined, we can derive other expectations. AM induces directional covariance between the effects of all causal variants. We quantify this as $g$ for the increase in genetic (co)variance for the haplotypic PGSs, $h$ for the increase in genetic (co)variance for the haplotypic LGS, and $i$ for the covariance between these two constructs. The $t$ and $c$ subscripts stand for "trans" (across mates) and "cis" (within-person) respectively. 

$ g_t = [N]T_p $ to $ [N]T_m $ 
$= \Omega^T_p \mu \Omega_m $

$ g_c ([N]T_p \rightarrow T_p) \text{ or } (T_p \rightarrow [N]T_p) \text{ or } ([N]T_m \rightarrow T_m) \text{ or } (T_m \rightarrow [N]T_m) $
$=\text{cov}(NT_{p/m}, T_{p/m})={1\over2}(g_t+g_t^T) $

$ h_t (L[N]T_p \rightarrow L[N]T_m) $
$ = \text{cov}(L[N]T_p, L[N]T_m) = \Gamma^T_p \mu \Gamma_m $

$ h_c(L[N]T_p \rightarrow LT_p) \text{ or } (LT_p \rightarrow L[N]T_p) \text{ or } (L[N]T_m \rightarrow LT_m) \text{ or } (LT_m \rightarrow L[N]T_m) $
$=\text{cov}(LNT_{p/m}, LT_{p/m})={1\over2}(h_t+h_t^T) $

\vspace{1em}
Unlike $g$ and $h$, the covariance between the the haplotypic PGSs and the haplotypic LGS, $i$, is a full matrix (the covariance between the PGS of trait 1 and the LGS of trait 2 is conceptually different than the covariance between the PGS of trait 2 and the LGS of trait 1) and so requires special attention regarding when it is transposed. For $i_t$, we adopt the convention (arbitrarily but consistently with $\mu$ in the model) that the paternal variable ($L[N]T_p$ and $[N]T_p$) is downstream and the maternal ($L[N]T_m$ and $[N]T_m$) variable is upstream, and denote the order in the $i$ subscripts. For $i_c$, we adopt a convention that LGSs are the downstream and PGSs are the upstream. 

$ i_{t_{LO}} (L[N]T_p \rightarrow [N]T_m) $
$ = \text{cov}(L[N]T_p, [N]T_m) = \Gamma^T_p \mu \Omega_m $

$ i_{t_{LO}}^T ([N]T_m \rightarrow L[N]T_p) $
$ = \text{cov}([N]T_m, L[N]T_p) = \Omega_m^T \mu^T \Gamma_p $

$ i_{t_{OL}} ([N]T_p \rightarrow L[N]T_m) $
$ = \text{cov}([N]T_p, L[N]T_m) = \Omega^T_p \mu \Gamma_m $

$ i_{t_{OL}}^T (L[N]T_m \rightarrow [N]T_p) $
$ = \text{cov}(L[N]T_m, [N]T_p) = \Gamma^T_m \mu^T \Omega_p $

$ i_{c}(L[N]T_p \rightarrow [N]T_p) \text{ or } (L[N]T_m \rightarrow [N]T_m) $
$=\text{cov}(L[N]T_{p/m}, [N]T_{p/m})={1\over2}(i_{t_{LO}}+i_{t_{OL}}^T) $

$ i_{c}^T([N]T_p \rightarrow L[N]T_p) \text{ or } ([N]T_m \rightarrow L[N]T_m) $
$=\text{cov}([N]T_{p/m}, L[N]T_{p/m} )^T={1\over2}(i_{t_{OL}}+i_{t_{LO}}^T) $

\vspace{1em}
The gene-environment (G-E) covariance matrices $\mathbf{w}$ and $\mathbf{v}$ are full because they are derived from $\mathbf{f}$, which is itself a full matrix as a consequence of vertical transmission (VT). Conceptually, the covariance between family environment for trait 2 and PGS/LGS for trait 1 can differ from the covariance between family environment for trait 1 and PGS/LGS for trait 2. 

$ w_{p/m} (F_{p/m} \rightarrow [N]T_{p/m}) = w_o(F_{o} \rightarrow [N]T_{o} )$
$ = \text{cov}(F_{o}, NT_{p} + NT_{m}) = \text{cov}(F_{o}, T_{p} + T_{m}) = f_{p}\Omega_p + f_{m}\Omega_m + f_{p}V_{Yp}\mu\Omega_m + f_{m}V_{Ym}\mu^T\Omega_p $

$ v_{p/m} (F_{p/m} \rightarrow L[N]T_{p/m})= v_o(F_{o} \rightarrow L[N]T_{o} ) $
$ = \text{cov}(F_{o}, LNT_{p} + LNT_{m}) = \text{cov}(F_{o}, LT_{p} + LT_{m}) = f_{p}\Gamma_p + f_{m}\Gamma_m + f_{p}V_{Yp}\mu\Gamma_{m} + f_{m}V_{Ym}\mu^T\Gamma_{p} $

Here, we note a minor correction to the equations for $\theta_{NT}$ and $\theta_{LNT}$ presented in the supplement of \textcite{balbona_estimation_2021}. In the previous paper, the covariance terms $\mathbf{w}$ and $\mathbf{v}$ were given a scalar coefficient of 2, whereas the correct coefficient is 1, as reflected in our equations above.

$ \theta_{NTp/m} (Y_o \rightarrow [N]T_{p/m}) $
$ = 4 \delta_{p/m} g_c + 4a_{p/m} i_c + w_o $

$ \theta_{Tp/m} (Y_o \rightarrow T_{p/m}) $
$ = \text{cov}(Y_o, T_{p/m}) = 2\delta_{p/m} k + \theta_{NTp/m} $

$ \theta_{LNTp/m} (Y_o \rightarrow L[N]T_{p/m}) $
$ = \text{cov}(Y_o, LNT_{p/m}) =  4 a_{p/m} h_c + 4 \delta_{p/m} i_c^T + v_o$

$ \theta_{LTp/m} (Y_o \rightarrow LT_{p/m}) $
$ = \text{cov}(Y_o, LT_{p/m}) = 2a_{p/m} j + \theta_{LNTp/m} $

$ V_{Yp} = V_{Ym} = V_{Yo} = 2\Omega_{p/m}\delta_{p/m} + 2\Gamma_{p/m} a_{p/m} + \delta_{p/m} w^T_{p/m} + a_{p/m} v^T_{p/m} + V_{Fp/m} + V_{\epsilon p/m} $

$ V_{Fp} = V_{Fm} = V_{Fo} = f_{p} V_{Yp} f_{p}^T+ f_{m} V_{Ym} f_{m}^T + f_{p} V_{Yp} \mu V_{Ym} f_{m}^T + f_{m} V_{Ym} \mu^T V_{Yp} f_{p}^T $

$ V_{G_{Obs,p/m}} = 2\delta_{p/m} k \delta^T_{p/m} + 4\delta_{p/m} g_c \delta^T_{p/m} $

$ V_{G_{Lat,p/m}} = 2a_{p/m} j a_{p/m}^T + 4 a_{p/m} h_c a^T_{p/m} $

% $ COV_{G_{Obs,Lat,p/m}} = 4 \delta_{p/m} i_c^T a_{p/m}^T + 4 a_{p/m} i_c \delta_{p/m}^T $

% $ V_{G_{tot,p/m}} = 2\delta_{p/m} k \delta^T_{p/m} + 4\delta_{p/m} g_c \delta^T_{p/m} + 4 \delta_{p/m} i_c^T a_{p/m}^T  + 2a_{p/m} j a_{p/m}^T + 4 a_{p/m} h_c a^T_{p/m} + 4 a_{p/m} i_c \delta_{p/m}^T $

% $ COV_{GE_{p/m}} = \delta_{p/m} w_{p/m}^T + w_{p/m} \delta_{p/m}^T + a_{p/m} v_{p/m}^T + v_{p/m} a_{p/m}^T $

\subsubsection{Assuming Disequilibrium AM}
The following equations are parameter expectations in a model that has only one generation of AM, where AM occurs only in the parental generation but not before.  Note that other types of disequilibrium AM are also possible, and we believe the model could be adjusted to incorporate these.

$ \Omega_{p/m} = \delta_{p/m} k + {1\over2}w_{p/m} $

$ \Gamma_{p/m} = a_{p/m} j + {1\over2}v_{p/m} $

$ g_t = \Omega^T_p \mu \Omega_m $

$ h_t = \Gamma^T_p \mu \Gamma_m $

$ i_{t_{LO}} = \Gamma^T_p \mu \Omega_m $

$ i_{t_{OL}} = \Omega^T_p \mu \Gamma_m $

$ g_c = h_c = i_c = 0 $

$ w_{p/m} = f_{p/m}\Omega_{p/m} + f_{m/p}\Omega_{m/p} $

$ v_{p/m} = f_{p}\Gamma_p + f_{m}\Gamma_m $

$ \theta_{NTp/m} = a_{p/m} i_{t_{LO}} + a_{p/m} i^T_{t_{OL}} + \delta_{p/m} g_t + \delta_{p/m} g^T_t + w_{p/m} $

$ \theta_{Tp/m} = 2\delta_{p/m} k + \theta_{NTp/m} $

$ \theta_{LNTp/m} = a_{p/m} h_t + a_{p/m} h^T_t + \delta_{p/m} i_{t_{OL}} + \delta_{p/m} i^T_{t_{LO}} + v_{p/m} $

$ \theta_{LTp/m}= 2a_{p/m} j + \theta_{LNTp/m} $

$ V_{Fp/m} = f_{p} V_{Yp} f_{p}^T + f_{m} V_{Ym} f_{m}^T $

$ V_{Yp/m} = 2\Omega_{p/m}\delta_{p/m} + 2\Gamma_{p/m} a_{p/m} + \delta_{p/m} w^T_{p/m} + a_{p/m} v^T_{p/m} + V_{Fp/m} + V_{\epsilon p/m} $

\subsection{Supplementary Figures}

\renewcommand{\thefigure}{S\arabic{figure}}
\setcounter{figure}{0}

\begin{figure}[H]
    \centering
    \includegraphics[width=1\linewidth]{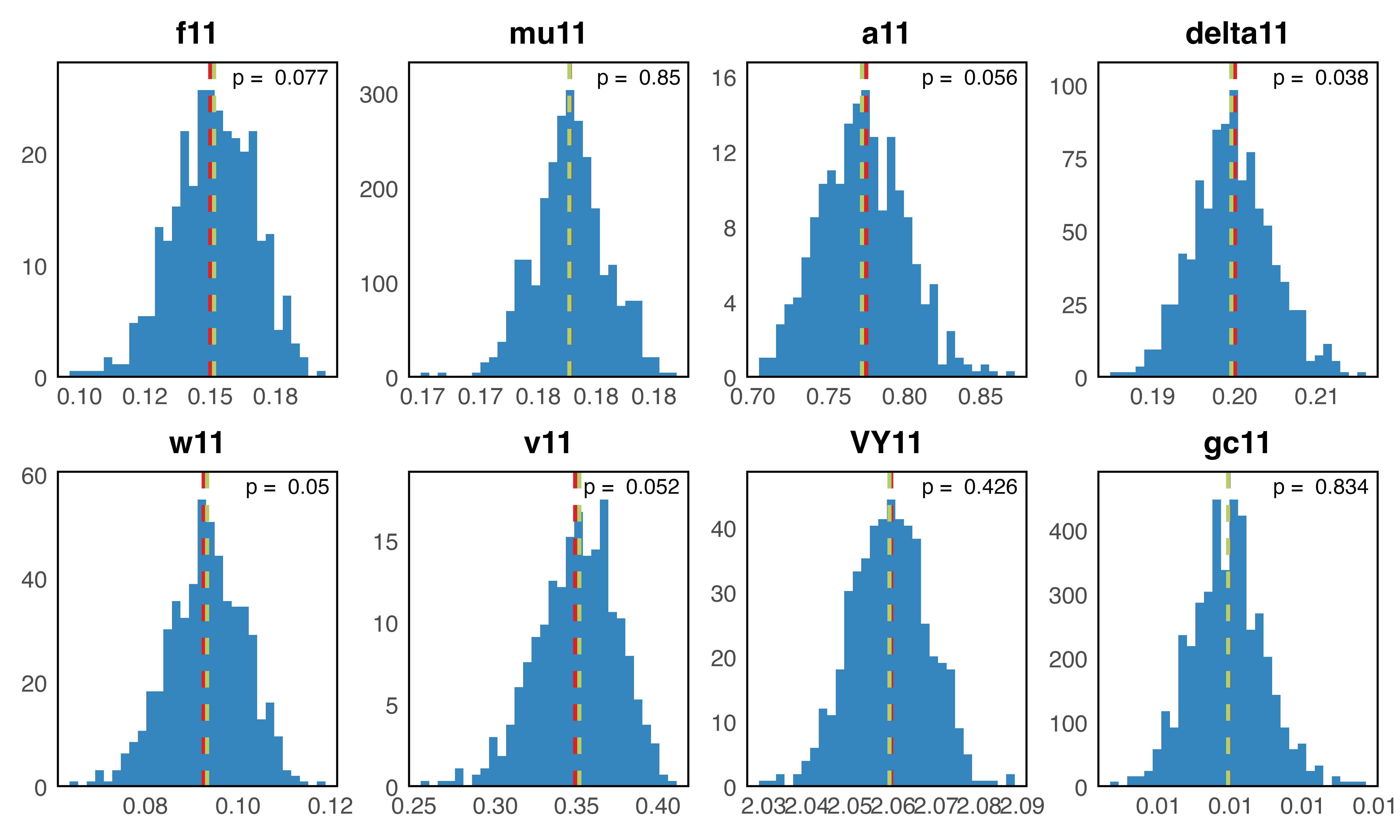}
    \caption{Histograms of within-trait parameter estimates obtained by fitting the model to 500 datasets simulated from a multivariate normal distribution. The red dashed line indicates the true parameter value, while the blue dashed line represents the median of the 500 estimates. P-values from bootstrap tests assessing whether the median significantly deviates from the true value are shown in the top-right corner of each panel.}
    \label{fig:within_mvn}
\end{figure}

\begin{figure}[H]
    \centering
    \includegraphics[width=1\linewidth]{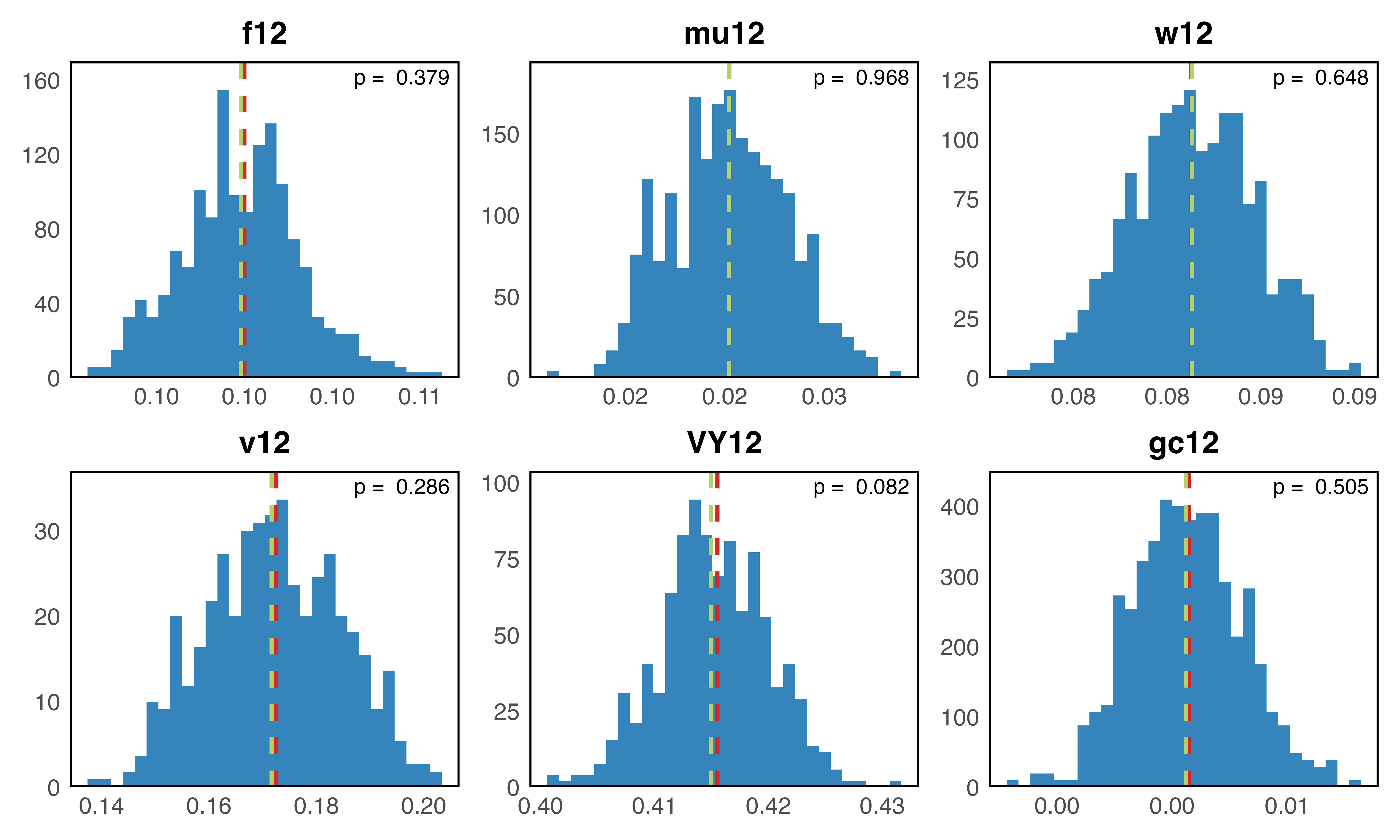}
    \caption{Histograms of cross-trait parameter estimates obtained by fitting the model to 500 datasets simulated from a multivariate normal distribution. The red dashed line indicates the true parameter value, while the blue dashed line represents the median of the 500 estimates. P-values from bootstrap tests assessing whether the median significantly deviates from the true value are shown in the top-right corner of each panel.}
    \label{fig:cross_mvn}
\end{figure}

\begin{figure}[H]
    \centering
    \includegraphics[width=1\linewidth]{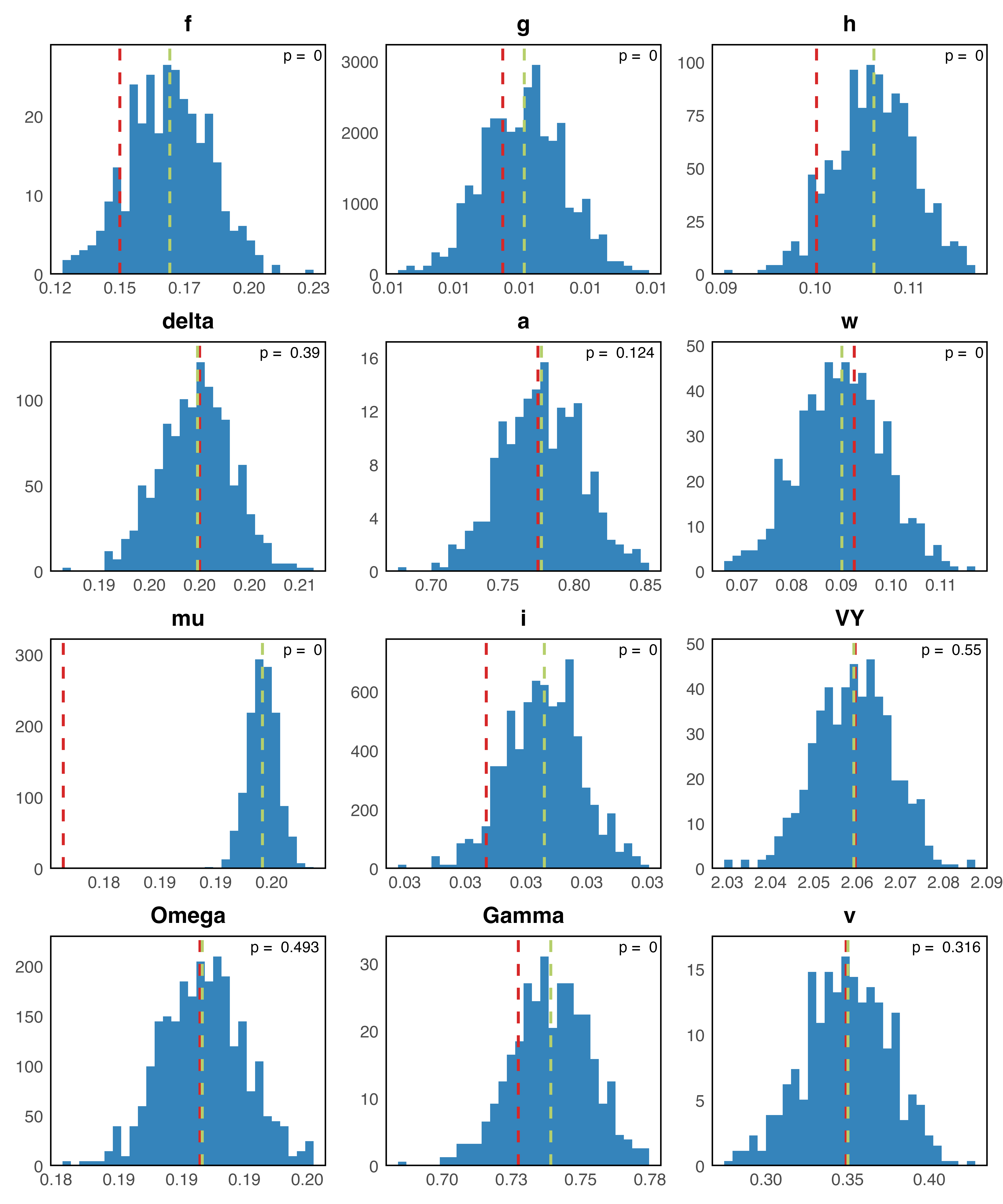}
    \caption{Histograms of 500 parameter estimates from a univariate SEM-PGS model when the data is simulated with strong within-trait effects and weak cross-trait effects. The red dashed line indicates the true parameter value. The green dashed line indicates the median of the 500 estimates. P values of the bootstrap test for each parameter are shown for all parameters.}
    \label{fig:uni_bias_2}
\end{figure}

\begin{figure}[H]
    \centering
    \includegraphics[width=1\linewidth]{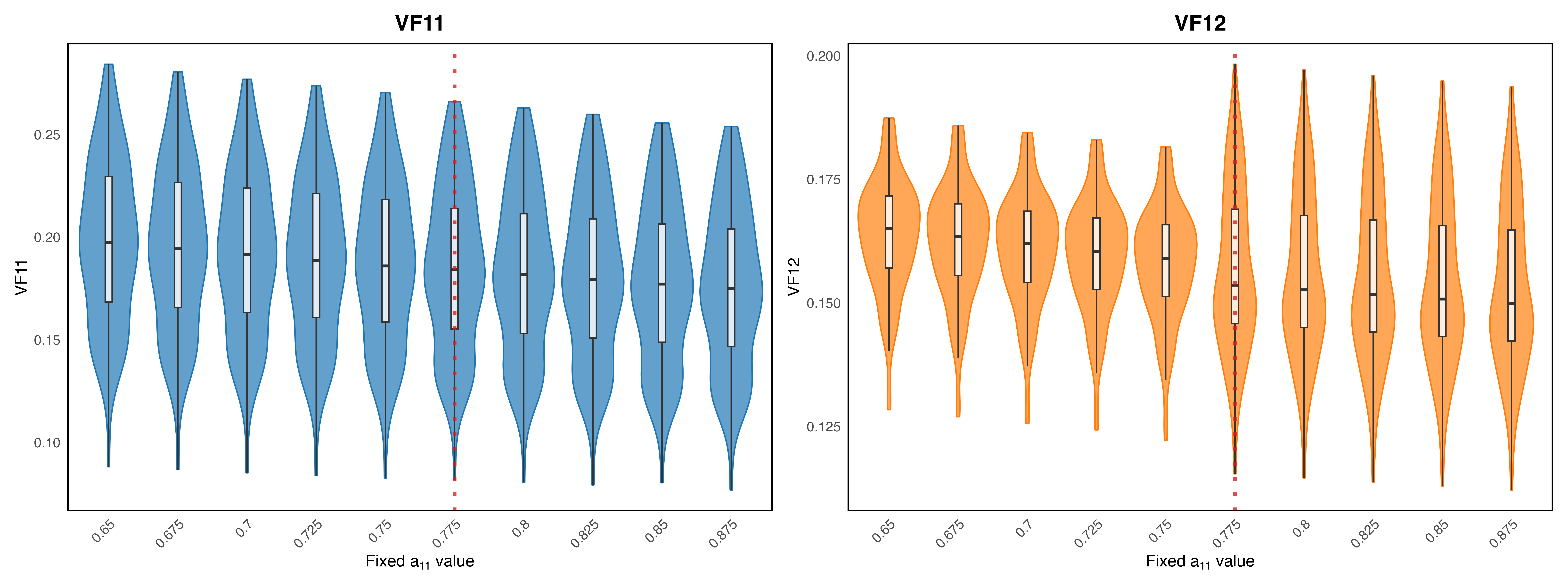}
    \caption{Distribution of $V_{F11}$ and $V_{F12}$ from a sensitivity analysis where the direct genetic effect, $a_{11}$, was fixed to various incorrect values. Each violin plot shows the distribution of estimates across 100 replications fitted in OpenMx. The red dashed line highlights the results obtained when $a_{11}$ was fixed to its true simulated value. The underlying data were generated with a sample size of $N_{trio} = 32k$ and $r^2_{pgs1} = 4\%$.}
    \label{fig:sensiVF}
\end{figure}

\subsection{Suggested Approach for Fitting the multivariate SEM-PGS Model}
Given the complexity of the Multivariate SEM-PGS model, we have outlined a step-by-step procedure to guide researchers in its application. Below is a concise and clear workflow to ensure accurate implementation and interpretation of the model.

\begin{enumerate}
\item \textbf{Prepare the genomic data.} Begin by performing quality control (QC) to identify Mendelian errors and verify that the parent-offspring relatedness ($\hat{\pi}$) is approximately 0.5, ensuring accurate kinship specification. Next, separate the PGS of offspring and parents into four distinct haplotypic PGSs. %This can be achieved with software in 
 
\item \textbf{Compute $r^2_{pgs}$ for all traits.} Ensure that the sample size ($N$) is sufficient relative to the observed $r^2_{pgs}$ by consulting supplementary tables or conducting simulations. 

\item Scale the haplotypic PGSs. We recommend dividing each haplotypic PGS by the variance of the sum of the two haplotypic PGSs within an individual. With this approach, the scaling factors $k$ and $j$ are set equal to ${1 \over 2} - 2g$ and ${1 \over 2} - 2h$, respectively. This parameterization is preferable to simply fixing the variance of each haplotypic PGS to $1/2$, because it allows the haplotypic variances, and not just their covariances, to carry information about AM.

\item \textbf{Determine the estimation of parameter $\mathbf{a}$.} Based on the trait and whether the residual parent-offspring covariance is likely to be due only to additive genetics, decide whether to estimate $a$ within the model or fix it using external methods. In general, we recommend estimating $\mathbf{a}$ outside the model. An unbiased estimation of $\mathbf{V}_{A-Base}$ can be achieved through approaches such as Relatedness Disequilibrium Regression (RDR) \autocite{young_relatedness_2018} or sibling-based methods \autocite{visscher_assumption-free_2006}. Then, set $\mathbf{V}\mathbf{A}_{Base} = (\mathbf{a}^2 + \boldsymbol{\delta}^2)(\mathbf{a}^2 + \boldsymbol{\delta}^2 + \mathbf{V}_E)$.   

\item \textbf{Check the PGS Predictability.} Compare $r^2_{pgs-p/m}$ (Parental PGS Predictibility) with $r^2_{pgs-o}$ (Offspring PGS Predictibility) by running a regression $Y \sim PGS + Covariates$ for both parental and offspring traits. If $r^2_{pgs-p/m}$ is substantially different from $r^2_{pgs-o}$, consider modeling distinct values of $\mathbf{a}$ and $\boldsymbol{\delta}$ for parents and offspring. Different $\boldsymbol{\delta}$ values for parents and offspring can be modeled directly, as they come from distinct information. However, the model presented in Figure \ref{fig:BiSEMPGS} can not identify both $\mathbf{a_{parent}}$ and $\mathbf{a_{offspring}}$ at the same time. Therefore, fixing at least one of $\mathbf{a_{parent}}$ or $\mathbf{a_{offspring}}$ using the approaches mentioned in the last bullet point is required for model identification.

\item \textbf{Evaluate assortative mating equilibrium.} Verify whether $\mathbf{g}_c$ (genetic covariance) approximates $0.5(\mathbf{g}_{t} + \mathbf{g}_{t}^T)$. If not, this suggests that AM has not reached equilibrium. Refer to the equations provided in Appendix I, which account for deviations from AM equilibrium. 

\item \textbf{Evaluate the mechanism of assortative mating.} Compare the observed average $\mathbf{g}$ with its expected value under phenotypic AM. If inconsistencies are found, introduce latent AM factors to represent different AM mechanisms and conduct model comparisons to identify the best fit \autocite{keller_modeling_2009}. 

\item \textbf{Constrain or free parental effects.} Determine whether $\mathbf{f}_p$ (paternal effect) and $\mathbf{f}_m$ (maternal effect) can be constrained to equality based on statistical or theoretical grounds. If necessary, estimate these parameters independently.

\item \textbf{Fit the model using OpenMx.} Implement the model using an OpenMx script that aligns with the decisions made in previous steps. Utilize the MxTryhard function to explore multiple starting values, adjust feasibility and optimality tolerance settings, and increase the number of trials.

\item \textbf{Evaluate model fit.} Examine the OpenMx output for the following:
\begin{enumerate}
    \item Confirm that the OpenMx Status Code is 1 or 0, indicating successful convergence.
    \item Ensure that the estimated $\mathbf{V}_Y$ (variance of the outcome) aligns with the observed $\mathbf{V}_Y$ in the covariance matrix. If there are discrepancies, interpret parameter estimates with caution, and/or consider modifying the model to account for the discrepancies.
    \item Check for any parameter estimates that deviate greatly from the typical range of parameter values in an SEM model.
\end{enumerate}
If issues arise, adjust starting values or MxTryhard parameters and rerun the model. Persistent problems may require revisiting earlier steps. For additional support, researchers are encouraged to open an issue on the paper’s GitHub page (https://github.com/Xuanyu-Lyu/BiSEMPGS) or contact the corresponding authors via email.

\item \textbf{Report and interpret results.} Present the final estimates and provide a detailed interpretation of the findings, ensuring alignment with the model’s theoretical framework.
\end{enumerate}
}

\newpage
\section{Appendix I}

Descriptive statistics are provided for all parameters for the condition $r^2_{pgs1} = .04$ values and all sample sizes. The p-value from the Wilcoxon rank-sum test (W-R test) is used to compare the sample median with the true value of the parameter. Systematic variance refers to the proportion of total variance that can be attributed to the systematic bias arising from the imperfect simulation of assortative mating.

\subsection{Effect size $r^2_{pgs1}$ = .04 with Freely Estimated $a$}

\begin{longtable}{rrrrrr}
\toprule
sample\_sizes & median & MAD & trueValue & p\_value & proportion\_systematic\\
\midrule
\addlinespace[0.3em]
\multicolumn{6}{l}{\textbf{a11}}\\
\hspace{1em}\cellcolor{gray!10}{4000} & \cellcolor{gray!10}{0.7535} & \cellcolor{gray!10}{0.1721} & \cellcolor{gray!10}{0.7746} & \cellcolor{gray!10}{0.0016} & \cellcolor{gray!10}{0.0009}\\
\hspace{1em}8000 & 0.7709 & 0.1229 & 0.7746 & 0.6776 & 0.0025\\
\hspace{1em}\cellcolor{gray!10}{16000} & \cellcolor{gray!10}{0.7697} & \cellcolor{gray!10}{0.0762} & \cellcolor{gray!10}{0.7746} & \cellcolor{gray!10}{0.1578} & \cellcolor{gray!10}{0.0106}\\
\hspace{1em}32000 & 0.7703 & 0.0580 & 0.7746 & 0.0668 & 0.0016\\
\hspace{1em}\cellcolor{gray!10}{48000} & \cellcolor{gray!10}{0.7697} & \cellcolor{gray!10}{0.0479} & \cellcolor{gray!10}{0.7746} & \cellcolor{gray!10}{0.0654} & \cellcolor{gray!10}{0.0068}\\
\hspace{1em}64000 & 0.7730 & 0.0442 & 0.7746 & 0.3520 & 0.0108\\
\addlinespace[0.3em]
\multicolumn{6}{l}{\textbf{a22}}\\
\hspace{1em}\cellcolor{gray!10}{4000} & \cellcolor{gray!10}{0.5159} & \cellcolor{gray!10}{0.1350} & \cellcolor{gray!10}{0.5367} & \cellcolor{gray!10}{0.0064} & \cellcolor{gray!10}{0.0023}\\
\hspace{1em}8000 & 0.5154 & 0.0954 & 0.5367 & 0.0000 & 0.0568\\
\hspace{1em}\cellcolor{gray!10}{16000} & \cellcolor{gray!10}{0.5244} & \cellcolor{gray!10}{0.0703} & \cellcolor{gray!10}{0.5367} & \cellcolor{gray!10}{0.0000} & \cellcolor{gray!10}{0.0417}\\
\hspace{1em}32000 & 0.5263 & 0.0472 & 0.5367 & 0.0008 & 0.0280\\
\hspace{1em}\cellcolor{gray!10}{48000} & \cellcolor{gray!10}{0.5300} & \cellcolor{gray!10}{0.0467} & \cellcolor{gray!10}{0.5367} & \cellcolor{gray!10}{0.0024} & \cellcolor{gray!10}{0.0084}\\
\hspace{1em}64000 & 0.5305 & 0.0410 & 0.5367 & 0.0028 & 0.0243\\
\addlinespace[0.3em]
\multicolumn{6}{l}{\textbf{delta11}}\\
\hspace{1em}\cellcolor{gray!10}{4000} & \cellcolor{gray!10}{0.1995} & \cellcolor{gray!10}{0.0207} & \cellcolor{gray!10}{0.2000} & \cellcolor{gray!10}{0.6958} & \cellcolor{gray!10}{0.0000}\\
\hspace{1em}8000 & 0.2002 & 0.0139 & 0.2000 & 0.7576 & 0.0000\\
\hspace{1em}\cellcolor{gray!10}{16000} & \cellcolor{gray!10}{0.1999} & \cellcolor{gray!10}{0.0102} & \cellcolor{gray!10}{0.2000} & \cellcolor{gray!10}{0.8358} & \cellcolor{gray!10}{0.0005}\\
\hspace{1em}32000 & 0.1996 & 0.0067 & 0.2000 & 0.1760 & 0.0130\\
\hspace{1em}\cellcolor{gray!10}{48000} & \cellcolor{gray!10}{0.1995} & \cellcolor{gray!10}{0.0060} & \cellcolor{gray!10}{0.2000} & \cellcolor{gray!10}{0.1810} & \cellcolor{gray!10}{0.0063}\\
\hspace{1em}64000 & 0.1994 & 0.0052 & 0.2000 & 0.0528 & 0.0080\\
\addlinespace[0.3em]
\multicolumn{6}{l}{\textbf{delta22}}\\
\hspace{1em}\cellcolor{gray!10}{4000} & \cellcolor{gray!10}{0.2666} & \cellcolor{gray!10}{0.0210} & \cellcolor{gray!10}{0.2683} & \cellcolor{gray!10}{0.1856} & \cellcolor{gray!10}{0.0142}\\
\hspace{1em}8000 & 0.2669 & 0.0155 & 0.2683 & 0.0238 & 0.0044\\
\hspace{1em}\cellcolor{gray!10}{16000} & \cellcolor{gray!10}{0.2681} & \cellcolor{gray!10}{0.0102} & \cellcolor{gray!10}{0.2683} & \cellcolor{gray!10}{0.5824} & \cellcolor{gray!10}{0.0039}\\
\hspace{1em}32000 & 0.2672 & 0.0071 & 0.2683 & 0.0016 & 0.0222\\
\hspace{1em}\cellcolor{gray!10}{48000} & \cellcolor{gray!10}{0.2678} & \cellcolor{gray!10}{0.0056} & \cellcolor{gray!10}{0.2683} & \cellcolor{gray!10}{0.2264} & \cellcolor{gray!10}{0.0181}\\
\hspace{1em}64000 & 0.2671 & 0.0049 & 0.2683 & 0.0000 & 0.0330\\
\addlinespace[0.3em]
\multicolumn{6}{l}{\textbf{f11}}\\
\hspace{1em}\cellcolor{gray!10}{4000} & \cellcolor{gray!10}{0.1541} & \cellcolor{gray!10}{0.0935} & \cellcolor{gray!10}{0.1500} & \cellcolor{gray!10}{0.5142} & \cellcolor{gray!10}{0.0005}\\
\hspace{1em}8000 & 0.1471 & 0.0695 & 0.1500 & 0.5860 & 0.0002\\
\hspace{1em}\cellcolor{gray!10}{16000} & \cellcolor{gray!10}{0.1451} & \cellcolor{gray!10}{0.0408} & \cellcolor{gray!10}{0.1500} & \cellcolor{gray!10}{0.0404} & \cellcolor{gray!10}{0.0013}\\
\hspace{1em}32000 & 0.1461 & 0.0319 & 0.1500 & 0.0246 & 0.0234\\
\hspace{1em}\cellcolor{gray!10}{48000} & \cellcolor{gray!10}{0.1460} & \cellcolor{gray!10}{0.0266} & \cellcolor{gray!10}{0.1500} & \cellcolor{gray!10}{0.0190} & \cellcolor{gray!10}{0.0116}\\
\hspace{1em}64000 & 0.1434 & 0.0242 & 0.1500 & 0.0000 & 0.0254\\
\addlinespace[0.3em]
\multicolumn{6}{l}{\textbf{f12}}\\
\hspace{1em}\cellcolor{gray!10}{4000} & \cellcolor{gray!10}{0.0991} & \cellcolor{gray!10}{0.0218} & \cellcolor{gray!10}{0.1000} & \cellcolor{gray!10}{0.4442} & \cellcolor{gray!10}{0.0076}\\
\hspace{1em}8000 & 0.1020 & 0.0155 & 0.1000 & 0.0098 & 0.0008\\
\hspace{1em}\cellcolor{gray!10}{16000} & \cellcolor{gray!10}{0.1006} & \cellcolor{gray!10}{0.0130} & \cellcolor{gray!10}{0.1000} & \cellcolor{gray!10}{0.2878} & \cellcolor{gray!10}{0.0001}\\
\hspace{1em}32000 & 0.1003 & 0.0075 & 0.1000 & 0.2172 & 0.0034\\
\hspace{1em}\cellcolor{gray!10}{48000} & \cellcolor{gray!10}{0.1001} & \cellcolor{gray!10}{0.0070} & \cellcolor{gray!10}{0.1000} & \cellcolor{gray!10}{0.8936} & \cellcolor{gray!10}{0.0001}\\
\hspace{1em}64000 & 0.0999 & 0.0058 & 0.1000 & 0.5556 & 0.0064\\
\addlinespace[0.3em]
\multicolumn{6}{l}{\textbf{f21}}\\
\hspace{1em}\cellcolor{gray!10}{4000} & \cellcolor{gray!10}{0.0503} & \cellcolor{gray!10}{0.0108} & \cellcolor{gray!10}{0.0500} & \cellcolor{gray!10}{0.4982} & \cellcolor{gray!10}{0.0072}\\
\hspace{1em}8000 & 0.0502 & 0.0086 & 0.0500 & 0.7474 & 0.0001\\
\hspace{1em}\cellcolor{gray!10}{16000} & \cellcolor{gray!10}{0.0498} & \cellcolor{gray!10}{0.0060} & \cellcolor{gray!10}{0.0500} & \cellcolor{gray!10}{0.5632} & \cellcolor{gray!10}{0.0038}\\
\hspace{1em}32000 & 0.0498 & 0.0042 & 0.0500 & 0.1336 & 0.0088\\
\hspace{1em}\cellcolor{gray!10}{48000} & \cellcolor{gray!10}{0.0499} & \cellcolor{gray!10}{0.0036} & \cellcolor{gray!10}{0.0500} & \cellcolor{gray!10}{0.7520} & \cellcolor{gray!10}{0.0100}\\
\hspace{1em}64000 & 0.0499 & 0.0033 & 0.0500 & 0.3830 & 0.0013\\
\addlinespace[0.3em]
\multicolumn{6}{l}{\textbf{f22}}\\
\hspace{1em}\cellcolor{gray!10}{4000} & \cellcolor{gray!10}{0.1114} & \cellcolor{gray!10}{0.0664} & \cellcolor{gray!10}{0.1000} & \cellcolor{gray!10}{0.0002} & \cellcolor{gray!10}{0.0132}\\
\hspace{1em}8000 & 0.1092 & 0.0468 & 0.1000 & 0.0000 & 0.0376\\
\hspace{1em}\cellcolor{gray!10}{16000} & \cellcolor{gray!10}{0.1043} & \cellcolor{gray!10}{0.0354} & \cellcolor{gray!10}{0.1000} & \cellcolor{gray!10}{0.0158} & \cellcolor{gray!10}{0.0290}\\
\hspace{1em}32000 & 0.1046 & 0.0232 & 0.1000 & 0.0000 & 0.0216\\
\hspace{1em}\cellcolor{gray!10}{48000} & \cellcolor{gray!10}{0.1012} & \cellcolor{gray!10}{0.0227} & \cellcolor{gray!10}{0.1000} & \cellcolor{gray!10}{0.6346} & \cellcolor{gray!10}{0.0063}\\
\hspace{1em}64000 & 0.1013 & 0.0204 & 0.1000 & 0.1988 & 0.0018\\
\addlinespace[0.3em]
\multicolumn{6}{l}{\textbf{Gamma11}}\\
\hspace{1em}\cellcolor{gray!10}{4000} & \cellcolor{gray!10}{0.7162} & \cellcolor{gray!10}{0.0841} & \cellcolor{gray!10}{0.7094} & \cellcolor{gray!10}{0.0504} & \cellcolor{gray!10}{0.0008}\\
\hspace{1em}8000 & 0.7191 & 0.0582 & 0.7094 & 0.0130 & 0.0000\\
\hspace{1em}\cellcolor{gray!10}{16000} & \cellcolor{gray!10}{0.7213} & \cellcolor{gray!10}{0.0384} & \cellcolor{gray!10}{0.7094} & \cellcolor{gray!10}{0.0000} & \cellcolor{gray!10}{0.0130}\\
\hspace{1em}32000 & 0.7213 & 0.0283 & 0.7094 & 0.0000 & 0.0270\\
\hspace{1em}\cellcolor{gray!10}{48000} & \cellcolor{gray!10}{0.7213} & \cellcolor{gray!10}{0.0245} & \cellcolor{gray!10}{0.7094} & \cellcolor{gray!10}{0.0000} & \cellcolor{gray!10}{0.0257}\\
\hspace{1em}64000 & 0.7221 & 0.0210 & 0.7094 & 0.0000 & 0.0338\\
\addlinespace[0.3em]
\multicolumn{6}{l}{\textbf{Gamma12}}\\
\hspace{1em}\cellcolor{gray!10}{4000} & \cellcolor{gray!10}{0.1591} & \cellcolor{gray!10}{0.0340} & \cellcolor{gray!10}{0.1622} & \cellcolor{gray!10}{0.1434} & \cellcolor{gray!10}{0.0017}\\
\hspace{1em}8000 & 0.1604 & 0.0290 & 0.1622 & 0.1278 & 0.0004\\
\hspace{1em}\cellcolor{gray!10}{16000} & \cellcolor{gray!10}{0.1625} & \cellcolor{gray!10}{0.0189} & \cellcolor{gray!10}{0.1622} & \cellcolor{gray!10}{0.7512} & \cellcolor{gray!10}{0.0003}\\
\hspace{1em}32000 & 0.1640 & 0.0140 & 0.1622 & 0.0022 & 0.0045\\
\hspace{1em}\cellcolor{gray!10}{48000} & \cellcolor{gray!10}{0.1642} & \cellcolor{gray!10}{0.0131} & \cellcolor{gray!10}{0.1622} & \cellcolor{gray!10}{0.0000} & \cellcolor{gray!10}{0.0045}\\
\hspace{1em}64000 & 0.1637 & 0.0102 & 0.1622 & 0.0002 & 0.0114\\
\addlinespace[0.3em]
\multicolumn{6}{l}{\textbf{Gamma21}}\\
\hspace{1em}\cellcolor{gray!10}{4000} & \cellcolor{gray!10}{0.1281} & \cellcolor{gray!10}{0.0210} & \cellcolor{gray!10}{0.1272} & \cellcolor{gray!10}{0.6696} & \cellcolor{gray!10}{0.0121}\\
\hspace{1em}8000 & 0.1276 & 0.0156 & 0.1272 & 0.7500 & 0.0018\\
\hspace{1em}\cellcolor{gray!10}{16000} & \cellcolor{gray!10}{0.1285} & \cellcolor{gray!10}{0.0124} & \cellcolor{gray!10}{0.1272} & \cellcolor{gray!10}{0.0092} & \cellcolor{gray!10}{0.0221}\\
\hspace{1em}32000 & 0.1297 & 0.0085 & 0.1272 & 0.0000 & 0.0355\\
\hspace{1em}\cellcolor{gray!10}{48000} & \cellcolor{gray!10}{0.1294} & \cellcolor{gray!10}{0.0074} & \cellcolor{gray!10}{0.1272} & \cellcolor{gray!10}{0.0000} & \cellcolor{gray!10}{0.0418}\\
\hspace{1em}64000 & 0.1297 & 0.0058 & 0.1272 & 0.0000 & 0.0395\\
\addlinespace[0.3em]
\multicolumn{6}{l}{\textbf{Gamma22}}\\
\hspace{1em}\cellcolor{gray!10}{4000} & \cellcolor{gray!10}{0.3594} & \cellcolor{gray!10}{0.0687} & \cellcolor{gray!10}{0.3683} & \cellcolor{gray!10}{0.0324} & \cellcolor{gray!10}{0.0216}\\
\hspace{1em}8000 & 0.3608 & 0.0498 & 0.3683 & 0.0018 & 0.0212\\
\hspace{1em}\cellcolor{gray!10}{16000} & \cellcolor{gray!10}{0.3650} & \cellcolor{gray!10}{0.0340} & \cellcolor{gray!10}{0.3683} & \cellcolor{gray!10}{0.0398} & \cellcolor{gray!10}{0.0232}\\
\hspace{1em}32000 & 0.3680 & 0.0246 & 0.3683 & 0.7626 & 0.0054\\
\hspace{1em}\cellcolor{gray!10}{48000} & \cellcolor{gray!10}{0.3687} & \cellcolor{gray!10}{0.0237} & \cellcolor{gray!10}{0.3683} & \cellcolor{gray!10}{0.6242} & \cellcolor{gray!10}{0.0113}\\
\hspace{1em}64000 & 0.3698 & 0.0210 & 0.3683 & 0.1448 & 0.0005\\
\addlinespace[0.3em]
\multicolumn{6}{l}{\textbf{gc11}}\\
\hspace{1em}\cellcolor{gray!10}{4000} & \cellcolor{gray!10}{0.0063} & \cellcolor{gray!10}{0.0042} & \cellcolor{gray!10}{0.0069} & \cellcolor{gray!10}{0.0036} & \cellcolor{gray!10}{0.0287}\\
\hspace{1em}8000 & 0.0068 & 0.0032 & 0.0069 & 0.3452 & 0.0117\\
\hspace{1em}\cellcolor{gray!10}{16000} & \cellcolor{gray!10}{0.0064} & \cellcolor{gray!10}{0.0023} & \cellcolor{gray!10}{0.0069} & \cellcolor{gray!10}{0.0018} & \cellcolor{gray!10}{0.0579}\\
\hspace{1em}32000 & 0.0065 & 0.0016 & 0.0069 & 0.0002 & 0.0626\\
\hspace{1em}\cellcolor{gray!10}{48000} & \cellcolor{gray!10}{0.0065} & \cellcolor{gray!10}{0.0015} & \cellcolor{gray!10}{0.0069} & \cellcolor{gray!10}{0.0000} & \cellcolor{gray!10}{0.0779}\\
\hspace{1em}64000 & 0.0065 & 0.0014 & 0.0069 & 0.0000 & 0.0975\\
\addlinespace[0.3em]
\multicolumn{6}{l}{\textbf{gc12}}\\
\hspace{1em}\cellcolor{gray!10}{4000} & \cellcolor{gray!10}{0.0028} & \cellcolor{gray!10}{0.0040} & \cellcolor{gray!10}{0.0030} & \cellcolor{gray!10}{0.0970} & \cellcolor{gray!10}{0.0000}\\
\hspace{1em}8000 & 0.0030 & 0.0030 & 0.0030 & 0.9122 & 0.0000\\
\hspace{1em}\cellcolor{gray!10}{16000} & \cellcolor{gray!10}{0.0028} & \cellcolor{gray!10}{0.0019} & \cellcolor{gray!10}{0.0030} & \cellcolor{gray!10}{0.0450} & \cellcolor{gray!10}{0.0029}\\
\hspace{1em}32000 & 0.0030 & 0.0015 & 0.0030 & 0.6840 & 0.0020\\
\hspace{1em}\cellcolor{gray!10}{48000} & \cellcolor{gray!10}{0.0029} & \cellcolor{gray!10}{0.0012} & \cellcolor{gray!10}{0.0030} & \cellcolor{gray!10}{0.2274} & \cellcolor{gray!10}{0.0031}\\
\hspace{1em}64000 & 0.0030 & 0.0011 & 0.0030 & 0.7142 & 0.0011\\
\addlinespace[0.3em]
\multicolumn{6}{l}{\textbf{gc22}}\\
\hspace{1em}\cellcolor{gray!10}{4000} & \cellcolor{gray!10}{0.0076} & \cellcolor{gray!10}{0.0038} & \cellcolor{gray!10}{0.0078} & \cellcolor{gray!10}{0.3422} & \cellcolor{gray!10}{0.0051}\\
\hspace{1em}8000 & 0.0075 & 0.0029 & 0.0078 & 0.1188 & 0.0414\\
\hspace{1em}\cellcolor{gray!10}{16000} & \cellcolor{gray!10}{0.0074} & \cellcolor{gray!10}{0.0023} & \cellcolor{gray!10}{0.0078} & \cellcolor{gray!10}{0.0002} & \cellcolor{gray!10}{0.0411}\\
\hspace{1em}32000 & 0.0073 & 0.0016 & 0.0078 & 0.0000 & 0.0754\\
\hspace{1em}\cellcolor{gray!10}{48000} & \cellcolor{gray!10}{0.0075} & \cellcolor{gray!10}{0.0015} & \cellcolor{gray!10}{0.0078} & \cellcolor{gray!10}{0.0000} & \cellcolor{gray!10}{0.0716}\\
\hspace{1em}64000 & 0.0075 & 0.0013 & 0.0078 & 0.0000 & 0.1040\\
\addlinespace[0.3em]
\multicolumn{6}{l}{\textbf{hc11}}\\
\hspace{1em}\cellcolor{gray!10}{4000} & \cellcolor{gray!10}{0.1014} & \cellcolor{gray!10}{0.0243} & \cellcolor{gray!10}{0.0982} & \cellcolor{gray!10}{0.0284} & \cellcolor{gray!10}{0.0099}\\
\hspace{1em}8000 & 0.1024 & 0.0167 & 0.0982 & 0.0000 & 0.0055\\
\hspace{1em}\cellcolor{gray!10}{16000} & \cellcolor{gray!10}{0.1024} & \cellcolor{gray!10}{0.0114} & \cellcolor{gray!10}{0.0982} & \cellcolor{gray!10}{0.0000} & \cellcolor{gray!10}{0.0155}\\
\hspace{1em}32000 & 0.1023 & 0.0085 & 0.0982 & 0.0000 & 0.0208\\
\hspace{1em}\cellcolor{gray!10}{48000} & \cellcolor{gray!10}{0.1021} & \cellcolor{gray!10}{0.0072} & \cellcolor{gray!10}{0.0982} & \cellcolor{gray!10}{0.0000} & \cellcolor{gray!10}{0.0194}\\
\hspace{1em}64000 & 0.1027 & 0.0061 & 0.0982 & 0.0000 & 0.0249\\
\addlinespace[0.3em]
\multicolumn{6}{l}{\textbf{hc12}}\\
\hspace{1em}\cellcolor{gray!10}{4000} & \cellcolor{gray!10}{0.0241} & \cellcolor{gray!10}{0.0086} & \cellcolor{gray!10}{0.0244} & \cellcolor{gray!10}{0.5102} & \cellcolor{gray!10}{0.0166}\\
\hspace{1em}8000 & 0.0245 & 0.0067 & 0.0244 & 0.7578 & 0.0019\\
\hspace{1em}\cellcolor{gray!10}{16000} & \cellcolor{gray!10}{0.0250} & \cellcolor{gray!10}{0.0049} & \cellcolor{gray!10}{0.0244} & \cellcolor{gray!10}{0.0536} & \cellcolor{gray!10}{0.0138}\\
\hspace{1em}32000 & 0.0252 & 0.0030 & 0.0244 & 0.0000 & 0.0390\\
\hspace{1em}\cellcolor{gray!10}{48000} & \cellcolor{gray!10}{0.0253} & \cellcolor{gray!10}{0.0032} & \cellcolor{gray!10}{0.0244} & \cellcolor{gray!10}{0.0000} & \cellcolor{gray!10}{0.0413}\\
\hspace{1em}64000 & 0.0253 & 0.0023 & 0.0244 & 0.0000 & 0.0550\\
\addlinespace[0.3em]
\multicolumn{6}{l}{\textbf{hc22}}\\
\hspace{1em}\cellcolor{gray!10}{4000} & \cellcolor{gray!10}{0.0322} & \cellcolor{gray!10}{0.0131} & \cellcolor{gray!10}{0.0326} & \cellcolor{gray!10}{0.4604} & \cellcolor{gray!10}{0.0081}\\
\hspace{1em}8000 & 0.0316 & 0.0098 & 0.0326 & 0.0240 & 0.0065\\
\hspace{1em}\cellcolor{gray!10}{16000} & \cellcolor{gray!10}{0.0325} & \cellcolor{gray!10}{0.0063} & \cellcolor{gray!10}{0.0326} & \cellcolor{gray!10}{0.6030} & \cellcolor{gray!10}{0.0033}\\
\hspace{1em}32000 & 0.0328 & 0.0045 & 0.0326 & 0.8262 & 0.0034\\
\hspace{1em}\cellcolor{gray!10}{48000} & \cellcolor{gray!10}{0.0331} & \cellcolor{gray!10}{0.0045} & \cellcolor{gray!10}{0.0326} & \cellcolor{gray!10}{0.0408} & \cellcolor{gray!10}{0.0027}\\
\hspace{1em}64000 & 0.0332 & 0.0038 & 0.0326 & 0.0078 & 0.0063\\
\addlinespace[0.3em]
\multicolumn{6}{l}{\textbf{ic11}}\\
\hspace{1em}\cellcolor{gray!10}{4000} & \cellcolor{gray!10}{0.0265} & \cellcolor{gray!10}{0.0039} & \cellcolor{gray!10}{0.0260} & \cellcolor{gray!10}{0.0328} & \cellcolor{gray!10}{0.0205}\\
\hspace{1em}8000 & 0.0265 & 0.0027 & 0.0260 & 0.0052 & 0.0229\\
\hspace{1em}\cellcolor{gray!10}{16000} & \cellcolor{gray!10}{0.0266} & \cellcolor{gray!10}{0.0018} & \cellcolor{gray!10}{0.0260} & \cellcolor{gray!10}{0.0000} & \cellcolor{gray!10}{0.0001}\\
\hspace{1em}32000 & 0.0267 & 0.0012 & 0.0260 & 0.0000 & 0.0161\\
\hspace{1em}\cellcolor{gray!10}{48000} & \cellcolor{gray!10}{0.0267} & \cellcolor{gray!10}{0.0010} & \cellcolor{gray!10}{0.0260} & \cellcolor{gray!10}{0.0000} & \cellcolor{gray!10}{0.0109}\\
\hspace{1em}64000 & 0.0267 & 0.0009 & 0.0260 & 0.0000 & 0.0278\\
\addlinespace[0.3em]
\multicolumn{6}{l}{\textbf{ic12}}\\
\hspace{1em}\cellcolor{gray!10}{4000} & \cellcolor{gray!10}{0.0100} & \cellcolor{gray!10}{0.0026} & \cellcolor{gray!10}{0.0071} & \cellcolor{gray!10}{0.0000} & \cellcolor{gray!10}{0.2677}\\
\hspace{1em}8000 & 0.0101 & 0.0018 & 0.0071 & 0.0000 & 0.5153\\
\hspace{1em}\cellcolor{gray!10}{16000} & \cellcolor{gray!10}{0.0102} & \cellcolor{gray!10}{0.0013} & \cellcolor{gray!10}{0.0071} & \cellcolor{gray!10}{0.0000} & \cellcolor{gray!10}{0.3867}\\
\hspace{1em}32000 & 0.0103 & 0.0009 & 0.0071 & 0.0000 & 0.3869\\
\hspace{1em}\cellcolor{gray!10}{48000} & \cellcolor{gray!10}{0.0102} & \cellcolor{gray!10}{0.0008} & \cellcolor{gray!10}{0.0071} & \cellcolor{gray!10}{0.0000} & \cellcolor{gray!10}{0.4429}\\
\hspace{1em}64000 & 0.0102 & 0.0007 & 0.0071 & 0.0000 & 0.3801\\
\addlinespace[0.3em]
\multicolumn{6}{l}{\textbf{ic21}}\\
\hspace{1em}\cellcolor{gray!10}{4000} & \cellcolor{gray!10}{0.0069} & \cellcolor{gray!10}{0.0026} & \cellcolor{gray!10}{0.0100} & \cellcolor{gray!10}{0.0000} & \cellcolor{gray!10}{0.2682}\\
\hspace{1em}8000 & 0.0070 & 0.0020 & 0.0100 & 0.0000 & 0.4008\\
\hspace{1em}\cellcolor{gray!10}{16000} & \cellcolor{gray!10}{0.0071} & \cellcolor{gray!10}{0.0014} & \cellcolor{gray!10}{0.0100} & \cellcolor{gray!10}{0.0000} & \cellcolor{gray!10}{0.5893}\\
\hspace{1em}32000 & 0.0072 & 0.0009 & 0.0100 & 0.0000 & 0.6847\\
\hspace{1em}\cellcolor{gray!10}{48000} & \cellcolor{gray!10}{0.0072} & \cellcolor{gray!10}{0.0009} & \cellcolor{gray!10}{0.0100} & \cellcolor{gray!10}{0.0000} & \cellcolor{gray!10}{0.7466}\\
\hspace{1em}64000 & 0.0072 & 0.0007 & 0.0100 & 0.0000 & 0.7801\\
\addlinespace[0.3em]
\multicolumn{6}{l}{\textbf{ic22}}\\
\hspace{1em}\cellcolor{gray!10}{4000} & \cellcolor{gray!10}{0.0154} & \cellcolor{gray!10}{0.0038} & \cellcolor{gray!10}{0.0159} & \cellcolor{gray!10}{0.0008} & \cellcolor{gray!10}{0.0372}\\
\hspace{1em}8000 & 0.0156 & 0.0027 & 0.0159 & 0.0052 & 0.0603\\
\hspace{1em}\cellcolor{gray!10}{16000} & \cellcolor{gray!10}{0.0159} & \cellcolor{gray!10}{0.0019} & \cellcolor{gray!10}{0.0159} & \cellcolor{gray!10}{0.9520} & \cellcolor{gray!10}{0.0602}\\
\hspace{1em}32000 & 0.0160 & 0.0012 & 0.0159 & 0.1344 & 0.0339\\
\hspace{1em}\cellcolor{gray!10}{48000} & \cellcolor{gray!10}{0.0160} & \cellcolor{gray!10}{0.0011} & \cellcolor{gray!10}{0.0159} & \cellcolor{gray!10}{0.0978} & \cellcolor{gray!10}{0.0278}\\
\hspace{1em}64000 & 0.0161 & 0.0009 & 0.0159 & 0.0006 & 0.0057\\
\addlinespace[0.3em]
\multicolumn{6}{l}{\textbf{itlo11}}\\
\hspace{1em}\cellcolor{gray!10}{4000} & \cellcolor{gray!10}{0.0267} & \cellcolor{gray!10}{0.0037} & \cellcolor{gray!10}{0.0262} & \cellcolor{gray!10}{0.0952} & \cellcolor{gray!10}{0.0007}\\
\hspace{1em}8000 & 0.0267 & 0.0026 & 0.0262 & 0.0016 & 0.0027\\
\hspace{1em}\cellcolor{gray!10}{16000} & \cellcolor{gray!10}{0.0267} & \cellcolor{gray!10}{0.0017} & \cellcolor{gray!10}{0.0262} & \cellcolor{gray!10}{0.0000} & \cellcolor{gray!10}{0.0154}\\
\hspace{1em}32000 & 0.0269 & 0.0011 & 0.0262 & 0.0000 & 0.0222\\
\hspace{1em}\cellcolor{gray!10}{48000} & \cellcolor{gray!10}{0.0268} & \cellcolor{gray!10}{0.0010} & \cellcolor{gray!10}{0.0262} & \cellcolor{gray!10}{0.0000} & \cellcolor{gray!10}{0.0198}\\
\hspace{1em}64000 & 0.0269 & 0.0009 & 0.0262 & 0.0000 & 0.0020\\
\addlinespace[0.3em]
\multicolumn{6}{l}{\textbf{itlo12}}\\
\hspace{1em}\cellcolor{gray!10}{4000} & \cellcolor{gray!10}{0.0141} & \cellcolor{gray!10}{0.0028} & \cellcolor{gray!10}{0.0141} & \cellcolor{gray!10}{0.5530} & \cellcolor{gray!10}{0.0000}\\
\hspace{1em}8000 & 0.0141 & 0.0018 & 0.0141 & 0.7688 & 0.0026\\
\hspace{1em}\cellcolor{gray!10}{16000} & \cellcolor{gray!10}{0.0143} & \cellcolor{gray!10}{0.0015} & \cellcolor{gray!10}{0.0141} & \cellcolor{gray!10}{0.1348} & \cellcolor{gray!10}{0.0128}\\
\hspace{1em}32000 & 0.0143 & 0.0009 & 0.0141 & 0.0074 & 0.0231\\
\hspace{1em}\cellcolor{gray!10}{48000} & \cellcolor{gray!10}{0.0142} & \cellcolor{gray!10}{0.0009} & \cellcolor{gray!10}{0.0141} & \cellcolor{gray!10}{0.4018} & \cellcolor{gray!10}{0.0065}\\
\hspace{1em}64000 & 0.0142 & 0.0008 & 0.0141 & 0.0646 & 0.0148\\
\addlinespace[0.3em]
\multicolumn{6}{l}{\textbf{itlo21}}\\
\hspace{1em}\cellcolor{gray!10}{4000} & \cellcolor{gray!10}{0.0051} & \cellcolor{gray!10}{0.0018} & \cellcolor{gray!10}{0.0050} & \cellcolor{gray!10}{0.5828} & \cellcolor{gray!10}{0.0123}\\
\hspace{1em}8000 & 0.0051 & 0.0014 & 0.0050 & 0.3370 & 0.0012\\
\hspace{1em}\cellcolor{gray!10}{16000} & \cellcolor{gray!10}{0.0051} & \cellcolor{gray!10}{0.0011} & \cellcolor{gray!10}{0.0050} & \cellcolor{gray!10}{0.4570} & \cellcolor{gray!10}{0.0004}\\
\hspace{1em}32000 & 0.0052 & 0.0007 & 0.0050 & 0.0000 & 0.0002\\
\hspace{1em}\cellcolor{gray!10}{48000} & \cellcolor{gray!10}{0.0052} & \cellcolor{gray!10}{0.0006} & \cellcolor{gray!10}{0.0050} & \cellcolor{gray!10}{0.0000} & \cellcolor{gray!10}{0.0378}\\
\hspace{1em}64000 & 0.0052 & 0.0005 & 0.0050 & 0.0000 & 0.0244\\
\addlinespace[0.3em]
\multicolumn{6}{l}{\textbf{itlo22}}\\
\hspace{1em}\cellcolor{gray!10}{4000} & \cellcolor{gray!10}{0.0157} & \cellcolor{gray!10}{0.0036} & \cellcolor{gray!10}{0.0161} & \cellcolor{gray!10}{0.0012} & \cellcolor{gray!10}{0.0049}\\
\hspace{1em}8000 & 0.0158 & 0.0026 & 0.0161 & 0.0018 & 0.0104\\
\hspace{1em}\cellcolor{gray!10}{16000} & \cellcolor{gray!10}{0.0161} & \cellcolor{gray!10}{0.0019} & \cellcolor{gray!10}{0.0161} & \cellcolor{gray!10}{0.5170} & \cellcolor{gray!10}{0.0173}\\
\hspace{1em}32000 & 0.0162 & 0.0012 & 0.0161 & 0.0616 & 0.0005\\
\hspace{1em}\cellcolor{gray!10}{48000} & \cellcolor{gray!10}{0.0163} & \cellcolor{gray!10}{0.0011} & \cellcolor{gray!10}{0.0161} & \cellcolor{gray!10}{0.0142} & \cellcolor{gray!10}{0.0021}\\
\hspace{1em}64000 & 0.0163 & 0.0009 & 0.0161 & 0.0010 & 0.0036\\
\addlinespace[0.3em]
\multicolumn{6}{l}{\textbf{itol11}}\\
\hspace{1em}\cellcolor{gray!10}{4000} & \cellcolor{gray!10}{0.0264} & \cellcolor{gray!10}{0.0036} & \cellcolor{gray!10}{0.0258} & \cellcolor{gray!10}{0.0020} & \cellcolor{gray!10}{0.0012}\\
\hspace{1em}8000 & 0.0264 & 0.0025 & 0.0258 & 0.0004 & 0.0000\\
\hspace{1em}\cellcolor{gray!10}{16000} & \cellcolor{gray!10}{0.0264} & \cellcolor{gray!10}{0.0017} & \cellcolor{gray!10}{0.0258} & \cellcolor{gray!10}{0.0000} & \cellcolor{gray!10}{0.0082}\\
\hspace{1em}32000 & 0.0265 & 0.0011 & 0.0258 & 0.0000 & 0.0197\\
\hspace{1em}\cellcolor{gray!10}{48000} & \cellcolor{gray!10}{0.0265} & \cellcolor{gray!10}{0.0010} & \cellcolor{gray!10}{0.0258} & \cellcolor{gray!10}{0.0000} & \cellcolor{gray!10}{0.0248}\\
\hspace{1em}64000 & 0.0265 & 0.0009 & 0.0258 & 0.0000 & 0.0269\\
\addlinespace[0.3em]
\multicolumn{6}{l}{\textbf{itol12}}\\
\hspace{1em}\cellcolor{gray!10}{4000} & \cellcolor{gray!10}{0.0087} & \cellcolor{gray!10}{0.0029} & \cellcolor{gray!10}{0.0092} & \cellcolor{gray!10}{0.0002} & \cellcolor{gray!10}{0.0081}\\
\hspace{1em}8000 & 0.0089 & 0.0023 & 0.0092 & 0.0000 & 0.0119\\
\hspace{1em}\cellcolor{gray!10}{16000} & \cellcolor{gray!10}{0.0091} & \cellcolor{gray!10}{0.0016} & \cellcolor{gray!10}{0.0092} & \cellcolor{gray!10}{0.3754} & \cellcolor{gray!10}{0.0007}\\
\hspace{1em}32000 & 0.0092 & 0.0011 & 0.0092 & 0.3286 & 0.0002\\
\hspace{1em}\cellcolor{gray!10}{48000} & \cellcolor{gray!10}{0.0092} & \cellcolor{gray!10}{0.0010} & \cellcolor{gray!10}{0.0092} & \cellcolor{gray!10}{0.4994} & \cellcolor{gray!10}{0.0001}\\
\hspace{1em}64000 & 0.0092 & 0.0008 & 0.0092 & 0.9872 & 0.0013\\
\addlinespace[0.3em]
\multicolumn{6}{l}{\textbf{itol21}}\\
\hspace{1em}\cellcolor{gray!10}{4000} & \cellcolor{gray!10}{0.0061} & \cellcolor{gray!10}{0.0027} & \cellcolor{gray!10}{0.0059} & \cellcolor{gray!10}{0.1182} & \cellcolor{gray!10}{0.0096}\\
\hspace{1em}8000 & 0.0061 & 0.0017 & 0.0059 & 0.0560 & 0.0154\\
\hspace{1em}\cellcolor{gray!10}{16000} & \cellcolor{gray!10}{0.0062} & \cellcolor{gray!10}{0.0012} & \cellcolor{gray!10}{0.0059} & \cellcolor{gray!10}{0.0000} & \cellcolor{gray!10}{0.0292}\\
\hspace{1em}32000 & 0.0062 & 0.0008 & 0.0059 & 0.0000 & 0.0355\\
\hspace{1em}\cellcolor{gray!10}{48000} & \cellcolor{gray!10}{0.0062} & \cellcolor{gray!10}{0.0007} & \cellcolor{gray!10}{0.0059} & \cellcolor{gray!10}{0.0000} & \cellcolor{gray!10}{0.0282}\\
\hspace{1em}64000 & 0.0062 & 0.0006 & 0.0059 & 0.0000 & 0.0325\\
\addlinespace[0.3em]
\multicolumn{6}{l}{\textbf{itol22}}\\
\hspace{1em}\cellcolor{gray!10}{4000} & \cellcolor{gray!10}{0.0154} & \cellcolor{gray!10}{0.0035} & \cellcolor{gray!10}{0.0157} & \cellcolor{gray!10}{0.0970} & \cellcolor{gray!10}{0.0400}\\
\hspace{1em}8000 & 0.0154 & 0.0026 & 0.0157 & 0.0606 & 0.0631\\
\hspace{1em}\cellcolor{gray!10}{16000} & \cellcolor{gray!10}{0.0158} & \cellcolor{gray!10}{0.0018} & \cellcolor{gray!10}{0.0157} & \cellcolor{gray!10}{0.2802} & \cellcolor{gray!10}{0.0598}\\
\hspace{1em}32000 & 0.0158 & 0.0012 & 0.0157 & 0.4394 & 0.0427\\
\hspace{1em}\cellcolor{gray!10}{48000} & \cellcolor{gray!10}{0.0158} & \cellcolor{gray!10}{0.0010} & \cellcolor{gray!10}{0.0157} & \cellcolor{gray!10}{0.1210} & \cellcolor{gray!10}{0.0328}\\
\hspace{1em}64000 & 0.0159 & 0.0009 & 0.0157 & 0.0014 & 0.0108\\
\addlinespace[0.3em]
\multicolumn{6}{l}{\textbf{k12}}\\
\hspace{1em}\cellcolor{gray!10}{4000} & \cellcolor{gray!10}{0.0501} & \cellcolor{gray!10}{0.0055} & \cellcolor{gray!10}{0.0500} & \cellcolor{gray!10}{0.7376} & \cellcolor{gray!10}{0.0002}\\
\hspace{1em}8000 & 0.0502 & 0.0040 & 0.0500 & 0.4498 & 0.0018\\
\hspace{1em}\cellcolor{gray!10}{16000} & \cellcolor{gray!10}{0.0500} & \cellcolor{gray!10}{0.0027} & \cellcolor{gray!10}{0.0500} & \cellcolor{gray!10}{0.6910} & \cellcolor{gray!10}{0.0009}\\
\hspace{1em}32000 & 0.0500 & 0.0020 & 0.0500 & 0.8954 & 0.0010\\
\hspace{1em}\cellcolor{gray!10}{48000} & \cellcolor{gray!10}{0.0501} & \cellcolor{gray!10}{0.0016} & \cellcolor{gray!10}{0.0500} & \cellcolor{gray!10}{0.1380} & \cellcolor{gray!10}{0.0020}\\
\hspace{1em}64000 & 0.0501 & 0.0014 & 0.0500 & 0.7538 & 0.0002\\
\addlinespace[0.3em]
\multicolumn{6}{l}{\textbf{mu11}}\\
\hspace{1em}\cellcolor{gray!10}{4000} & \cellcolor{gray!10}{0.2015} & \cellcolor{gray!10}{0.0070} & \cellcolor{gray!10}{0.1996} & \cellcolor{gray!10}{0.0000} & \cellcolor{gray!10}{0.0637}\\
\hspace{1em}8000 & 0.2011 & 0.0051 & 0.1996 & 0.0000 & 0.0858\\
\hspace{1em}\cellcolor{gray!10}{16000} & \cellcolor{gray!10}{0.2012} & \cellcolor{gray!10}{0.0030} & \cellcolor{gray!10}{0.1996} & \cellcolor{gray!10}{0.0000} & \cellcolor{gray!10}{0.1982}\\
\hspace{1em}32000 & 0.2010 & 0.0023 & 0.1996 & 0.0000 & 0.3059\\
\hspace{1em}\cellcolor{gray!10}{48000} & \cellcolor{gray!10}{0.2009} & \cellcolor{gray!10}{0.0018} & \cellcolor{gray!10}{0.1996} & \cellcolor{gray!10}{0.0000} & \cellcolor{gray!10}{0.3349}\\
\hspace{1em}64000 & 0.2010 & 0.0015 & 0.1996 & 0.0000 & 0.4418\\
\addlinespace[0.3em]
\multicolumn{6}{l}{\textbf{mu12}}\\
\hspace{1em}\cellcolor{gray!10}{4000} & \cellcolor{gray!10}{-0.0018} & \cellcolor{gray!10}{0.0094} & \cellcolor{gray!10}{0.0005} & \cellcolor{gray!10}{0.0000} & \cellcolor{gray!10}{0.0549}\\
\hspace{1em}8000 & -0.0013 & 0.0066 & 0.0005 & 0.0000 & 0.0837\\
\hspace{1em}\cellcolor{gray!10}{16000} & \cellcolor{gray!10}{-0.0014} & \cellcolor{gray!10}{0.0045} & \cellcolor{gray!10}{0.0005} & \cellcolor{gray!10}{0.0000} & \cellcolor{gray!10}{0.1434}\\
\hspace{1em}32000 & -0.0012 & 0.0028 & 0.0005 & 0.0000 & 0.2971\\
\hspace{1em}\cellcolor{gray!10}{48000} & \cellcolor{gray!10}{-0.0014} & \cellcolor{gray!10}{0.0022} & \cellcolor{gray!10}{0.0005} & \cellcolor{gray!10}{0.0000} & \cellcolor{gray!10}{0.4074}\\
\hspace{1em}64000 & -0.0014 & 0.0016 & 0.0005 & 0.0000 & 0.5278\\
\addlinespace[0.3em]
\multicolumn{6}{l}{\textbf{mu21}}\\
\hspace{1em}\cellcolor{gray!10}{4000} & \cellcolor{gray!10}{-0.0657} & \cellcolor{gray!10}{0.0097} & \cellcolor{gray!10}{-0.0673} & \cellcolor{gray!10}{0.0006} & \cellcolor{gray!10}{0.0157}\\
\hspace{1em}8000 & -0.0666 & 0.0065 & -0.0673 & 0.1242 & 0.0272\\
\hspace{1em}\cellcolor{gray!10}{16000} & \cellcolor{gray!10}{-0.0663} & \cellcolor{gray!10}{0.0047} & \cellcolor{gray!10}{-0.0673} & \cellcolor{gray!10}{0.0184} & \cellcolor{gray!10}{0.0583}\\
\hspace{1em}32000 & -0.0663 & 0.0032 & -0.0673 & 0.0000 & 0.1165\\
\hspace{1em}\cellcolor{gray!10}{48000} & \cellcolor{gray!10}{-0.0661} & \cellcolor{gray!10}{0.0021} & \cellcolor{gray!10}{-0.0673} & \cellcolor{gray!10}{0.0000} & \cellcolor{gray!10}{0.2235}\\
\hspace{1em}64000 & -0.0660 & 0.0016 & -0.0673 & 0.0000 & 0.3270\\
\addlinespace[0.3em]
\multicolumn{6}{l}{\textbf{mu22}}\\
\hspace{1em}\cellcolor{gray!10}{4000} & \cellcolor{gray!10}{0.2317} & \cellcolor{gray!10}{0.0124} & \cellcolor{gray!10}{0.2311} & \cellcolor{gray!10}{0.6164} & \cellcolor{gray!10}{0.0035}\\
\hspace{1em}8000 & 0.2330 & 0.0081 & 0.2311 & 0.0000 & 0.0261\\
\hspace{1em}\cellcolor{gray!10}{16000} & \cellcolor{gray!10}{0.2324} & \cellcolor{gray!10}{0.0054} & \cellcolor{gray!10}{0.2311} & \cellcolor{gray!10}{0.0000} & \cellcolor{gray!10}{0.0325}\\
\hspace{1em}32000 & 0.2322 & 0.0033 & 0.2311 & 0.0000 & 0.1109\\
\hspace{1em}\cellcolor{gray!10}{48000} & \cellcolor{gray!10}{0.2324} & \cellcolor{gray!10}{0.0024} & \cellcolor{gray!10}{0.2311} & \cellcolor{gray!10}{0.0000} & \cellcolor{gray!10}{0.2251}\\
\hspace{1em}64000 & 0.2324 & 0.0017 & 0.2311 & 0.0000 & 0.2151\\
\addlinespace[0.3em]
\multicolumn{6}{l}{\textbf{Omega11}}\\
\hspace{1em}\cellcolor{gray!10}{4000} & \cellcolor{gray!10}{0.1874} & \cellcolor{gray!10}{0.0088} & \cellcolor{gray!10}{0.1877} & \cellcolor{gray!10}{0.6838} & \cellcolor{gray!10}{0.0001}\\
\hspace{1em}8000 & 0.1881 & 0.0069 & 0.1877 & 0.0694 & 0.0000\\
\hspace{1em}\cellcolor{gray!10}{16000} & \cellcolor{gray!10}{0.1875} & \cellcolor{gray!10}{0.0048} & \cellcolor{gray!10}{0.1877} & \cellcolor{gray!10}{0.3084} & \cellcolor{gray!10}{0.0001}\\
\hspace{1em}32000 & 0.1875 & 0.0040 & 0.1877 & 0.3126 & 0.0004\\
\hspace{1em}\cellcolor{gray!10}{48000} & \cellcolor{gray!10}{0.1879} & \cellcolor{gray!10}{0.0036} & \cellcolor{gray!10}{0.1877} & \cellcolor{gray!10}{0.4404} & \cellcolor{gray!10}{0.0000}\\
\hspace{1em}64000 & 0.1878 & 0.0032 & 0.1877 & 0.2934 & 0.0000\\
\addlinespace[0.3em]
\multicolumn{6}{l}{\textbf{Omega12}}\\
\hspace{1em}\cellcolor{gray!10}{4000} & \cellcolor{gray!10}{0.0656} & \cellcolor{gray!10}{0.0089} & \cellcolor{gray!10}{0.0652} & \cellcolor{gray!10}{0.5782} & \cellcolor{gray!10}{0.0022}\\
\hspace{1em}8000 & 0.0657 & 0.0057 & 0.0652 & 0.1886 & 0.0014\\
\hspace{1em}\cellcolor{gray!10}{16000} & \cellcolor{gray!10}{0.0654} & \cellcolor{gray!10}{0.0044} & \cellcolor{gray!10}{0.0652} & \cellcolor{gray!10}{0.5380} & \cellcolor{gray!10}{0.0066}\\
\hspace{1em}32000 & 0.0653 & 0.0037 & 0.0652 & 0.8948 & 0.0001\\
\hspace{1em}\cellcolor{gray!10}{48000} & \cellcolor{gray!10}{0.0653} & \cellcolor{gray!10}{0.0032} & \cellcolor{gray!10}{0.0652} & \cellcolor{gray!10}{0.5536} & \cellcolor{gray!10}{0.0000}\\
\hspace{1em}64000 & 0.0652 & 0.0029 & 0.0652 & 0.8620 & 0.0001\\
\addlinespace[0.3em]
\multicolumn{6}{l}{\textbf{Omega21}}\\
\hspace{1em}\cellcolor{gray!10}{4000} & \cellcolor{gray!10}{0.0428} & \cellcolor{gray!10}{0.0063} & \cellcolor{gray!10}{0.0420} & \cellcolor{gray!10}{0.0134} & \cellcolor{gray!10}{0.0104}\\
\hspace{1em}8000 & 0.0421 & 0.0048 & 0.0420 & 0.6444 & 0.0009\\
\hspace{1em}\cellcolor{gray!10}{16000} & \cellcolor{gray!10}{0.0420} & \cellcolor{gray!10}{0.0036} & \cellcolor{gray!10}{0.0420} & \cellcolor{gray!10}{0.9988} & \cellcolor{gray!10}{0.0002}\\
\hspace{1em}32000 & 0.0420 & 0.0027 & 0.0420 & 0.9626 & 0.0002\\
\hspace{1em}\cellcolor{gray!10}{48000} & \cellcolor{gray!10}{0.0420} & \cellcolor{gray!10}{0.0022} & \cellcolor{gray!10}{0.0420} & \cellcolor{gray!10}{0.8236} & \cellcolor{gray!10}{0.0000}\\
\hspace{1em}64000 & 0.0420 & 0.0020 & 0.0420 & 0.7228 & 0.0002\\
\addlinespace[0.3em]
\multicolumn{6}{l}{\textbf{Omega22}}\\
\hspace{1em}\cellcolor{gray!10}{4000} & \cellcolor{gray!10}{0.1828} & \cellcolor{gray!10}{0.0071} & \cellcolor{gray!10}{0.1834} & \cellcolor{gray!10}{0.0958} & \cellcolor{gray!10}{0.0048}\\
\hspace{1em}8000 & 0.1828 & 0.0047 & 0.1834 & 0.0308 & 0.0015\\
\hspace{1em}\cellcolor{gray!10}{16000} & \cellcolor{gray!10}{0.1834} & \cellcolor{gray!10}{0.0034} & \cellcolor{gray!10}{0.1834} & \cellcolor{gray!10}{0.7694} & \cellcolor{gray!10}{0.0001}\\
\hspace{1em}32000 & 0.1833 & 0.0027 & 0.1834 & 0.6190 & 0.0006\\
\hspace{1em}\cellcolor{gray!10}{48000} & \cellcolor{gray!10}{0.1834} & \cellcolor{gray!10}{0.0022} & \cellcolor{gray!10}{0.1834} & \cellcolor{gray!10}{0.8868} & \cellcolor{gray!10}{0.0023}\\
\hspace{1em}64000 & 0.1832 & 0.0022 & 0.1834 & 0.2454 & 0.0085\\
\addlinespace[0.3em]
\multicolumn{6}{l}{\textbf{v11}}\\
\hspace{1em}\cellcolor{gray!10}{4000} & \cellcolor{gray!10}{0.3271} & \cellcolor{gray!10}{0.1325} & \cellcolor{gray!10}{0.3374} & \cellcolor{gray!10}{0.3388} & \cellcolor{gray!10}{0.0367}\\
\hspace{1em}8000 & 0.3286 & 0.1018 & 0.3374 & 0.0960 & 0.0387\\
\hspace{1em}\cellcolor{gray!10}{16000} & \cellcolor{gray!10}{0.3307} & \cellcolor{gray!10}{0.0680} & \cellcolor{gray!10}{0.3374} & \cellcolor{gray!10}{0.0988} & \cellcolor{gray!10}{0.0058}\\
\hspace{1em}32000 & 0.3350 & 0.0526 & 0.3374 & 0.2812 & 0.0055\\
\hspace{1em}\cellcolor{gray!10}{48000} & \cellcolor{gray!10}{0.3353} & \cellcolor{gray!10}{0.0452} & \cellcolor{gray!10}{0.3374} & \cellcolor{gray!10}{0.2658} & \cellcolor{gray!10}{0.0026}\\
\hspace{1em}64000 & 0.3310 & 0.0406 & 0.3374 & 0.0022 & 0.0060\\
\addlinespace[0.3em]
\multicolumn{6}{l}{\textbf{v12}}\\
\hspace{1em}\cellcolor{gray!10}{4000} & \cellcolor{gray!10}{0.1573} & \cellcolor{gray!10}{0.0500} & \cellcolor{gray!10}{0.1684} & \cellcolor{gray!10}{0.0000} & \cellcolor{gray!10}{0.0073}\\
\hspace{1em}8000 & 0.1613 & 0.0399 & 0.1684 & 0.0002 & 0.0127\\
\hspace{1em}\cellcolor{gray!10}{16000} & \cellcolor{gray!10}{0.1644} & \cellcolor{gray!10}{0.0287} & \cellcolor{gray!10}{0.1684} & \cellcolor{gray!10}{0.0118} & \cellcolor{gray!10}{0.0166}\\
\hspace{1em}32000 & 0.1661 & 0.0185 & 0.1684 & 0.1176 & 0.0172\\
\hspace{1em}\cellcolor{gray!10}{48000} & \cellcolor{gray!10}{0.1662} & \cellcolor{gray!10}{0.0176} & \cellcolor{gray!10}{0.1684} & \cellcolor{gray!10}{0.0558} & \cellcolor{gray!10}{0.0193}\\
\hspace{1em}64000 & 0.1662 & 0.0159 & 0.1684 & 0.0072 & 0.0018\\
\addlinespace[0.3em]
\multicolumn{6}{l}{\textbf{v21}}\\
\hspace{1em}\cellcolor{gray!10}{4000} & \cellcolor{gray!10}{0.1354} & \cellcolor{gray!10}{0.0472} & \cellcolor{gray!10}{0.1392} & \cellcolor{gray!10}{0.0528} & \cellcolor{gray!10}{0.0005}\\
\hspace{1em}8000 & 0.1430 & 0.0343 & 0.1392 & 0.0034 & 0.0067\\
\hspace{1em}\cellcolor{gray!10}{16000} & \cellcolor{gray!10}{0.1417} & \cellcolor{gray!10}{0.0274} & \cellcolor{gray!10}{0.1392} & \cellcolor{gray!10}{0.1412} & \cellcolor{gray!10}{0.0179}\\
\hspace{1em}32000 & 0.1421 & 0.0185 & 0.1392 & 0.0012 & 0.0263\\
\hspace{1em}\cellcolor{gray!10}{48000} & \cellcolor{gray!10}{0.1408} & \cellcolor{gray!10}{0.0166} & \cellcolor{gray!10}{0.1392} & \cellcolor{gray!10}{0.0378} & \cellcolor{gray!10}{0.0217}\\
\hspace{1em}64000 & 0.1420 & 0.0148 & 0.1392 & 0.0004 & 0.0300\\
\addlinespace[0.3em]
\multicolumn{6}{l}{\textbf{v22}}\\
\hspace{1em}\cellcolor{gray!10}{4000} & \cellcolor{gray!10}{0.1171} & \cellcolor{gray!10}{0.0352} & \cellcolor{gray!10}{0.1202} & \cellcolor{gray!10}{0.2320} & \cellcolor{gray!10}{0.0798}\\
\hspace{1em}8000 & 0.1179 & 0.0283 & 0.1202 & 0.3476 & 0.0375\\
\hspace{1em}\cellcolor{gray!10}{16000} & \cellcolor{gray!10}{0.1184} & \cellcolor{gray!10}{0.0233} & \cellcolor{gray!10}{0.1202} & \cellcolor{gray!10}{0.1396} & \cellcolor{gray!10}{0.0369}\\
\hspace{1em}32000 & 0.1217 & 0.0159 & 0.1202 & 0.0798 & 0.0037\\
\hspace{1em}\cellcolor{gray!10}{48000} & \cellcolor{gray!10}{0.1200} & \cellcolor{gray!10}{0.0147} & \cellcolor{gray!10}{0.1202} & \cellcolor{gray!10}{0.7382} & \cellcolor{gray!10}{0.0298}\\
\hspace{1em}64000 & 0.1205 & 0.0137 & 0.1202 & 0.5714 & 0.0157\\
\addlinespace[0.3em]
\multicolumn{6}{l}{\textbf{VY11}}\\
\hspace{1em}\cellcolor{gray!10}{4000} & \cellcolor{gray!10}{2.0145} & \cellcolor{gray!10}{0.0331} & \cellcolor{gray!10}{2.0220} & \cellcolor{gray!10}{0.0002} & \cellcolor{gray!10}{0.0294}\\
\hspace{1em}8000 & 2.0167 & 0.0293 & 2.0220 & 0.0006 & 0.0425\\
\hspace{1em}\cellcolor{gray!10}{16000} & \cellcolor{gray!10}{2.0146} & \cellcolor{gray!10}{0.0221} & \cellcolor{gray!10}{2.0220} & \cellcolor{gray!10}{0.0000} & \cellcolor{gray!10}{0.1170}\\
\hspace{1em}32000 & 2.0169 & 0.0177 & 2.0220 & 0.0000 & 0.1316\\
\hspace{1em}\cellcolor{gray!10}{48000} & \cellcolor{gray!10}{2.0166} & \cellcolor{gray!10}{0.0178} & \cellcolor{gray!10}{2.0220} & \cellcolor{gray!10}{0.0000} & \cellcolor{gray!10}{0.1330}\\
\hspace{1em}64000 & 2.0168 & 0.0170 & 2.0220 & 0.0000 & 0.1497\\
\addlinespace[0.3em]
\multicolumn{6}{l}{\textbf{VY12}}\\
\hspace{1em}\cellcolor{gray!10}{4000} & \cellcolor{gray!10}{0.4769} & \cellcolor{gray!10}{0.0205} & \cellcolor{gray!10}{0.4777} & \cellcolor{gray!10}{0.5092} & \cellcolor{gray!10}{0.0030}\\
\hspace{1em}8000 & 0.4779 & 0.0141 & 0.4777 & 0.7318 & 0.0006\\
\hspace{1em}\cellcolor{gray!10}{16000} & \cellcolor{gray!10}{0.4765} & \cellcolor{gray!10}{0.0103} & \cellcolor{gray!10}{0.4777} & \cellcolor{gray!10}{0.0382} & \cellcolor{gray!10}{0.0206}\\
\hspace{1em}32000 & 0.4771 & 0.0085 & 0.4777 & 0.2212 & 0.0196\\
\hspace{1em}\cellcolor{gray!10}{48000} & \cellcolor{gray!10}{0.4770} & \cellcolor{gray!10}{0.0072} & \cellcolor{gray!10}{0.4777} & \cellcolor{gray!10}{0.0620} & \cellcolor{gray!10}{0.0334}\\
\hspace{1em}64000 & 0.4766 & 0.0061 & 0.4777 & 0.0000 & 0.0512\\
\addlinespace[0.3em]
\multicolumn{6}{l}{\textbf{VY22}}\\
\hspace{1em}\cellcolor{gray!10}{4000} & \cellcolor{gray!10}{1.2734} & \cellcolor{gray!10}{0.0195} & \cellcolor{gray!10}{1.2769} & \cellcolor{gray!10}{0.0010} & \cellcolor{gray!10}{0.0125}\\
\hspace{1em}8000 & 1.2741 & 0.0148 & 1.2769 & 0.0002 & 0.0121\\
\hspace{1em}\cellcolor{gray!10}{16000} & \cellcolor{gray!10}{1.2744} & \cellcolor{gray!10}{0.0101} & \cellcolor{gray!10}{1.2769} & \cellcolor{gray!10}{0.0002} & \cellcolor{gray!10}{0.0441}\\
\hspace{1em}32000 & 1.2742 & 0.0064 & 1.2769 & 0.0000 & 0.1421\\
\hspace{1em}\cellcolor{gray!10}{48000} & \cellcolor{gray!10}{1.2743} & \cellcolor{gray!10}{0.0060} & \cellcolor{gray!10}{1.2769} & \cellcolor{gray!10}{0.0000} & \cellcolor{gray!10}{0.1985}\\
\hspace{1em}64000 & 1.2739 & 0.0053 & 1.2769 & 0.0000 & 0.1473\\
\addlinespace[0.3em]
\multicolumn{6}{l}{\textbf{w11}}\\
\hspace{1em}\cellcolor{gray!10}{4000} & \cellcolor{gray!10}{0.0922} & \cellcolor{gray!10}{0.0485} & \cellcolor{gray!10}{0.0915} & \cellcolor{gray!10}{0.8262} & \cellcolor{gray!10}{0.0006}\\
\hspace{1em}8000 & 0.0885 & 0.0350 & 0.0915 & 0.0192 & 0.0001\\
\hspace{1em}\cellcolor{gray!10}{16000} & \cellcolor{gray!10}{0.0885} & \cellcolor{gray!10}{0.0227} & \cellcolor{gray!10}{0.0915} & \cellcolor{gray!10}{0.0126} & \cellcolor{gray!10}{0.0012}\\
\hspace{1em}32000 & 0.0893 & 0.0165 & 0.0915 & 0.0030 & 0.0241\\
\hspace{1em}\cellcolor{gray!10}{48000} & \cellcolor{gray!10}{0.0890} & \cellcolor{gray!10}{0.0148} & \cellcolor{gray!10}{0.0915} & \cellcolor{gray!10}{0.0004} & \cellcolor{gray!10}{0.0123}\\
\hspace{1em}64000 & 0.0887 & 0.0124 & 0.0915 & 0.0008 & 0.0251\\
\addlinespace[0.3em]
\multicolumn{6}{l}{\textbf{w12}}\\
\hspace{1em}\cellcolor{gray!10}{4000} & \cellcolor{gray!10}{0.0762} & \cellcolor{gray!10}{0.0176} & \cellcolor{gray!10}{0.0772} & \cellcolor{gray!10}{0.2620} & \cellcolor{gray!10}{0.0001}\\
\hspace{1em}8000 & 0.0771 & 0.0129 & 0.0772 & 0.7620 & 0.0004\\
\hspace{1em}\cellcolor{gray!10}{16000} & \cellcolor{gray!10}{0.0768} & \cellcolor{gray!10}{0.0083} & \cellcolor{gray!10}{0.0772} & \cellcolor{gray!10}{0.2872} & \cellcolor{gray!10}{0.0002}\\
\hspace{1em}32000 & 0.0763 & 0.0060 & 0.0772 & 0.0008 & 0.0223\\
\hspace{1em}\cellcolor{gray!10}{48000} & \cellcolor{gray!10}{0.0765} & \cellcolor{gray!10}{0.0055} & \cellcolor{gray!10}{0.0772} & \cellcolor{gray!10}{0.0590} & \cellcolor{gray!10}{0.0191}\\
\hspace{1em}64000 & 0.0759 & 0.0050 & 0.0772 & 0.0000 & 0.0286\\
\addlinespace[0.3em]
\multicolumn{6}{l}{\textbf{w21}}\\
\hspace{1em}\cellcolor{gray!10}{4000} & \cellcolor{gray!10}{0.0407} & \cellcolor{gray!10}{0.0125} & \cellcolor{gray!10}{0.0390} & \cellcolor{gray!10}{0.0004} & \cellcolor{gray!10}{0.0017}\\
\hspace{1em}8000 & 0.0404 & 0.0089 & 0.0390 & 0.0102 & 0.0197\\
\hspace{1em}\cellcolor{gray!10}{16000} & \cellcolor{gray!10}{0.0395} & \cellcolor{gray!10}{0.0066} & \cellcolor{gray!10}{0.0390} & \cellcolor{gray!10}{0.1842} & \cellcolor{gray!10}{0.0088}\\
\hspace{1em}32000 & 0.0394 & 0.0044 & 0.0390 & 0.0328 & 0.0033\\
\hspace{1em}\cellcolor{gray!10}{48000} & \cellcolor{gray!10}{0.0390} & \cellcolor{gray!10}{0.0042} & \cellcolor{gray!10}{0.0390} & \cellcolor{gray!10}{0.9548} & \cellcolor{gray!10}{0.0000}\\
\hspace{1em}64000 & 0.0390 & 0.0036 & 0.0390 & 0.9826 & 0.0003\\
\addlinespace[0.3em]
\multicolumn{6}{l}{\textbf{w22}}\\
\hspace{1em}\cellcolor{gray!10}{4000} & \cellcolor{gray!10}{0.0623} & \cellcolor{gray!10}{0.0306} & \cellcolor{gray!10}{0.0575} & \cellcolor{gray!10}{0.0000} & \cellcolor{gray!10}{0.0108}\\
\hspace{1em}8000 & 0.0610 & 0.0226 & 0.0575 & 0.0184 & 0.0380\\
\hspace{1em}\cellcolor{gray!10}{16000} & \cellcolor{gray!10}{0.0600} & \cellcolor{gray!10}{0.0165} & \cellcolor{gray!10}{0.0575} & \cellcolor{gray!10}{0.0004} & \cellcolor{gray!10}{0.0285}\\
\hspace{1em}32000 & 0.0595 & 0.0117 & 0.0575 & 0.0044 & 0.0198\\
\hspace{1em}\cellcolor{gray!10}{48000} & \cellcolor{gray!10}{0.0577} & \cellcolor{gray!10}{0.0108} & \cellcolor{gray!10}{0.0575} & \cellcolor{gray!10}{0.6112} & \cellcolor{gray!10}{0.0051}\\
\hspace{1em}64000 & 0.0581 & 0.0102 & 0.0575 & 0.0838 & 0.0014\\
\bottomrule
\end{longtable}

\end{document}